\documentclass[12pt]{article}
\usepackage{geometry}
\usepackage{caption}
\usepackage[greek,english]{babel}

\usepackage{stmaryrd}

\usepackage[usenames,dvipsnames]{color}
\newcommand{\red}[1]{\textbf{\color{red}#1}}

\usepackage{amsfonts}
\usepackage{amsmath}
\usepackage{amssymb}
\usepackage{amsthm}
\usepackage{mathrsfs}
\usepackage{bbm}
\usepackage{bm}
\usepackage{cancel}
\usepackage{graphicx}
\usepackage{caption}
\usepackage{subcaption}
\usepackage{fancyhdr}
\usepackage{mathrsfs}
\usepackage{wrapfig}

\def\lmp{\langle\hspace{-3pt}\langle}
\def\rmp{\rangle\hspace{-3pt}\rangle}

\def\ldm{\{\hspace{-4pt}\{}
\def\rdm{\}\hspace{-4pt}\}}

\def\lsymb{(\hspace{-3pt}(}
\def\rsymb{)\hspace{-3pt})}

\def\dimp{\pi} 

\def\p{\omega}  

\newcommand{\st}[1]{\text{\tiny \rm #1}}

\def\e{\emph}
\def\nb{$N$-body problem~}
\def\nbn{$N$-body problem}
\def\tb{3-body problem~}
\def\tbn{3-body problem}

\def\Q{{\sf Q}}
\def\QR{{\sf Q}_\textrm R}
\def\shs{{\sf S}}
\def\pshs{{\sf PS}}

\def\Sim{{\sf Sim}}

\def\Riem{{\sf Riem}}
\def\Sup{{\sf Sup}}
\def\CSup{{\sf CS}}

\def\Dil{{\sf Dil}}
\def\Transl{{\sf Transl}}

\def\cs{$C_\st{S}$~}
\def\csn{$C_\st{S}$}
\def\bq{\begin{equation}}
\def\ee{\end{equation}}
\def\m{$I_\st{cm}$~}
\def\mn{$I_\st{cm}$}
\def\vs{$V_\st{S}$~}
\def\vsn{$V_\st{S}$}
\def\iv{$V_\st{New}$~}
\def\ivn{$V_\st{New}$}
\def\id{, i.e., }

\def\dimt{\zeta} 
\def\yorkt{\tau} 

\usepackage{hyperref}

\hypersetup{
   colorlinks=true,         
   linkcolor=blue,          
   citecolor=red,        
   urlcolor=Violet            
}

\usepackage{tocloft}

\begin{document}

\title{\Large \textbf{A Gravitational Origin of the Arrows of Time} }

\author{{Julian Barbour,$^{1,2}$
Tim~Koslowski,$^{3}$ and Flavio~Mercati,$^4$
{\rm }
}\vspace{12pt} \\
\it \small $^1$College Farm, South Newington, Banbury, Oxon, OX15 4JG UK,\\
\it \small $^2$Visiting Professor in Physics at the University of Oxford, UK.\\
\it \small  $^3$ University of New Brunswick, Fredericton, NB, E3B 5A3 Canada,\\
\it \small $^4$Perimeter Institute for Theoretical Physics, 31 Caroline Street North,\\
\it \small Waterloo, ON, N2L 2Y5 Canada,}
\date{\today}

\maketitle

\begin{abstract}
The only widely accepted explanation for the various arrows of time that everywhere and at all epochs point in the same direction is the `past hypothesis': the Universe had a very special low-entropy initial state.  We present the first evidence for an alternative conjecture: the arrows exist in all solutions of the gravitational law that governs the Universe and arise because the space of its true 
degrees of freedom (shape space) is asymmetric. We prove our conjecture for arrows of complexity and information in the Newtonian $N$-body problem. Except for a set of measure zero, all of its solutions for non-negative energy divide at a uniquely defined point into two halves. In each a well-defined measure of complexity fluctuates but grows irreversibly between rising bounds from that point. Structures that store dynamical information are created as the complexity grows. Recognition of the division is a key novelty of our approach. Each solution can be viewed as having  
a single past and two distinct futures emerging from it. Any internal observer must be in one half of the solution and will only be aware of one past and one future. The `paradox' of a time-symmetric law that leads to observationally irreversible behaviour is fully resolved. 
General Relativity shares enough architectonic structure with the \nb for us to prove the existence of analogous complexity arrows in the vacuum Bianchi IX model. In the absence of non-trivial solutions with matter we cannot prove that arrows of dynamical information will arise in GR, though they have in our Universe. Finally, we indicate how the other arrows of time could arise.

\end{abstract}

\newpage

\tableofcontents

\newpage

\section{Introduction}

\begin{quote}\small

{\e{``It seems to me that the idea of trying to obtain a universe in the form we know it by applying time-symmetric physics to a generic unconstrained initial state is basically misconceived.''}} R. Penrose \cite{Penrose1989}.

\end{quote}\normalsize

\subsection{Doubts about existing approaches \label{doubts}}

Most discussions of the various arrows of time concentrate on the growth of entropy. This is natural; the entropy arrow was the first that attracted widespread interest. It is also the one most readily observed, as when we drop a glass and know it cannot be reassembled. However, we question whether the entropy concept and statistical mechanics, which are undoubtedly excellent for characterizing and understanding subsystems of the Universe, are appropriate for the Universe  as a whole. These are some of our reasons:

\begin{enumerate}

\item Boltzmann and Gibbs developed the entropy concept to describe \emph{non-gravitating} particles in a \emph{confined space}, e.g., a box, and based it on phase-space \emph{volume}. Are the italicized aspects appropriate for the Universe? Gravity dominates it, no walls confine it, and volume presupposes an external scale. 

\item By assuring probability conservation, Liouville's theorem for ensembles in phase space provides the foundation of statistical mechanics for dynamical systems evolving wrt an external time. But the Universe is a unique system with a unique history and all physical clocks are subsystems of it.

\item Self-gravitating systems are `anti-thermodynamic'. They have \e{negative} heat capacity and cannot equilibrate. Instead of making non-uniform systems more uniform, gravity fosters clustering, i.e., complexity. This has long been recognized, but to the best of our knowledge no one has hitherto quantified the effect. We shall.

\item Although black holes and other solutions of Einstein's equations with horizons have remarkable thermodynamic properties and suggest an intimate connection between gravity, entropy and quantum dynamics, the general covariance of GR has defeated attempts to define gravitational entropy in generic situations.

\item The entropy concept is often illustrated in configuration space {\sf Q} alone, e.g., atoms initially confined to a small region then spread out over a complete box. In the great majority of the naturally occurring far-from-equilibrium objects in the Universe, the momenta are effectively random and the disequilibrium is manifested almost entirely in {\sf Q}, which may therefore be more relevant than phase space. 
 
\end{enumerate}

These are the main reasons why we seek a new way to understand the arrows of time. Specifically, we suggest that three-dimensional (3D) \e{{scale-invariant} configurational complexity} is the fundamental concept that should be studied in the first place. We define such a concept for Newtonian gravity and show that it exhibits striking irreversible behaviour. We  also propose a candidate analogue for GR and indicate how our framework could explain not only the Universe's manifest complexity arrow but also the other arrows. {Our approach has similarities to the Weyl curvature hypothesis of Penrose \cite{Penrose1979, Penrose1989}, differing mainly in seeking a notion of 3D complexity rather than 4D entropy. We do also think his belief quoted above may be too pessimistic.}

\subsection{Our conceptual framework and main results \label{intro2}}

Our fundamental assumptions are:

\begin{enumerate}

\item \emph{The Universe is a closed dynamical system.} In Newtonian gravity (NG), this means an `island universe' of $N$ point particles. In general relativity (GR), the Universe must be spatially closed. This is not in conflict with current observations.

\item \emph{A notion of universal simultaneity exists.} This is built into NG. In GR, we rely on the theory of Shape Dynamics (SD), discussed below, to supply this notion.

\item \emph{Since all measurements are relational, only shapes are physical.} We will define the \emph{complexity}, denoted $C_\st{S}$, of any complete shape of the Universe. It is a pure number.

\end{enumerate}

Our basic arena is \emph{shape space}, denoted $\shs$. In NG, it is obtained by quotienting the standard Newtonian configuration space {\sf Q} by Euclidean translations, rotations and dilatations, i.e., wrt the similarity group. Then for the \nb its $3N$-dimensional {\sf Q} is replaced by the $3N-7$-dimensional $\shs$. Shape space for GR will be introduced later.

The guiding principle of SD is to abstract from dynamics all external structures. In NG, these are position, orientation and size in an inertial frame
and also an external time. A dynamical history is then simply an unparametrized curve {\sf c} in $\shs$. The equations of both NG and GR are time-reversal symmetric, so they define no orientation on {\sf c}. This is why the entropy and other arrows are a problem. However, for NG we show that its generic\footnote{Throughout the paper, we use generic to denote the sets of solutions that are not of measure zero.} solutions for non-negative energy, $E\ge 0$, exhibit `two-sided' time asymmetry, namely, they divide into two halves in each of which $C_\st{S}$ fluctuates but overall grows irreversibly from a common minimum. Moreover, structures that store dynamical information are created as $C_\st{S}$ grows. In our Universe, we identify the direction to the future with the arrows of increasing complexity and information. On this basis, since the minimum divides the solution effectively into distinct halves, two `directions of time', pointing away from the minimum, exist in all generic NG solutions. We are not aware that this `one-past--two-futures' structure has hitherto been noted or related to the arrows of time.

We first demonstrate the structure in conventional Newtonian terms and then derive the equations that determine the evolution curve in $\shs$. We believe this provides strong evidence that shape space is the arena in which to study all the arrows of time. Our long-term aim is to show that the \e{past hypothesis} (that the arrows of time can only be explained by an exceptionally low-entropy birth of the Universe) is unnecessary. Instead, we shall suggest that the arrows all have their origin in \emph{an asymmetry of shape space} {\sf S}.

We see support for our conjecture in the form of the evolution in {\sf S} in each of the above halves, which is \e{asymmetric} in time and \e{dissipative}\footnote{Irreversibility of the dynamics of the true conformal degrees of freedom of GR, manifested as monotonic decrease of the reduced Hamiltonian, has been noticed before, and  was exploited by Fischer and Moncrief in a study of attractors of the motion \cite{Fischer-Moncrief}. In vacuum GR we actually find anti-dissipation (Sec.~\ref{sec:TimeAsymmetryInDynamicalGeometry}).} in the naturally defined direction of increasing complexity. This is so despite the absence of `hidden' microscopic degrees of freedom of the kind that normally give rise to irreversible behaviour. We think the time-asymmetric evolution could be related to the deterministic laws of black hole thermodynamics found in classical GR in the late 1960s. It might also be a manifestation of hidden degrees of freedom and an entropic origin of gravity. Whatever the truth, the dissipation is a mathematical fact and a direct consequence of our fundamental ontology, in accordance with which only shape evolution is physical. 

We should like to emphasize here that in both NG and GR  the physical degrees of freedom (dofs) with which we are concerned are heterogeneous in nature. There are purely dimensionless shape dofs and \emph{one} dimensionful scale dof.\footnote{Machian arguments allow us to eliminate translational and rotational degrees of freedom.}
Moreover, under a physically reasonable restriction, its conjugate momentum, unlike all the momenta of the shape dofs, is monotonic along the solution curve. Its existence as a unique Lyapunov function in the \nb has long been known. It has an equally striking counterpart in GR, called the York time. The direction of increase of these two Lyapunov functions is conventional, so their existence does not conflict with the time-reversal symmetry of the laws that define them.

However, since the scale dof is unique and, being dimensionful, can only be given a value if an external scale is present, it literally `cries out' for a role distinct from the shape dofs. Through the operation of deparametrization, which we explained in detail in \cite{BKMpaper}, we transform the scale variable into the Hamiltonian and its monotonic conjugate momentum into the evolution parameter. We are left with the \e{minimal} set of variables needed to describe the Universe objectively. Any attempt to remove more would bring down the whole structure.

The transition to this optimal (fully reduced) description automatically removes from the system the scale kinetic energy present in the conventional description and explains why the dynamics in $\shs$ is dissipative. The potential significance of our result stands or falls with our ontology. We ask readers who suspect we have created an artefact by tampering with hallowed principles to bear in mind that Einstein was led to create GR precisely in order to eliminate external background structures from physics. We are suggesting that one last step needs to be taken: the elimination of external scale.

In Sec.~\ref{sec:TimeAsymmetryInDynamicalGeometry}, we consider GR. For reasons that we shall spell out, we cannot as yet obtain results as definitive as for the \nbn. However, we find enough similarities to encourage us to believe that in this much more realistic context the route to an understanding of the arrows of time is through study of the problem in a suitably defined shape space. 
The main argument and the novel aspects of our approach can be understood without reading Sec.~\ref{sec:TimeAsymmetryInDynamicalGeometry}
which discusses the application to dynamical geometry. However, for readers unfamiliar with GR in its Hamiltonian formulation, 
we briefly introduce some background in Sec.~\ref{ShapeSpaceForDynamicalGeometry}.

\section{Time Asymmetry in Particle Dynamics}
\label{sec:TimeAsymmetryInParticleShapeDynamics}

\subsection{Generic solutions\label{TheNewtonianSolutionsSubSubSection}}

We here review facts about the \nbn\footnote{{See Chenciner's \cite{Chenciner1998} for a rigorous review. Marchal \cite{Marchal1990}} treats the \tb in detail. {Sundman \cite{Sundman1913}} first established the 3-body behaviour described below.} as formulated in the `scaffolding' of an inertial reference frame, external clock and reference scale. We call this the \e{coordinatized description} and contrast it with the \e{objective description}. This latter is obtained by abstracting away everything that is not unambiguously intrinsic to the system. All that remains are the dimensionless mass ratios of the particles and the successive shapes through which the system passes in shape space.\footnote{This is the conceptual framework of Shape Dynamics. For details, including its origin in Machian considerations, see \cite{barbour:nature,barbourbertotti:mach,barbour_el_al:rel_wo_rel,Barbour_Niall:first_cspv,barbour_el_al:physical_dof,gryb:shape_dyn, Gomes:linking_paper,Koslowski:ObservableEquivalence,BKMpaper,JuliansReview,
FlaviosSDtutorial}.} The elimination of the strictly redundant part of the coordinatized description reveals effective asymmetry in $\shs$.

We begin with the qualitative behaviour of the 3-body problem in the coordinatized description. Since we use this as a toy model for the 
Universe, we limit ourselves on Machian grounds to the zero-angular momentum, ${\bf J}_\st{tot} =0$, and zero-energy, $E=0$, case. By Galilean invariance, we can always assume that the momentum $\bf P_\st{tot}$ vanishes.\footnote{\label{Mprin}\e{Best matching} (see \cite{JuliansReview,FlaviosSDtutorial} for details) shows that the conditions ${\bf P}_\st{tot}={\bf J}_\st{tot}=0$ ensure overall translation and rotation of the Universe make no contribution to its action, as Mach's principle requires.}

\begin{figure}[h]
\begin{center}
\includegraphics[width=0.64\textwidth]{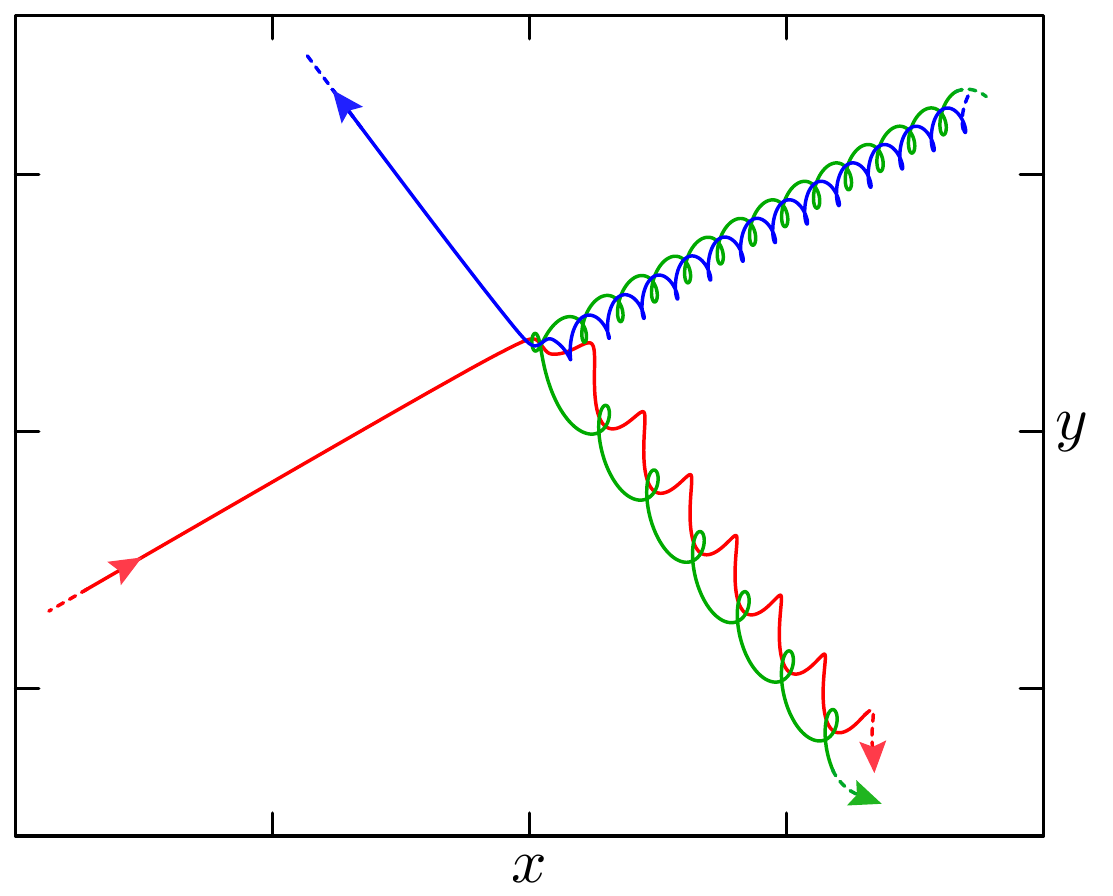}
\end{center}
\caption{\it \it \small Orbits of the three particles in a pair exchange process in the centre-of-mass inertial frame. Because we assume ${\bf J}_\st{tot}=0$ and consider the \tbn, the motion is planar. The orientation of the axes and the 
units are arbitrary. Time-reversal symmetry makes it possible to read two `pair-swapping' stories in one picture. With the option shown by the arrows, a boy (red, coming from bottom left) meets a Kepler pair dancing in from the top right, grabs the girl (green) and goes off with her bottom right, ensuring momentum conservation if not happiness for the jilted boy (blue). But reverse the arrows, and the blue boy gets the girl. {The cameo story shows how illusory it is to say some initial condition `causes' what happens later. For a time-symmetric system, the solutions do not have an initial condition.} They each have their own overall structure encoded equally well in any phase-space point along the solution.}\label{PairExchangeCenterOfMomentumFrame}
\end{figure}

All generic three-body solutions with $E={\bf J}_\st{tot} =0$ have a period of nontrivial three-body interaction that develops asymptotically in both time directions into \e{hyperbolic--elliptic escape} in which a pair of particles (not necessarily the same in the two time directions) separates from the
third. As the pair becomes more and more isolated, its motion is 
ever better approximated by elliptical Keplerian motion. In the meantime, the third particle, the `escapee', tends to increasingly undisturbed inertial motion directed away from the pair. This behaviour is illustrated in Fig.~\ref{PairExchangeCenterOfMomentumFrame}.

Being time-reversal symmetric, Newton's equations do not define any temporal ordering on the {complete} orbit. However, we shall show that there exists a unique point on it at which the \e{dilatational momentum}\footnote{Coined in \cite{barbour:scale_inv_particles} by analogy with angular momentum, which has the same dimensions. It has not been named in the $N$-body literature, but is generally denoted by $J$, probably for Jacobi.}
\begin{equation}
D = \sum_{a=1}^3 {\bf r}_a^\st{cm} \cdot {\bf p}^a_\st{cm} \,,\label{dilmtm}
\end{equation}
where ${\bf r}_a^\st{cm}$ and ${\bf p}^a_\st{cm}$ are the centre-of-mass coordinate and momentum vectors of the particles, vanishes. This point divides the orbit into two halves and, as we shall show, serves as a `past' for each half, both of which have an infinitely distant `future' as the Kepler pair and escapee drift forever apart.\footnote{Since only ratios have meaning in shape space, the objective fact is that the ratio between 
the semi-major axis of the pair and the distance of the third particle from the pair tends to infinity.} Thus every generic $E={\bf J}_\st{tot}=0$ solution has `one past and two futures'. This is also true of all generic $N$-body solutions with $E={\bf J}_\st{tot}=0$. We defer discussion of the residual measure-zero solutions, which exhibit quite different behaviour. 
 
\subsection{Definition and growth of the complexity \label{exc}}

We wish to define the complexity $C_\st{S}$ as a scale-invariant, and hence dimensionsless, function on {\sf S}. A simple way is to make \cs the ratio of two `democratically' mass-weighted lengths. If  $m_a$ is the mass of particle $a$, and $\bf r_\st{a}$ is its position vector, an obvious candidate for one is the \e{root-mean-square length} $\ell_\st{rms}$:
\bq
\ell_\st{rms}:={1\over m_\st{tot}}\sqrt{\sum_{a<b}m_am_b\,r_{ab}^2},~~m_\st{tot}=\sum_{a=1}^Nm_a,~~r_{ab}=\|{\bf r}_b-{\bf r}_a\|.\label{rms}
\ee
Another is the \e{mean harmonic length} $\ell_\st{mhl}$:
\bq
{1\over\ell_\st{mhl}}:={1\over m_\st{tot}^2}{\sum_{r<a}{m_am_b\over r_{ab}}}.\label{mhl}
\ee
Then the \e{complexity}, a pure number that depends only on $N$ and the mass ratios, is
\bq
C_\st{S}:={\ell_\st{rms}\over\ell_\st{mhl}}.\label{comp}
\ee

We are not aware of an earlier proposal for this purpose, but it is easy to see that $C_\st{S}$ is a good measure of non-uniformity and hence complexity. Even for relatively small $N$, $\ell_\st{rms}$ (\ref{rms}) changes little if two particles approach each other or even coincide. In contrast, $\ell_\st{mhl}$ is sensitive to any clustering and tends to zero if that happens. Moreover, while $C_\st{S}$ grows with clustering, Battye et al's  \cite{BattyeGibbons} numerical calculations show that the minima of $C_\st{S}$ up to $N\approx 10^4$ correspond to extraordinarily uniform (super-Poissonian) shapes.\footnote{Conceptually at least, our definition bears no obvious resemblance to Kolmogorov complexity defined by the number of binary digits needed in an algorithm to generate a given distribution. It is obvious that one could define more sophisticated measures of complexity than (\ref{comp}), e.g., ones that take into account alignments, but (\ref{comp}) appears to be the most appropriate as a measure for a self-gravitating universe.}

\begin{figure}[t]
 \begin{center}
 {\includegraphics[width=0.6\textwidth]{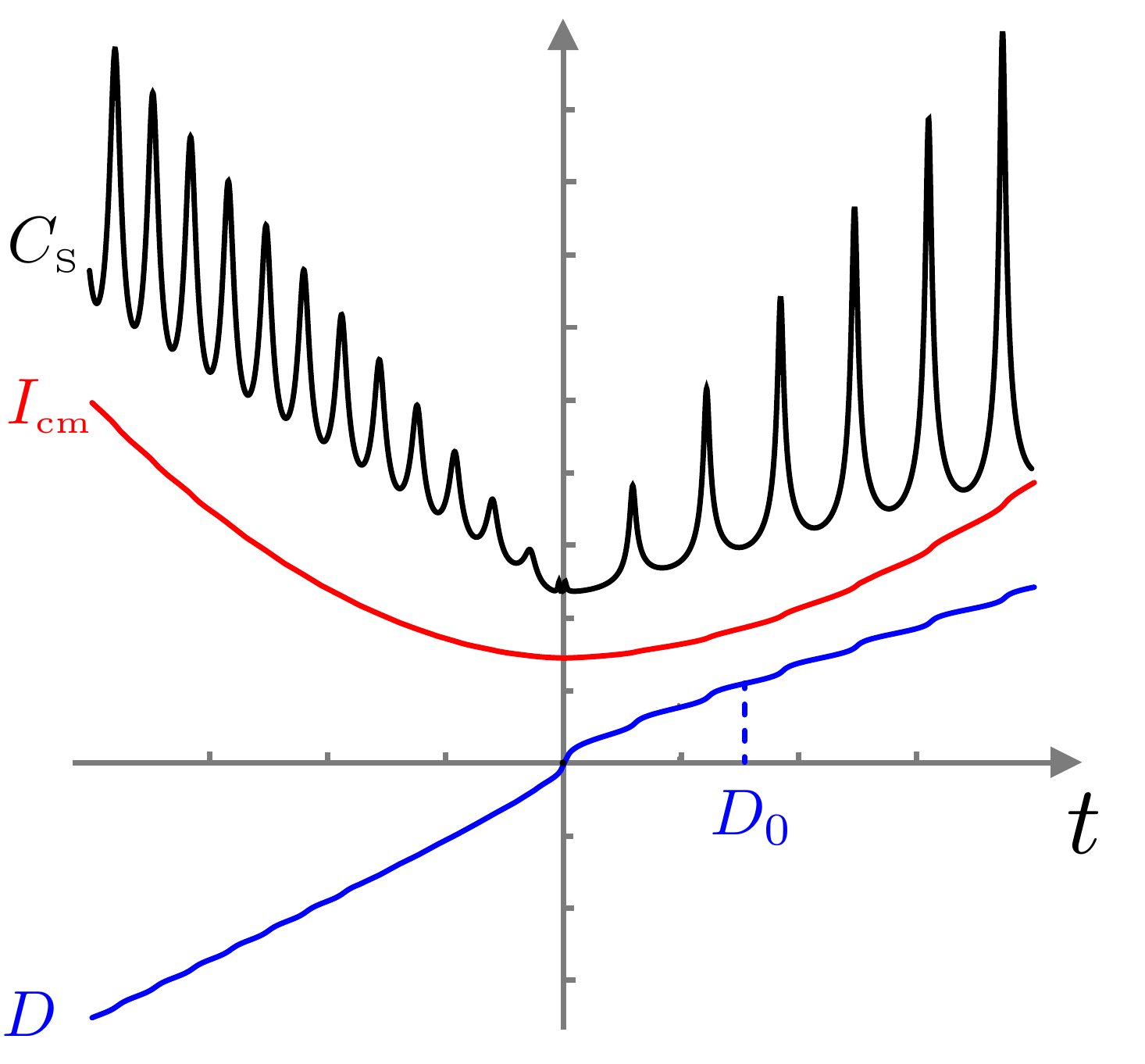}}
                
                \label{Cs-I-D-Newtoniantime}
                \end{center}
\caption {\it \small The moment of inertia $I_\st{cm}$ (red), dilatational momentum $D$ (blue), and complexity $C_\st{S}$ (black) as functions of the Newtonian time for the solution of the \tb shown in Fig.~\ref{PairExchangeCenterOfMomentumFrame}. The different heights of the fluctuations on the two sides of the figure reflect the different orbital elements of the corresponding Kepler pairs and escapees. Note that in the asymptotic regions, where the orbital parameters stabilize, numerical calculation of the evolution in either time direction is relatively easy but becomes much harder when the three-body interactions become non-trivial. Given asymptotic data on one side, it is not easy to predict with any accuracy the behaviour on the other. The two sides are effectively different worlds.}
\label{Figure3}
\end{figure}


Apart from the division by $1/m_\st{tot}^2$ and the absence of the constant $G$, $\ell_\st{mhl}$ is, of course, the Newton potential. Less obvious is that $\ell_\st{rms}$ is, the normalization apart, the square root of the centre-of-mass moment of inertia $I_\st{cm}$. This follows from the identity
\bq
{1\over m_\st{tot}}\sum_{r<a}m_am_br_{ab}^2\equiv \sum_{a=1}^Nm_a \|  \mathbf r_a - \mathbf r_\st{cm} \|^2  \,:=I_\st{cm}, ~~  \mathbf r_\st{cm}
= \sum_{a=1}^N  {\textstyle  \frac{m_a}{m_\st{tot}}}\bf r_a.
\ee
Thus, the complexity is formed from the two most fundamental quantities in Newtonian gravitational dynamics. Note also that $D$ (\ref{dilmtm}) is \e{half the time derivative of} $I_\st{cm}$.

Figure~\ref{Figure3} gives a first hint why we study $C_\st{S}$. It fluctuates but has a clear tendency to increase between growing bounds either side of the central region of minimal \mn. It is easy to see why: first, the escapee's increasing separation leads to asymptotic linear growth of \mn; second, the Kepler pair forms with eccentricity, and the varying separation of its constituents causes \iv to fluctuate. Behind the behaviour of $C_\st{S}=\sqrt{I_\st{cm}}V_\st{New}$ we directly see the cause. It is not some initial condition but the effect of law.

\begin{figure}[t]\begin{center}
\includegraphics[width=0.6\textwidth]{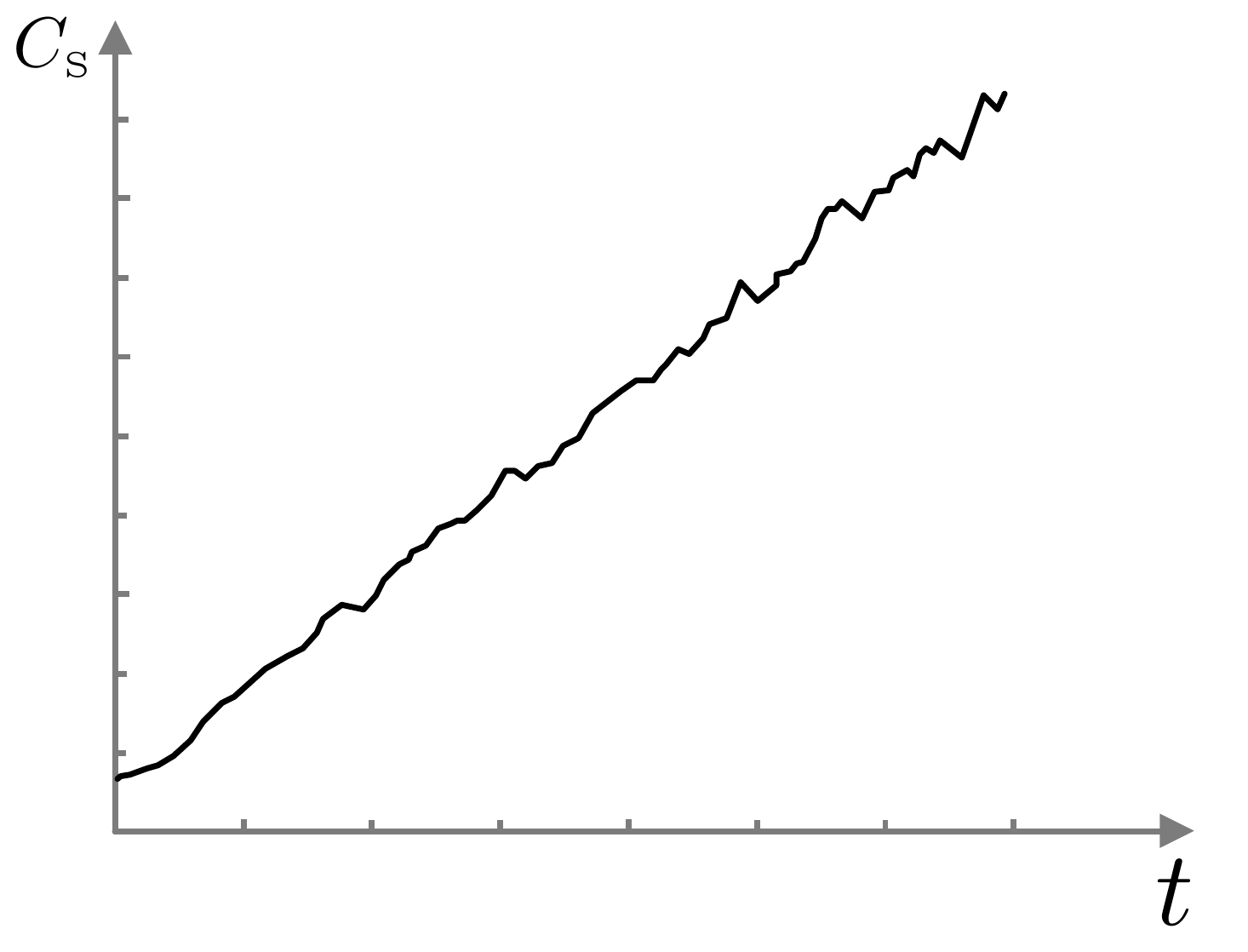}
\caption{\it \small The complexity $C_\st{S}$ vs.~Newtonian time for a typical solution with $N\sim1000$. {The initial state was a Gaussian distribution of all coordinates and velocities around the origin.} It clearly exhibits a linear growth, and, {as was to be expected from the increase in $N$,} the fluctuations due to the eccentricity of the Keplerian orbits average each other, giving a smoother curve. Time reversal (implemented by reversing the initial velocities but not shown) leads to qualitatively similar behaviour on either side of the minimum of $C_\st{S}$ (as in Fig.~\ref{Figure3}).}
\label{1000particles}
\end{center}
\end{figure}

The fluctuations in \cs for the \nb with large $N$ are much weaker (Fig.~\ref{1000particles}). An initial cluster of particles `evaporates' (in both time directions), forming quasi-bound few-particle systems and some stable Kepler pairs. As the system disperses, $\ell_\st{rms}$ grows steadily, while the Kepler pairs and quasi-bound systems, whose phases are uncorrelated, ensure that $|V_\st{New}|=\ell_\st{mhl}^{-1}$ declines to a more or less stable asymptotic value.\footnote{The \e{deterministic} manifestation of `two-sided' arrows of time described here should be compared with Boltzmann's suggestion, made in a non-gravitational context, of rare deep fluctuations out of statistical equilibrium, in which intelligent beings can only exist near the bottom of an entropy fluctuation. If present on both sides, each would regard the entropy minimum as lying to their past. Note that the Boltzmann fluctuations reoccur infinitely often, whereas there is just one pair of arrows of time in each of the deterministic solutions we consider. Moreover, the entropy arrow points to a `heat-death' future, whereas the complexity arrow points in the direction of greater structure.}

Marchal and Saari \cite{Marchial:1976fi} obtained rigorous results which show that this must happen. Subject to certain caveats (see Appendix~\ref{MS}), the basic reason is that the $N$-body system breaks up into subsystems whose centres of mass separate linearly with the Newtonian time $t$ in the asymptotic limit. Each subsystem consists of individual particles and clusters whose constituents remain close to each other. The separations within a subsystem are bounded by $O(t^{2/3})$. Thus, if particle $a$ belongs to cluster $\mathcal J$, we have\,\footnote{\label{xias} If $N>3$, there can also be super-hyperbolic solutions, a well known example of which is Xia's 5-body solution \cite{Xia1992}, in which four particles reach infinity in finite time. However, as Edward Anderson pointed out to us, these are not compatible with special relativity, so we do not consider them.}
       \begin{equation}\label{MSA}
         {\bf  r}_a= {\bf c}_{\mathcal J}\,t\,+\,\mathcal O (t^{2/3}), \qquad \text{if}~a \in \mathcal J\,,
       \end{equation}
where ${\bf c}_{\mathcal J} \in \mathbbm{R}^3$ is a constant vector. It follows immediately that $\sqrt{I_\st{cm}}$ must grow linearly with $t$, while \iv declines not faster than $t^{-2/3}$, so that their product \cs grows on average at least as $t^{1/3}$. In fact, the formation of at least one asymptotically stable Kepler pair will ensure that \iv asymptotes to a constant average, so that \cs grows linearly.

Although we have as yet only shown how $C_\st{S}$ grows asymptotically in Newtonian gravity, it seems to be an excellent candidate measure of complexity in our actual Universe, in which non-gravitational forces also play an important role. The protons and other nuclei together with possible dark matter particles can be taken to represent the bulk of the inhomogeneously distributed matter in the Universe. Suppose that at each instant of cosmic time since last scattering of the CMB the shortest spatial geodesic distances between all of them were determined  in spacelike hypersurfaces\footnote{Assumed to foliate the complete Universe, taken to be spatially closed.} in which the CMB is at 
rest on average. Let the obtained values be inserted as the inter-particle distances $r_{ab}$ in the expression (\ref{comp}) for the complexity \csn. Even allowing for uncertainty about the fate of matter in black holes, it will surely be the case that this \cs for the Universe will have increased on average monotonically to an extremely good accuracy from last scattering to the present epoch.

\subsection{Definition and growth of information}

We now want to suggest that an \e{arrow of information growth} also emerges
generically. Of course, we must first define information. We adhere to the ideas sketched by one of us in the essay \cite{BitFromIt} and assume that all kinds of information (factual, semantic and Shannon) have a physical basis and that their existence is tied to the availability of an adequately rich physical substrate. Here we are concerned with deterministic processes, so we leave the discussion of Shannon information, which is about probabilities, for later studies. 

For the purposes of this paper, the results that we have so far obtained lead us naturally to an information-theoretical identification of complexity as `the necessary condition for a system to store recognizable local information'. This enables us to synthesize the concepts of complexity and information, in the sense that `complexity is potential information'. The \nb provides a chance to test this conjecture using our intuitive (and quantitative!) notion of \emph{configurational} complexity as defined by $C_\st{S}$. Indeed, we now conjecture that there is an equally intuitive notion of \emph{dynamical} information. This is suggested by the dynamics of the system, which, evolving in the direction of our arrow of time, can spontaneously create subsystems that, as Marchal and Saari \cite{Marchial:1976fi} show, become increasingly isolated from the rest of the Universe. As this happens, the subsystems develop (approximate) Galilean symmetries with which
there are associated seven conserved quantities: the linear and angular momenta ${\bf P}_{\mathcal J}$, ${\bf J}_{\mathcal J}$, 
and the energy $E_{\mathcal J}$ of the clusters (the index $\mathcal J$ identifies the cluster). On Machian grounds, the whole Universe is constrained to have a vanishing value for these quantities in an inertial frame, but subsystems are not, so that any non-zero value of ${\bf P}_{\mathcal J}$, ${\bf J}_{\mathcal J}$ and $E_{\mathcal J}$ has to be compensated by an equal and opposite value of the same quantities for the rest of the Universe. 

The key observation is that when a subsystem becomes isolated, it develops the seven symmetries and corresponding conserved quantities $E_{\mathcal J}$, ${\bf P}_{\mathcal J}$ and ${\bf J}_{\mathcal J}$ that, as shown in \cite{Marchial:1976fi}, are more and more accurately conserved. With time both the number $n$ of clusters ${\mathcal J}=1,...,n$ and the number of `frozen' digits of $(E_1,{\bf P}_1,{\bf J}_1,...,E_n,{\bf P}_n,{\bf J}_n)$ increases, implying that the number of digits reliably stored in subsystems increases.  We propose that the total amount of data `saved' in the `frozen' digits measures the information content of the system. By the results of \cite{Marchial:1976fi}, this increases in time together with the measure  $C_\st{S}$ of configurational complexity.
 
Moreover, one can also say that physical rods and clocks emerge spontaneously in the form of Kepler pairs. If they are to have utility, rods must remain mutually congruent and clocks must remain in phase -- they must march in step \cite{Barbour:nature_of_time}. This is what happens when Kepler pairs form. Their semi-major axes become mutually fixed with ever greater precision and therefore serve as rods, while the areas swept out by the major axes measure time concordantly in accordance with Kepler's second law. Both information and the means to measure it emerge dynamically and generically.

Of course, Kepler pairs do not meet all the criteria of metrology since two such pairs would disrupt each other when in close proximity and the very essence of measurement is the bringing of a rod and the measured interval into overlap. Metrology now relies on quantum mechanics and the great weakness of gravity compared with the other forces. We return to this question in Sec.~\ref{erc}.

\subsection{Dynamical similarity}

We now want to understand, at the most basic level, the behaviour described in the previous subsections and shown in Figs.~2 and 3. The characteristic features are the U-shaped graph of \m and the fluctuating growth of \cs either side of $D=0$.

The behaviour of \m is easily explained and has long been known. As the first \e{qualitative} result in dynamics, Lagrange discovered it over 200 years ago. It relies on two architectonic properties of the Newton potential. 

The first is \e{homogeneity}: if for any dynamical system and any real constant $k$ the potential satisfies $V(\alpha \, {\bf r}_a)=\alpha^k \, V(\mathbf r_a)$, then it is homogeneous of degree $k$ and \e{dynamical similarity} holds: the equations of motion permit a series of geometrically similar paths (\cite{Landau-Lifshitz}, p.~22), in which the times between corresponding points satisfy $t'/t=(l'/l)^{1-k/2}$ if the distances are scaled as $l'/l$. The best known example of this is Kepler's third law,\footnote{Dynamical similarity is also the basis of the virial theorem \cite{Landau-Lifshitz}.} for which $1-k/2=3/2$ and the periods of planets of the same eccentricity (and therefore shape) but different semi-major axes $a$ scale as $a^{3/2}$. The dynamical similarity in the \nb will be crucial below: it shows that, if (as for a dynamically closed universe) external standards of duration and scale are unavailable, then a one-parameter family of solutions in the coordinatized description collapses to a single curve in $\shs$. 

The homogeneity of degree $k$ of any potential also leads to the relation
\begin{equation}
\ddot I_{\st{cm}}=4E-2(k+2)V \,,\label{lagjac}
\end{equation}
which is often called the Lagrange--Jacobi relation. Its derivation uses Newton's second law and Euler's homogeneous function theorem.

We now come to the second important property of \ivn. Besides having $k=-1$, it is also negative definite. These two properties enable us to particularize (\ref{lagjac}) as follows:
\begin{equation}
\ddot I_{\st{cm}}=4E-2(k+2)V\Longrightarrow 4E-2V_\st{New} > 0~\textrm{if}~E\ge 0.\label{concave}
\end{equation}

Thus, if $E\ge 0$ it follows that $\ddot I_{\st{cm}}=2\dot D$ is positive [$D$ is defined in (\ref{dilmtm})]. Then $I_{\st{cm}}$ is concave upward, $\dot D$ is positive and $D$, whose sign is conventional, is monotonic.\footnote{If $k=-2$, then $ \ddot I_{\st{cm}} = 0$. 
This case, studied in \cite{barbour:scale_inv_particles,BLMpaper}, also plays a role below.} 
Since Figs.~\ref{Figure3} and \ref{1000particles} are based on calculations with $E=0$, this immediately explains the U-shaped behaviour of \mn, which is also bound to occur if $E>0$. 

Deferring for a moment the exceptional case in which \m reaches zero, it follows from its upward concavity that \m must tend to infinity in both time directions. This requires either one particle to recede infinitely far from the other two, which leads to the hyperbolic--elliptic escape described above, or all inter-particle separations to tend to infinity at the same time. This is also an exceptional case and will be considered below. As for the behaviour of \cs in Fig.~\ref{Figure3}, we have seen that this is directly due to the formation of Kepler pairs and escape of the third particle: the generic behaviour of the \tb with $E\ge 0$ is inevitable. Appendix~\ref{MS} shows this is also true for the $N$-body problem.

Let us here say something about our Machian assumption that the Universe has $E={\bf{P}}_\st{tot}={\bf{J}}_\st{tot}=0$ (in its centre-of-mass inertial frame). We noted in footnote \ref{Mprin} that the conditions ${\bf{P}}_\st{tot}={\bf{J}}_\st{tot}=0$ ensure that translation and rotation of the Universe as a whole make no contribution to its action. Moreover, solutions with $E={\bf{J}}_\st{tot}=0$ are important in $N$-body theory because they are \e{scale invariant}: if $E$ or $\bf{J}_\st{tot}$ is non-vanishing, its value changes under a change of units, but zero is obviously invariant. Of greater relevance to us is a corresponding reduction in the number of degrees of freedom. The exact number is important, so we do a count. We start with $3N$. By Galilean relativity, the centre-of-mass coordinates have no effect on the inter-particle separations, so that brings us down to $3N-3$. Next ${\bf J}_\st{tot}=0$ eliminates two,\footnote{Not three because the rotation group is non-Abelian: there are only two commuting quantum angular-momentum observables.} so we reach $3N-5$. Dynamical similarity and the condition $E=0$ enable us to make the final reduction below to $3N-7$ shape dofs and a time variable based on the dilatational momentum.

One more comment. Newton's equations have the same form in ${\sf Q}$ independently of the values of $E$ and $\bf{J}_\st{tot}$, but the objective equations in $\shs$ are very different. 
The reader may think $E={\bf{J}}_\st{tot}=0$ is merely a special initial condition `put in by hand'. However, we treat the \nb as a model `island Universe'. It is important that the Universe, as opposed to subsystems of it, is unique. The equations that describe such a universe objectively in $\shs$ have different, significantly more complicated forms if $E,{\bf{J}}_\st{tot}\ne 0$ as compared with the case ${\bf J}_\st{tot} =E=0$. Above all, if $E$ and $\bf{J}_\st{tot}$ are non-vanishing the Universe in its evolution responds not only to the structure of $\shs$ but also to external structures. Moreover, this case matches the basic structure of closed-space vacuum GR (which we show in Sec.~\ref{sec:TimeAsymmetryInDynamicalGeometry}) and provides a reasonably realistic toy model for at least the matter-dominated evolution of our Universe (see, e.g., \cite{GibbonsEllis}).

\subsection{Homothetic solutions and the topography of shape space \label{top}}

\begin{figure}[t]
\begin{center}
\includegraphics[width=0.58\textwidth]{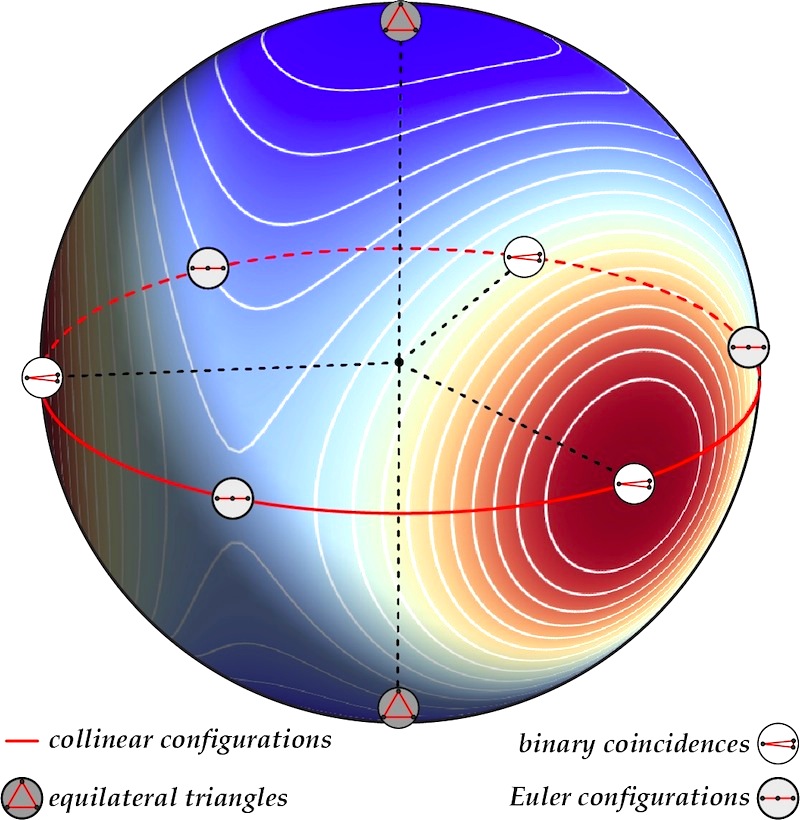}
\end{center}
\caption{\it \small The shape sphere of the three-body problem, with an elevation plot of $ V_\st{S}$ (\ref{shapepot}) on its surface. The equator corresponds to collinear configurations; at three points on it there are two-body coincidences, where $ V_\st{S}$ is singular with infinitely deep wells. Between them, at the azimuthal angles $\phi_{ab}$, there are the Euler configurations, saddle points of $V_{\st{S}}$. The triangles corresponding to points with the same longitude and opposite latitude are 
mirror images. For all values of the mass ratios, the absolute maximum of $ V_\st{S}$ is at the equilateral triangle. The figure shows the equal-mass case, for which $\phi_{ab}= \left( \pi,\frac \pi  3 , -\frac \pi 3 \right)$.}\label{Figura}
\end{figure}

Here we first wish to describe the structure of $\shs$. To this end we note that a mere sign change turns the complexity \cs into the \e{shape potential} \vs introduced in \cite{BLMpaper,BKMpaper}:
\bq
V_\st{S}:=\sqrt{I_\st{cm}}V_\st{New},~~V_\st{New}=-\sum_{a<b}{m_am_b\over r_{ab}}. \label{shapepot}
\ee
The absence of the Newton constant from \iv will be discussed below. We call \vs the \e{shape potential}\footnote{So far as we know, this name has not been used in $N$-body literature, probably because $N$-body equations are virtually always studied in Newtonian form even though in reality intrinsic change of shape is all that remains when the extrinsic scale and frame of reference are removed. Our \vs does figure prominently as the \e{configurational measure} in Saari's book \cite{SaariBook} and especially in his \cite{Saari2011}.} because its factor $\sqrt{I_\st{cm}}$ removes the scale dependence from \ivn, so forces derived from \vs can only change the shape of the system, not its size. We find it remarkable that the simplest obvious measure of clustering, or complexity, of a system of mass-weighted points is simultaneously the function that determines the objective behaviour of a universe subject to Newtonian gravity. Again, we are not aware that this has been noted, or at least emphasized. In the \tbn, for which shape space has two dimensions, \vs can be illustrated topographically as an elevation plot over a sphere (see Fig.~\ref{Figura}).

Figure \ref{Figura} exhibits the dominant features of all $N$-body shape spaces: the stationary points and infinitely deep wells of \vsn. The first are \e{central configurations} and play an important role in $N$-body theory.\footnote{The usual definition of central configurations, which explains their name, is that the resultant force exerted on each point by all the others is exactly in the direction of their common centre of mass. This is equivalent \cite{Saari2011,GibbonsEllis} to such configurations being stationary points of \vsn.} Since \vs is negative definite and regular except in the singular wells, it must have at least one absolute maximum $V_\st{S}^{\textrm{max}}$ as one of its stationary points. In the \tbn, $V_\st{S}^{\textrm{max}}$  is at the equilateral triangle for all mass values -- a result due to Lagrange.\footnote{In the 4-body problem, the maximum is at the regular tetrahedron, also for all mass values.} All non-maximal sationary points are saddles. In the \tbn, they are collinear, but for more than a few particles there are many non-collinear saddles. Our collaborator Jerome Barkley found the maxima and many saddles numerically for $N$ up to 20 in the equal-mass case; they can be seen at \cite{JeromeWebsite}. The number of central configurations increases rapidly with $N$. For $N=16$, Barkley readily found more than 70 000 in the equal-mass case; there must be vastly more when the masses are unequal. Saari's discussion of central configurations \cite{Saari2011} is very interesting.

Thus, the topographic features of $\shs$ are $V_\st{S}^{\textrm{max}}$, the saddles, and infinitely deep wells, whose number, as for the saddles, increases rapidly with $N$. Shape space is riddled with them. Interestingly, the values of \vs that Barkley found numerically for the saddles are not much lower than $V_\st{S}^{\textrm{max}}$. Thus, already for $N$  more than, say, 10, much of $\shs$ resembles a fairly gently undulating plateau with maxima not much higher than the plateau. The wells occupy a relatively small `area' of the plateau. Moreover, the shapes on the plateau have rather uniform, low complexity particle distributions. We have already mentioned the result of Battye et al \cite{BattyeGibbons} that for $N \approx 10^4$ the particle distribution at the minimum of \cs -- and thus maximum of \vs -- is extraordinarily smooth\footnote{Super-Poissonian. We believe study of the equal-mass case is justified because it best approximates field theory, modelling high field values by high particle densities.}  for equal-mass particles.

We say shape space is \e{asymmetric} because, through the level surfaces of \vsn, it acquires the structure with plateau and infinitely deep wells just described. It is certain that shape space with any potential will lack symmetry. 

The central configurations are important because they are associated with \e{central collisions}, when all the particles collide at once at their centre of mass and \m hits zero. This brings us to the zero-measure solutions of the \nb deferred earlier.

First, there are \e{homothetic} (unchanging shape) solutions: if the system is `held' at rest at a central configuration and released, it will fall homothetically until all the particles collide in a \e{central collison} at the centre of mass, beyond which the solution cannot be continued (a noteworthy result). 
The centre-of-mass position vector of each particle is 
\bq
{\bf r}_a^\st{cm}={\bf c}_af(t-t_0),~t>t_0,\label{homo}
\ee 
where ${\bf c}_a$ is a constant vector. Thus, all interparticle separations and their distances  from the centre of mass are proportional to a common function of the time $f(t-t_0)$ until the central collision.
In Newtonian terms, the complete set of ${\bf J}_\st{tot}= 0$ homothetic solutions corresponds to the system either hitting or exploding out of the centre of mass and behaving in one of three ways. If $E<0$, the system can only reach a finite size before collapsing back to a central collision; the solution exists only for a finite time interval. If $E=0$, the system can `just' escape to infinity, its scale increasing throughout as $(t-t_0)^{2/3}$. Finally, if $E>0$ the system `reaches infinity' with the scale increasing asymptotically as $t-t_0$; the system has escaped the effect of gravity and is `coasting' inertially.

All the homothetic solutions (\ref{homo}) exist as mere points in $\shs$. Much more interesting are the solutions that become homothetic only asymptotically, terminating at a central collision or escaping to infinity. In fact, central collisions can only occur if the solution does terminate at a central configuration {\cite{Chenciner1998}}. In $\shs$, these asymptotically homothetic solutions terminate at one end at a central configuration or, very exceptionally, at both ends. However, in all solutions that are asymptotically homothetic at one end the other end will be drawn forever down a well of \vsn. Since, as we noted, the complexity at saddles is typically low but tends to infinity in the wells, such solutions will exhibit clear complexity growth from one `past' to one `future'. As with the `two-sided' solutions, complexity arrows are due to the law alone. Moreover, all $E\ge 0$ solutions of the \nb except the fully homothetic ones, which are mere points in $\shs$, have the arrows. They are generic.

It might be argued that we have only been able to make this claim by an artificial `marriage' of two physically distinct things to make $C_\st{S}$: the (square root of the) moment of inertia and the Newton potential. But a closed dynamical system \e{must} be characterized in dimensionless terms: $C_\st{S}$ is the real thing and splitting it into $\sqrt{I_\st{cm}}$ and $V_\st{New}$ is artificial.  The objectively true arena in which a dynamically closed universe exists is like Fig.~{\ref{Figura}.

{This is the point for a preliminary summary, which we begin with a reiteration of the two main novelties of our approach. The first is consistent passage from the dimensionful coordinatized description in $\sf Q$ to the dimensionless $\shs$. Second, as we stressed in the caption to
Fig.~\ref{PairExchangeCenterOfMomentumFrame}, one should not be thinking about \e{initial} states but rather the structure of complete solutions, both the zero-measure and generic ones. Only the behaviour of a quantity like the complexity allows \e{pragmatic} identification of points on the solutions that can be termed `initial' or to lie in a `past'. In the generic $E\ge 0$ solutions to the \nbn, this criterion places the `initial' point \e{in the middle} of the solution.}

{ This leads us to
suggest that the puzzle of time asymmetry may have arisen because dynamics has been considered in the wrong arena. At least in the case of the \nbn, there is a time-symmetric dynamical law in the coordinatized representation, but in the \e{objective} description in $\shs$ a seemingly time-symmetric generic solution becomes, for all practical purposes, two time-asymmetric solutions that are independent. The fact is that any attempt to evolve, however accurately, asymptotic $N$-body data back in the direction of decreasing complexity will always lead to a more uniform state. Moreover, magnification of computational errors will mean that the overwhelming majority of retrodictions will make it seem that such a universe emerged from a very special, highly uniform state. Returning to Penrose's comment at the start of the paper, we see this as first tentative evidence that an explanation for the existence of ``a universe in the form we know it'' could be obtained provided we pass from the time-symmetric coordinatized representation in $\sf Q$ to the dimensionless, scale-invariant and time-asymmetric representation in $\shs$. Then no special initial condition -- no past hypothesis -- would be needed.}

We end this part of the paper with a question: since we can only observe and measure ratios, e.g., red shifts, why do we say the Universe is expanding? This is often illustrated by blowing up a balloon onto which coins, taken to represent galaxies, are glued: the distances between the coins grow relative to their diameters. {We find this} a misleading analogy, which limps on \e{two} crutches (`rigid' coins and `expansion' between them). We think the $N$-body problem provides a much more illuminating \e{dynamical} picture of crutch-free `expansion without expansion': once Kepler pairs form, the distances between them (and to other particles) increase relative to the semi-major axes, whose ratios remain unchanged. The behaviour of the true actors -- the ratios -- underlies the complexity growth and `expansion'. We will now show that deeper understanding of $N$-body dynamics is gained if we respond to the `cry' of change of scale to play a role distinct from that of shape.

\subsection{The 3-body problem in shape space\label{3BPsection}}

We have here one aim: to express everything intrinsically
on $\shs$. {This is appropriate if we treat the \tbn, the simplest nontrivial system, as a toy universe, for which external non-dynamical influences are manifestly questionable.} Following our aim consistently, we are led ineluctably to 
a dissipative structure on $\shs$ that exists identically in the $N$-body 
problem and in anti-dissipative form in vacuum GR.

We first mention a scale-invariant model \cite{barbour:scale_inv_particles} in which $V_\st{New}$ is replaced by a potential homogeneous of degree $-2$, $V = I_\st{cm}^{-1/ 2} \, V_\st{New}$. This simplest choice for dynamics on $\shs$ is \e{geodesic} and for large $N$ reproduces Newtonian gravity to good accuracy in small subsystems, but there is no secular growth of complexity, so the long-term Newtonian behaviour is not reproduced.\footnote{Dirac quantization of the model leads to an anomaly \cite{BLMpaper} that suggests holographic emergence of time.} The model serves as a useful reference to characterize the Newtonian dissipative behaviour by the deviation from a geodesic on $\shs$. 

We now make the \e{reduction to the 3-body shape space.} For three particles, $\shs$ is the two-dimensional space of triangle shapes. To arrive at it, we start with the 9D
space $\mathbbm R^9$ of particle positions $\mathbf r_a = (x_a,y_a,z_a)$, $a = 1,2,3$, and quotient wrt
the 3D similarity group $\Sim$ of rigid rotations, translations and rescalings.
Montgomery~\cite{InfinitelyManySygizes} gives the details, {Appendix~\ref{Appendix3bodyProblem}} the phase-space reduction in our notation.  The resulting space, to which the 3-body collision (not a shape) does not belong,\footnote{Denial of ontology to scale has consequences for the `Big Bang', modelled in the \tb by triple coincidence of the particles. The only candidate to replace it in $\shs$ is the equilateral triangle, the most uniform shape.} is topologically a sphere with three piercings as shown in Fig.~\ref{Figura} with centre at the origin of a Cartesian space with coordinates $\mathbf w = 
(w_1 , w_2 , w_3)$.\footnote{The $\mathbf w$ coordinates are nontrivially related to the
Cartesian coordinates $\mathbf r_a$, see Appendix~\ref{Appendix3bodyProblem}.} The square of the sphere's radius is 
\begin{equation}
||\mathbf w ||^2  = w_1^2 + w_2^2 + w_3^2 = \frac{I^2_\st{cm}}{4} \,.
\end{equation}
The three coordinates $\mathbf w$ permit full description of a Newtonian history,
including the changing size of the three-body triangle as measured by  $I_\st{cm}$. Quotienting wrt rescalings $\mathbf w \to \varphi \, \mathbf w$, $\varphi >0$, 
is the final step to $\shs$. 

To get there and exhibit the effect of the assumption $E=0$, we use Jacobi's principle, according to which (as {Lanczos \cite{Lanczos1949}} shows) the Newtonian \e{orbits} for each fixed value of $E$ and any potential $V$ are found as geodesics in ${\sf Q}$. The Jacobi action is
\bq
S_\st{Jacobi}=2\int\textrm ds\sqrt{(E-V)\sum_a{m_a\over 2}{\textrm d{\bf r}_a\over\textrm ds}\cdot {\textrm d{\bf r}_a\over\textrm ds}}.\label{jacobi}
\ee

This is a good first step: Newton's extraneous time is eliminated. But there is a problem since (\ref{jacobi}) is invariant under the reparametrization $s\rightarrow s'(s)$. In a generic geodesic principle, there is no obvious choice of a unique evolution parameter. Taking of one of the coordinates involves an arbitrary choice and in general will only work over a limited interval: generic dofs are not monotonic -- such a `clock' can stop and run backward.

This is where the split into scale and shape dofs is decisive \cite{BKMpaper}. There can be arbitrarily many shape dofs, but there is always only a single scale dof. Moreover, its derivative $D$ is monotonic in the \nb if $E\ge 0$. If we take it to be the independent variable, we `kill two birds with one stone'. We get a monotonic `time' and remove scale from among the dofs. Shape-dynamic purity is achieved.

At this point it is best to make the Legendre transformation from Lagrangian to Hamiltonian dofs and introduce canonical momenta. Because $S_\st{Jacobi}$ is reparametrization invariant, these are homogeneous of degree zero in the velocities and satisfy the constraint \cite{JuliansReview,FlaviosSDtutorial,barbourbertotti:mach,BLMpaper,BKMpaper}\,\footnote{At this point we set $E=0$. We will later show the effect on the equations in $\shs$ if it is retained. The \nb with $E=0$ is a good toy model of vacuum GR.}
\begin{equation}\label{ham}
H= \sum_a  \frac{\mathbf p^a \cdot \mathbf p^a  }{2 \, \mu_a} +   V_\st{New}  = \frac 1 2 \,  I_\st{cm}^{-1} \, \| \mathbf z \|^2 + V_\st{New}  = 0  \,, ~~~ \mu_a = {\textstyle \frac{m_a}{m_\st{tot}}} \,,
\end{equation}
where the  momenta $\mathbf z = (z^1 , z^2 , z^3)$ are conjugate to $\mathbf w$,
$\{ w_i , z_j \} = \delta_{ij} $, $\mathbf p^a = (p_x^a,p_y^a,p_z^a)$ are the 
Cartesian momenta and $V_\st{New} (\mathbf w)$ is the Newton potential
\begin{equation}
V_\st{New} = - \sum_{a<b} \frac{ (\mu_a \, \mu_b)^{\frac 3 2}(\mu_a + \mu_b)^{-\frac 1 2}}{\sqrt{ \| \mathbf w  \|  -  w_1  \, \cos \, \phi_{ab} -  w_2  \, \sin \, \phi_{ab}}} \,.
\end{equation}
The azimuthal angles $\phi_{ab}$ on the $w_3=0$ plane identify the direction 
of the two-body coincidences between particles $a$ and $b$. Their explicit expressions are
\begin{eqnarray}
&\phi_{12} = \pi \,, 
~~~
 \phi_{23}  =  \arctan {\textstyle \left(   2 \frac{\sqrt{m_1 \,m_2 \, m_3 (m_1+m_2+m_3) }}{m_2 (m_1+m_2+m_3) - m_1 \, m_3} \right) }\,, &\\
& \phi_{13}  = - \arctan {\textstyle \left(   2 \frac{\sqrt{m_1 \,m_2 \, m_3 (m_1+m_2+m_3) }}{m_1 (m_1+m_2+m_3) - m_2 \, m_3} \right) }\,,& \nonumber
\end{eqnarray}
which reduce to $\phi_{12} = \pi $, $ \phi_{23}  =  \frac \pi 3 $, $ \phi_{13}  = - \frac \pi 3$ in the equal-mass
case.

As we argued on Machian grounds,  \cite{JuliansReview,FlaviosSDtutorial} the Universe must have zero total linear $\textbf P_\st{tot} = \sum_a \textbf p^a$  and angular
$\textbf J_\st{tot} = \sum_a \textbf r_a \times \textbf p^a$ momentum. Quotienting the phase space wrt translations
and rotations we obtain $\textbf P_\st{tot} = \textbf J_\st{tot} = 0 $ as constraints on the coordinatized (extended) phase-space description. If now we define the {`mean square length'} $R^2 = \| \mathbf w \| = \frac 12 \, I_\st{cm}$ and use polar coordinates on the 2-sphere $\shs$,
\begin{equation}
w_1 = R^2 \, \sin \theta \, \cos \phi \,, ~ w_2 = R^2 \, \sin \theta \, \sin  \phi \,, ~ w_3 = R^2 \, \cos \theta \,,\label{three}
\end{equation}
the Hamiltonian constraint (\ref{ham}) becomes
\begin{equation}
H = \frac 1 2 \frac{p_\theta^2 + \sin^{-2} \theta \, p_\phi^2+ \frac 1 4 \, D^2 }{R^2}+ \frac{1}{R} \,  V_\st{S} (\theta,\phi)  \,,
\end{equation}
with \vs the shape potential (\ref{shapepot}), and the dilatational momentum (\ref{dilmtm}) takes the form 
\bq\label{dilm}
D = {\bf w}\cdot {\bf z} + {\bf r}_\st{cm} \cdot {\bf P}_\st{tot} \,,
\ee
where the  $\textbf P_\st{tot} = 0 $ constraint kills the second term.
As we noted, $D$ is half $\dot I_\st{cm}$; {it generates} dilatations in phase space. This is in the coordinatized representation; in $\shs$, a quantity related to $D$ will play the role of time.

\subsection{Dissipation in particle dynamics} 

We now proceed to the description on $\shs$ without \e{any} external element. This will give the most illuminating explanation of the dynamics described in Sec.~\ref{TheNewtonianSolutionsSubSubSection} (Fig.~1). As we recall, the generic 3-body solutions have a central region of strong three-body interaction. It corresponds to quasi-geodesic motion on $\shs$ well approximated by 
the shape kinetic metric (introduced later) of the scale-invariant
model \cite{barbour:scale_inv_particles,BLMpaper} mentioned at the start of Sec.~\ref{3BPsection}. In the asymptotic regions, the representative
point on $\shs$ spirals ever deeper into the potential wells of $ V_\st{S}$ (Fig.~\ref{PEss}). We will now show why this is inevitable. To that end, we must eliminate the residual dimensionful variables in the Hamiltonian constraint (\ref{ham}).

 \begin{figure}[t]
        \begin{center}
        
\includegraphics[width=0.6\textwidth]{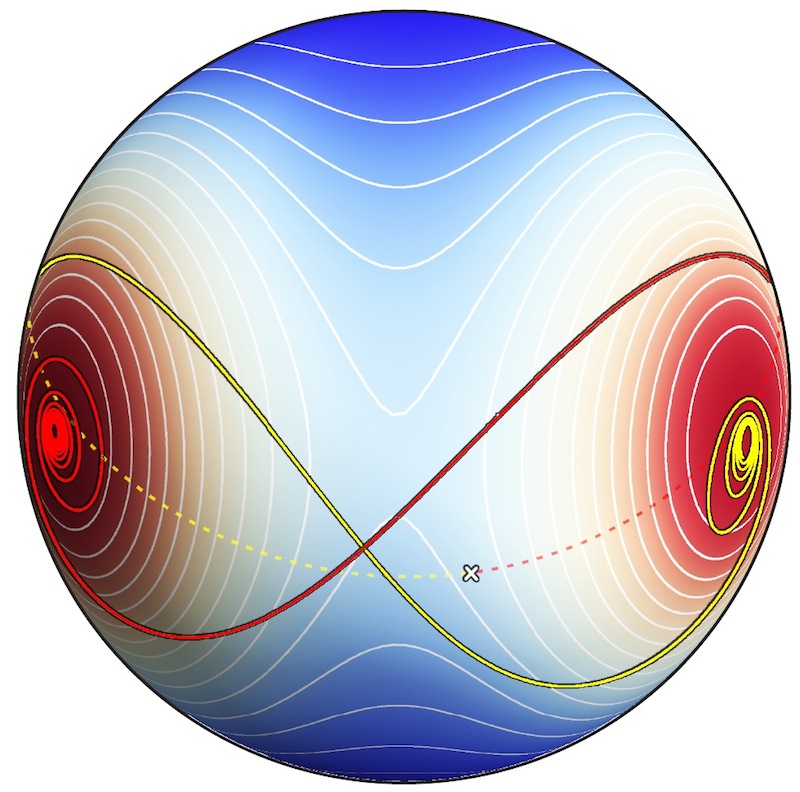}
                                
        \end{center}
\caption{\it \small The pair-exchange process of Fig. \ref{PairExchangeCenterOfMomentumFrame} as seen from shape space $\shs$. The red and the yellow part of the orbit are distinguished by belonging, respectively, to the (nominal) past and the 
future of the point at which the dilatational momentum $D$ is zero and the centre-of-mass moment of inertia $I_\st{cm}$ is at its 
minimum. This point is shown as an `x' at the back of $\shs$, where the orbit is dashed.\\
}\label{PEss}
\end{figure}

\hyphenation{de-para-me-tri-ze}

For this, we refer to Fig.~\ref{Figure3}, which exhibits the monotonicity of $D$.
{In \cite{BKMpaper} we exploited this to introduce a \e{dimensionless} time variable $\dimt$. We choose some point $D=D_0$, not at $D=0$, and define $\dimt=D/D_0$. As with $D$, the sign of $\dimt$ is conventional. Later, we will introduce the time $\lambda =\log\,\dimt$. {Up to the choice of its origin $\lambda = 0$ at $D_0$, $\lambda$ is uniquely defined}\,\footnote{The arbitrariness in the choice of $D_0$ maps to symmetry under shift of the origin of $\lambda$: $D_0 \to \alpha \, D_0$ corresponds to $\lambda \to \lambda - \log \alpha$.} and tends asymptotically to $-\infty$ as $D=0$ is approached. It grows without bound to $+\infty$ in the asymptotic region on the side of $D=0$ at which $D_0$ is chosen, which necessarily breaks the qualitative U-shaped symmetry of \mn.}

We now take the final step to explicitly \e{time-asymmetric} and \e{dimensionless} equations. We first find the dimensions of the relevant Newtonian variables, starting with the Cartesian coordinates $({\bf r}_a,{\bf p}^a)$, whose dimensions follow directly from the constraint (\ref{ham}):
\begin{equation}
[{\bf r}_a] = \ell \,, \qquad [{\bf p}^a] = \ell^{-\frac 1 2} \,.
\end{equation}
In their turn, the Poisson brackets have dimension [action]$^{-1}$, in our terms $[\{\cdot \,,\,\cdot\}]=\ell^{-\frac 1 2}$.\footnote{In (\ref{ham}) the Newton constant G has been absorbed into the definition of the momenta.  Both in NG and in vacuum GR, G is unphysical and absent if one attributes no dimensions to time.
See~\cite{FlaviosSDtutorial} for a derivation of these results from an action principle.} {The partially reduced coordinates} ${\bf w}$ {(\ref{three})} are manifestly translation- and rotation-invariant and, being quadratic in ${\bf r}_a$, they and their conjugate momenta ${\bf z}$ have dimensions
\begin{equation}
[{\bf w}] = \ell^2 \,, \qquad [{\bf z}] = \ell^{-\frac 3 2} \,.
\end{equation}
Finally, after the last phase-space splitting between dimensionless shape variables $(\theta,\phi)$ and the `rms length' $[R]=\ell$, the momenta have the dimensions
\begin{equation}
[p_\theta]=[p_\phi] =[D] = \ell^{\frac 1 2}\,.
\end{equation}
This last trace of `coordinatization', the dimensionality of the shape momenta $(p_\theta,p_\phi)$, will disappear below. Mass dimensions do not occur because (on rescaling of the action by $m_\st{tot}$ \cite{FlaviosSDtutorial}) we have only the dimensionless `geometrical' masses $\mu_a=\frac{m_a}{m_\st{tot}}$.

After these preliminaries, we first, as in \cite{BKMpaper}, express the dynamics purely on $\shs$ by using $\dimt = D/D_0$ as
a time label.\footnote{This `deparametrization' procedure, described in detail in \cite{BKMpaper}, consists of identifying the conjugate variable
to $D$, which is $\log R$ where $R = \sqrt{I_\st{cm}}$, and solving the Hamiltonian constraint  (\ref{ham}) for it. The resulting expression
for $\log R$ in terms of the other dofs (which are all shape dofs) is the Hamiltonian $\mathcal H$ which generates $D$-translations.} We also rescale the momenta $\dimp_i = p_i / D_0$ to make them dimensionless.
The shape Hamiltonian, generating evolution wrt $\dimt$, is
\begin{equation}
\mathcal H = \log \left( \frac 1 2 \frac{\dimp_\theta^2 + \sin^{-2} \theta \, \dimp_\phi^2 }{C_\st{S}  (\theta,\phi)} + \frac 1 8  \frac { \dimt^2 }{C_\st{S}  (\theta,\phi)} \right) \,, \label{ShapeHamiltonian}
\end{equation}
where we have introduced the complexity $C_\st{S}  =  - V_\st{S}  \geq 0$.
The {equations of motion are}
\begin{align}
&\frac{d \theta}{d \dimt} =  \frac{2 \, \dimp_\theta }{\pi_\theta^2  + \sin^{-2} \theta \, \dimp_\phi^2   + \frac 1 4 \, \dimt^2 } \,, &
&\frac{d \phi}{d \dimt} = \frac{2 \, \sin^{-2} \theta \, \dimp_\phi}{\dimp_\theta^2 + \sin^{-2} \theta \, \dimp_\phi^2  + \frac 1 4 \, \dimt^2 } \,, &
\\
&\frac{d \dimp_\theta}{d \dimt} =  \frac{  2 \, \sin^{-3} \theta \, \cos \theta \, \dimp_\phi^2  }{ \dimp_\theta^2 + \sin^{-2} \theta \, \dimp_\phi^2  + \frac 1 4 \, \dimt^2 } +  \frac{\partial \log C_\st{S}}{\partial \theta} \,,& 
&\frac{d \dimp_\phi}{d \dimt}   =    \frac{\partial \log C_\st{S}}{\partial \phi} \,. & \nonumber
\end{align}

As we noted, $C_{\st{S}}$ is a simple measure of shape complexity and, remarkably, as we see explicitly in these equations, $- \log C_\st{S}$ is the potential that governs Newtonian gravity represented objectively on $\shs$. At this point we want to compare the \e{time-dependent} Hamiltonian (\ref{ShapeHamiltonian}) of NG on $\shs$ with the geodesic model of \cite{barbour:scale_inv_particles}. To this end, consider the `complexity' metric, conformally related to the 
round metric, that assigns to a surface
element $d \cos \theta \, d\phi$ on $\shs$ the measure \cite{BLMpaper} $C_\st{S}(\theta ,\phi) \,d \cos \theta \, d\phi$.
This metric is
\begin{equation}\label{MetricOnS}
g^{ij} = 
\left( \begin{array}{cc}
C_\st{S}& 0 \\
0 & \sin^2 \theta \, C_\st{S}
\end{array}\right) \,.
\end{equation}
The non-reparametrization invariant action
\begin{equation}
S = \frac 1 2 \int d s \, g^{ij} \, \frac{d q_i}{d s}\, \frac{d q_j}{d s}\,,
\end{equation}
whose canonical Hamiltonian (with $g_{ij} $ the inverse metric) is
\begin{equation}
\mathcal H_\st{geo} = p^i \, \dot q_i - \mathcal L =  \frac 1 2   \, g_{ij} \, p^i \, p^j =  \frac 1 2 \frac{ p_\theta^2 + \sin^{-2} \theta \,   p_\phi^2 }{C_\st{S}  (\theta,\phi)}  \,, \label{geoham}
\end{equation}
 generates affinely parametrized geodesics wrt the metric (\ref{MetricOnS}). Comparison with (\ref{ShapeHamiltonian}) shows that its $\dimt^2$ term leads to
deviation from geodesics on $\shs$ -- and, as we shall now show, to dissipative behaviour in accordance with our formalism.  

To achieve a fully dimensionless description, initially for the \tbn, we introduce the logarithmic time $\lambda=\textrm{log}~\dimt$  and make a \e{non-canonical} transformation:
\begin{equation}
\lambda = \log \dimt \,,\qquad  \p_\theta = \frac {\dimp_\theta}{\dimt} \,, \qquad  \p_\phi = \frac{\dimp_\phi}{\dimt} \, .
\end{equation}
The equations of motion  for these variables are autonomous,\footnote{For the purposes of this paper, an \e{autonomous} dynamical system is one in which the equations of motion take the form $\dot x_\alpha = f_\alpha(x)$, where $x_\alpha$ are the phase-space
variables, and $f_\alpha$ does not depend explicitly on the independent variable.} 
\begin{align}\label{AutonomousEquations}
&\frac{d \theta}{d \lambda} =  \frac{2 \, \p_\theta }{\p_\theta^2  + \sin^{-2} \theta \, \p_\phi^2   + \frac 1 4 } \,,&
&\frac{d \phi}{d \lambda}  = \frac{2 \, \sin^{-2} \theta \, \p_\phi}{\p_\theta^2 + \sin^{-2}\theta \, \p_\phi^2  + \frac 1 4 } \,, &
\\
&\frac{d \p_\theta}{d \lambda}  = -  \p_\theta + \frac{  2 \, \sin^{-3} \theta \, \cos \theta \, \p_\phi^2  }{ \p_\theta^2 + \sin^{-2} \theta \, \p_\phi^2  + \frac 1 4 } + \frac{\partial \log C_\st{S}}{\partial \theta} \,, ~~ &
&\frac{d \p_\phi}{d \lambda}  =- \p_\phi+  \frac{\partial \log C_\st{S}}{\partial \phi} \,. &\nonumber
\end{align}
This system is dissipative: the equations for the momenta contain the terms $-\omega_\theta$ and $-\omega_\phi$ and therefore do not conserve phase-space volume (conserving phase-space volume is the necessary condition for non-dissipative dynamics).
The equations of motion can be thought as being generated by the time-independent Hamiltonian
\begin{equation}
H_0  = \log \left(   \frac{\p_\theta^2 + \sin^{-2} \theta \, \p_\phi^2 + \frac 1 4 }{C_\st{S}  (\theta,\phi)}   \right) \, \label{AutonomousShapeHamiltonian}
\end{equation}
and the dimensionless canonical structure
\begin{equation}
\ldm \theta , \p_\theta \rdm = 1 \,, ~~~ \ldm \theta , \p_\phi \rdm = 0 \,, ~~~ \ldm \phi , \p_\phi \rdm = 1 \,, ~~~ \ldm \phi , \p_\theta \rdm = 0 \,, ~~~
\end{equation}
 but with deformed, non-Hamiltonian {equations of motion}:
\begin{align}\label{DissipativeAutonomousEquations3BP}
&\frac{d \theta}{d \lambda} = \ldm \theta , H_0 \rdm \,,&
&\frac{d \phi}{d \lambda}  =\ldm \phi , H_0 \rdm\,, &
\\
&\frac{d \p_\theta}{d \lambda}  = \ldm \p_\theta , H_0 \rdm  -  \p_\theta   \,, ~~ &
&\frac{d \p_\phi}{d \lambda}  = \ldm \p_\phi  , H_0 \rdm - \p_\phi  \,. &\nonumber
\end{align}
The \emph{dimensionless} Poisson brackets, $[\ldm\, \cdot , \cdot \, \rdm] =1$, are related to the dimensionful
$\{ \cdot , \cdot \}$  by 
\begin{equation}
\ldm\, \cdot , \cdot \, \rdm  = \frac 1 {\dimt \, D_0} \{\, \cdot , \cdot \, \} \equiv D^{-1} \{\, \cdot , \cdot \}\, .
\end{equation}

We now extend our treatment to the \nbn. The 3-body model is particularly suited to build intuition because one can explicitly perform the configuration space reduction to $\shs$. This is not possible for $N>3$ because quotienting by rotations, unlike translations and scale, cannot be done explicitly. Luckily, the main interest arises from scale quotienting, and we can work with a partially reduced configuration space, `pre-shape space' $\pshs=\mathbbm{R}^{3N} / \Dil \! \times \! \Transl$. We will skip the intermediate step of obtaining the  shape momenta ${\bm \pi}^a$, which we derived in \cite{BKMpaper}. Instead we introduce directly the dimensionless and dissipative 
description with ${\bm \p}^a$, the dimensionless shape momenta: 
\begin{eqnarray}
&&\bm \sigma_a=\sqrt{ \mu_a} \, {{\bf r}_a^\st{cm} \over R} \,, \qquad  \bm \p^a= { 1 \over \sqrt\mu_a }\, \frac{R}{D} \,{\bf p}^a_\st{cm}  -  \bm \sigma_a \,,\\
&&R = I_{\st{cm}}^{1/2} \,, \qquad D = \sum_a{\bf r}_a^\st{cm} \cdot {\bf p}^a_\st{cm}  \,,  \nonumber \\
&&{\bf r}_a^\st{cm}  = {\bf r}_a - \sum_{b=1}^N \mu_b \, {\bf r}_b \,, \qquad  {\bf p}^a_\st{cm} = {\bf p}^a - \frac 1 N  \sum_{b=1}^N  {\bf p}^b \,. \nonumber
\end{eqnarray}
These coordinates on $\pshs$ satisfy the {constraints}
\begin{eqnarray}
\sum_a \bm \sigma_a \cdot \bm \sigma_a = 1 \,, \qquad \sum_a \bm \sigma_a \cdot \bm \p^a =0 \,,
\\
\sum_a  \sqrt{\mu_a}  \, \bm \sigma_a = 0 \,, \qquad \sum_a  \sqrt{\mu_a}  \, \bm \p_a =0 \,. \nonumber
\end{eqnarray}
The dimensionless Hamiltonian generating the dynamics on $\pshs$ is
\begin{equation}
H_0 = \log \left( \sum_{a=1}^N \bm \p^a \cdot \bm \p^a + 1\right) - \log C_\st{S}\,,\label{partham}
\end{equation}
and the equations of motion are 
\begin{equation}
\begin{array}{l}
\dot{\bm \sigma}_a = \ldm \bm \sigma_a , H_0 \rdm \,, 
\vspace{6pt}  \\
\dot{\bm \p}^a = \ldm \bm \p^a , H_0 \rdm  -   \bm \p^a  \,, 
\end{array}
\end{equation}
where the dimensionless Poisson brackets have the symplectic structure
\begin{equation}
\begin{array}{l}
\ldm \sigma_a^i ,  \p_b^j \rdm  =   {\delta^b}_a \, {\delta^i}_j -  \sigma_a^i \, \sigma_b^j   \,, 
\vspace{6pt}  \\
\ldm \p^a_i ,  \p^b_j \rdm
=    \sigma_a^i \, \p^b_j -\sigma_b^j \,\p^a_i   \,,  \label{PreShapeSpaceSymplStructure}
\vspace{6pt}  \\
\ldm \sigma^a_i ,  \sigma_b^j \rdm  = 0 \,. 
\end{array}
\end{equation}

An analogous dissipative representation exists for any shape-dynamic (particle or field-theoretic) model of the Universe provided the conditions assumed above hold: 1) the generator of scale transformations, here $D$ and in GR an analogous variable (the York time), is monotonic, allowing deparametrization wrt it;  2) the generator of dynamics, which takes the form of a Hamiltonian constraint, can be solved for $\log (R)$, which converts its conjugate momentum $D$ into a time variable and yields the physical Hamiltonian. 

In the introduction we noted that there is no obvious explanation (like microscopic degrees of freedom) for the dissipation we find in SD. We conjectured a possible connection with the deterministic laws of black-hole thermodynamics found in the late 1960s. Whatever the truth, we mention here that there exists a `metriplectic' formalism which makes it possible to introduce a formal entropy in cases when one has dissipative equations. We describe this formalism and apply it to shape-dynamic gravity in Appendix~A.4.

\subsection{Shape-dynamic explanation of 3- and $N$-body behaviour} 

Much of the long-known generic 3-body hyperbolic--elliptic behaviour when $E=0$ (Sec.~\ref{TheNewtonianSolutionsSubSubSection} and Fig.~\ref{PairExchangeCenterOfMomentumFrame}) can be directly `read off' the plot of $V_{\st{S}}$ in Fig.~\ref{Figura} knowing that the system is dissipative.
Equations (\ref{AutonomousEquations}) describe a particle moving in $\shs$ under the influence of the potential $V_\st{S}$ but subject to friction; the solutions have a transparent intuitive explanation. Locally the orbits are well approximated by geodesic motion wrt the  metric (\ref{MetricOnS}), but in the long run the momenta get depleted by friction, and
the orbits are drawn inescapably {ever deeper into the potential wells.}

The dissipative picture on $\shs$ provides an even more powerful intuition for the remaining measure-zero solutions (Sec.~\ref{top}). These end either with a central collision or escape of all three particles (no Kepler pair formed) and always tend asymptotically to \emph{homothetic} motion (the shape freezes); the final shape can only be a  central configuration. From the Newtonian coordinatized point of view, this behaviour is not obvious, but on $\shs$ it is. The central configurations are the stationary points of $V_\st{S}$, so the system can only end up not changing its shape if the derivatives of \vs vanish: at points where $\bm{\nabla} V_\st{S}=0$. Because of the dissipation, there will be orbits that reach these
stationary points with exactly zero velocity (wrt the logarithmic time $\lambda$). But all the stationary points of $V_\st{S}$ are
\emph{unstable} equilibria: the Euler configurations are saddles, the equilateral
triangle a maximum. That these are measure-zero solutions is therefore also
readily explained: the initial conditions must be doubly fine-tuned, to reach a stationary point and to arrive with zero velocity.

The difference between the solutions with $E=0$ and $E \ne 0$ is particularly interesting. As we show in Appendix~\ref{TimsAppendix}, the effect of $E>0$ on shape space $\shs$ is to add a time-dependent effective potential to $V_\st{S}$. This flattens the total potential and allows the solution curve in 
$\shs$ to asymptote as $\lambda \rightarrow\infty$ to points away from central configurations. Any scalene end shape of the triangle is possible. This shows that, in contrast to the case $E\ne 0$, the topography defined on $\shs$ by $V_\st{S}$ completely determines the $E=0$ solutions: they can asymptote only to its singularities (potential wells) or stationary points. There is a complete explanation for what happens in any $E=0$ solution, but not in the $E\ne 0$ case, which violates the principle of sufficient reason.\footnote{Einstein \cite{Einstein1916} made powerful use of this principle to argue against a dynamical role of absolute space in his famous example of two fluid bodies in relative rotation. No genuine cause could be given why one should be spherical, the other an ellipsoid of revolution.}

This is also true for the $E=0$ solutions of the \nb for all $N\ge 3$ and has a bearing on the `is-the-Universe-expanding' question raised at the end of Sec.~\ref{top}. As judged by the simplicity criterion of the amount of initial data needed to determine the evolution (a point and, respectively, direction or velocity), the two simplest theories on $\shs$ are the geodesic theory with Hamiltonian (\ref{geoham}) and Newtonian theory with $E={\bf J}_\st{tot}=0$. The latter is not quite `pure shape' in having an independent variable, but it is in giving all shapes independently specifiable velocities. In fact, as we have seen, the resulting dynamics `clings' to $\shs$ more perfectly than the geodesic dynamics, which allows the representative point in $\shs$ to `roam' more or less freely. We obtain a closed description of the Universe on $\shs$ without any external notion of scale or expansion. We next make some comments on their intrinsic emergence.

\subsection{On structure emergence \label{hylo}}

We have illustrated above a description of the dynamics in intrinsic terms, as a law generating curves on $\shs$,
without any external `props' (scale, location, orientation). This description contains all the physical information
that is contained in the Newtonian `coordinatized' description. The extra props needed for the standard Newtonian description
can actually be constructed from pure shape data as shown in \cite{JuliansReview,FlaviosSDtutorial}.
A notion of scale can be generated starting from the dynamical curve on $\shs$ by inverting the process that
brought us from the ${\bf r}_a$, ${\bf p}^a$ variables to the dimensionless ones, ${\bm \sigma}_a$, ${\bm \p}^a$,
which essentially consists in solving the Hamiltonian constraint for \mn.
Newtonian time can be abstracted from a measure of the change that the physical dofs undergo along the 
dynamical curve, called \emph{ephemeris time}. Similarly one can abstract a notion of \emph{equilocality}\footnote{`Equilocality'
means the ability to say a given point is at the same place at different times.} and inertial frames of reference
from the physical data. This is obtained through the mechanism of \emph{best-matching}, in which a preferred
orientation and location of the center of mass of the Universe is identified at each instant through a minimization process.

The result of this process is the definition of an invisible, purely mathematical framework, consisting of an inertial frame, 
in which the whole Universe has vanishing total momentum and angular momentum, evolving in a time parametrization which conserves the total energy, and Newton's laws hold. This will be true at any epoch in a given solution. The Machian conditions lead to what may be called \e{metrogenesis} (an emergent notion of scale) and \e{chronogenesis} (an emergent notion of duration).

However, away from the two asymptotic regions there will not be any clearly defined systems in which this structure created by the Machian law of the Universe is manifest. We have to `await' the asymptotic emergence of Kepler pairs and other bound systmes for that to be clearly revealed. Following Aristotle, let us call this process \e{hylogenesis} 
({\greektext <'ulh} (hyle) means `stuff', and to the extent that bound systems are stable they warrant such a designation).

\subsection{Remarks on time} 

Ellis and Gibbons \cite{GibbonsEllis} criticize our identification of dissipation in gravity as ``an artefact of an unphysical choice of the time parameter'', noting that one could obtain anti-dissipation
by a mere change of sign of our parameter $\lambda$ and that ``standard physics ... results only if one restricts oneself to affine transformations of the standard time function $t$''. But this confuses physics in the laboratory with the physics of the whole Universe, for which different criteria apply. Correctly interpreted in Machian terms, it is the Universe that, as we have just shown, creates local inertial frames, rods and clocks and with them `standard physics'. 

When we work in shape space, as opposed to the emergent inertial frames just described, we are led to replace $t$ by $\dimt$ by first principles: all external structures and dimensionful quantities are to be eliminated. To arrive at $\lambda$, we then require the Hamiltonian to be autonomous and time to increase with complexity. These are `gauge choices' but autonomy is the closest one can get to `standard physics' and distinguishes the choice $\lambda = \log \dimt$. And a `mere' reversal of its direction would make the Universe become less complex with time. Moreover, as we noted at the end of Sec.~\ref{intro2}, the real reason why dissipation appears is that in $\shs$ only shape kinetic energy is physical. We have used the Newtonian dilatational momentum to define our evolution parameter $\lambda$. This removes the corresponding kinetic energy from the Hamiltonian and explains why our equations are dissipative.


{The evolution parameter must be dimensionless in order to define the velocity of the representative point in the dimensionless $\shs$. Any such variable, based ultimately on the monotonicity of $D$, cannot `march in step' with Newtonian time, though it can on average in the asymptotic regimes and with better accuracy as $N$ increases. We find it particularly interesting that when a Kepler pair forms asymptotically it becomes a naturally created system that serves simultaneously as a rod (through its semi-major axis) and clock relative to which the escaping particle is found to be moving inertially with ever increasing accuracy.\footnote{{\label{auto}In this connection, Einstein admitted to a `sin' in his Autobiographical Notes \cite{EinsteinNote}: rods and clocks appear as independent external elements 
in GR and not as structures created through the equations of the theory. The spontaneous formation of Kepler pairs, seen clearly in the time-asymmetric behaviour in $\shs$, appears to be a first step to a satisfactory completion of Einstein's theory. We return to this in Sec.~\ref{erc}.}} 
Thus, there are two times in the theory: one dimensionless and fundamental, the other emergent. They only march in step in the asymptotic limit.} 

That the dimensionless time is fundamental is underlined by an analogy with standard Newtonian dynamics. In its variational formulation, one determines a solution by specifying two configurations and the \e{difference} $t_2-t_1$ between the times at them. In shape space, one specifies, as the absolutely minimal data, two shapes 1 and 2 and the \e{ratio} $D_2/D_1$ of the dilatational momenta at them.\footnote{{Note that $\frac{D_2}{D_1} = e^{\lambda_2-\lambda_1}$, which underlines the analogy and exhibits translational invariance wrt $\lambda$.}} Being dimensionful, Newtonian time cannot be specified in $\shs$. Since $D_2/D_1$ determines the objective behaviour, it is fundamental. Newtonian time is emergent in the behaviour of subsystems. 


\section{Time Asymmetry in Dynamical Geometry}
\label{sec:TimeAsymmetryInDynamicalGeometry}

In this section we will show that closed-space Einstein vacuum gravity exhibits several key similarities to the features of Newtonian gravity discussed in Sec.~2. First and foremost, it has both dynamical similarity and a monotonic time variable. In vacuum gravity, there is also, although not so unambiguously identifiable as in Newtonian gravity, a candidate for a measure of complexity; we believe it can be generalized to include matter. 

There is however a feature of Einstein gravity that cannot be ignored: in its coordinatized (spacetime) description the expansion-of-space kinetic energy has the opposite sign to the change-of-shape kinetic energy. This is a unique feature of GR and arises directly from the form of the Einstein--Hilbert action. Although we regard the size (volume) of the Universe as a gauge variable, this structural feature of the spacetime description appears prominently in the shape-space description: whereas Newtonian gravity is dissipative in the direction of increasing complexity in shape space, vacuum Einstein gravity is \e{anti-dissipative}. This does not take into account matter. The inclusion of matter is subtle: gravitational waves experience anti-dissipation, while matter degrees of freedom experience dissipation, as we already saw in the Newtonian limit. We thus still {expect} the arrow of time {to agree} with the direction of complexity growth in the matter sector. 

\subsection{Shape space for dynamical geometry} \label{ShapeSpaceForDynamicalGeometry}

For readers unfamiliar with the Hamiltonian formulation of vacuum GR due to Dirac and Arnowitt, Deser and Misner (ADM) \cite{DiracHamiltonianDynamics,Dirac:CMC_fixing}, we begin with a brief review of this important work. Einstein introduced spacetime as a \e{block} with \e{four-dimensional} metric $g_{\mu\nu}$ satisfying the field equations $G^{\mu\nu}={^\st{(4)}}R^{\mu\nu} - \frac 1 2 {^\st{(4)}}R \, g^{\mu\nu}=0$.\footnote{Here, $G^{\mu\nu}$ is the Einstein tensor, ${^\st{(4)}}R^{\mu\nu}$ is the 4D Ricci tensor, and ${^\st{(4)}}R$ is the 4D Ricci scalar (while here and henceforth $R$ denotes the 3D Ricci scalar curvature).} These apparently frozen equations are hyperbolic, and for dynamical purposes GR is better treated as the evolution of \e{three-dimensional} Riemannian metrics $g_{ab}$ (3-metrics). In the ADM formalism, these are regarded as canonical coordinates and have canonical momenta $p^{ab}$. In terms of them, Einstein's $G^{00}=0$ and $G^{0a}=0$ equations become the ADM constraints 
\begin{align}
&\textstyle{ 1 \over {\sqrt g}} ( p_{ab}\,p^{ab} - \textstyle{1\over 2}\,p^2 ) -  \sqrt g \, R = 0\,, \label{admham}
\\
&\nabla_b p^{ab} = 0\,,\label{admmtm}
\end{align}
where $p=g_{ab}p^{ab}$, $g =\textrm{det}~g_{ab}$ and $\nabla$ denotes covariant differentiation using the Levi-Civita connection of $g_{ab}$. The quadratic, or \e{Hamiltonian}, constraint (\ref{admham}) is analogous to the one that arises from Jacobi's principle when $E=0$, but crucially there is one such constraint at each space point (with consequences we come to in a moment). This is also true of the linear \e{momentum} constraint (\ref{admmtm}), which, like the conditions $ \textbf P_\st{tot}= \textbf J_\st{tot}=0$ in particle dynamics, can be derived as (Machian) constraints \cite{barbourbertotti:mach}.

The ADM system is fully constrained with total Hamiltonian
\bq
{\mathcal H}=\int_\Sigma {\textrm  d}^3x \left\{ N \, \left(\textstyle{ 1 \over {\sqrt g}} ( p_{ab}\,p^{ab} - \textstyle{1\over 2}\,p^2 ) -  \sqrt g \, R \right) -  2 \, N_a \, \nabla_b p^{ab}\right\} \label{admtotham}
\ee
with multipliers $N$ (the \e{lapse}) and $N_a$ (\e{shift}) that are arbitrary functions of the label time and \e{position}. Variation wrt them enforces the constraints (\ref{admham})--(\ref{admmtm}), but $N$ and $N_a$ are themselves freely specifiable in advance. If one has initial data that satisfy (\ref{admham})--(\ref{admmtm}), the evolution in accordance with (\ref{admtotham}) preserves them. The hard task, to which we shall come, is finding data that do satisfy (\ref{admham})--(\ref{admmtm}).

A spacetime $\mathcal M$ is built up as follows. The 3-manifold on which $g_{ab}$ and $p^{ab}$ are defined becomes a spacelike hypersurface embedded in $\mathcal M$. The momentum $p^{ab}$ is related to the extrinsic curvature $K^{ab}$ of the hypersurface by $p^{ab} = \sqrt g \, ( K \, g^{ab} - K^{ab} )$. The 4-dimensional line element is related to $g_{ab}$, the lapse $N$ and the shift $N_a$ by
\begin{equation} \label{4DLineElement}
\textrm d s^2 = g_{\mu\nu} \, \textrm d x^\mu \, \textrm d x^\nu =  
(-N^2 + g_{ab} \, N^a N^b ) \, (\textrm d x^0)^2 + 2 \, N_a \, \textrm d  x^a \, \textrm d  x^0 +  g_{ab} \, \textrm d  x^a \, \textrm d  x^b\,.
\end{equation}
Specification of the shift $N_a$ as a function of the label time and position determines how the coordinates will be laid out on the successive spacelike hypersurfaces of $\mathcal M$ as they are created by the dynamics. The critical issue is the role of the lapse $N$, which determines a foliation of $\mathcal M$. Choosing lapses with different dependences on the  label time and position, one creates the same $\mathcal M$ but with different foliations on it. 

Before we proceed, we introduce the three geometrodynamic spaces that correspond to the Cartesian $\Q$, the relational configuration space $\QR$ (the quotient of $\Q$  wrt translations and rotations), and shape space $\shs$.

Let $\Sigma$ be a 3D manifold (with manifold at this stage we mean a \emph{topological manifold}, without any metric structures on it)  that is compact (closed) without boundary. For simplicity,\footnote{One may also argue against topologically more complicated compact manifolds, constructed by identifications, on the grounds that they ``do not appear to be natural'', as Wald comments \cite{Wald1984}, p.~95.} we take this to be $S^3$. The space of all Riemannian 3-metrics defined on $\Sigma$ is $\Riem(\Sigma)$. This matches $\Q$. The quotient of $\Riem(\Sigma)$ wrt 3D diffeomorphisms is \e{superspace} $\Sup(\Sigma)$,\footnote{No relation to supersymmetry: the term `superspace'  has been coined by Wheeler \cite{MTW}.} each point of which is a 3-geometry.  This matches $\QR$. The final step is to quotient wrt 3D \e{conformal transformations} defined as follows:\footnote{\label{Rtran} The fourth power of $\phi$ in (\ref{contran}) is chosen for mathematical convenience to make the transformation of the 3D scalar curvature $R=g_{ab}R^{ab}$ take the simplest form, which is $R \to \phi^{-4} \, R - 8 \phi^{-5} \nabla^2 \phi$.}
\bq
g_{ij}\rightarrow \phi^4g_{ij},~\phi > 0, \label{contran}
\ee
where $\phi$ is a smooth function of position. The resulting space, the quotient of $\Riem$ wrt 3D diffeomorphisms and (\ref{contran}), is \e{conformal superspace} $\CSup(\Sigma)$. It is analogous to $\shs$. Each point of $\CSup(\Sigma)$ is a \e{conformal three-geometry}, represented as a joint diffeomorphism and conformal equivalence class of 3-metrics.

The passage to conformal 3-geometries changes the ontology of gravity. The determinant $g=\textrm{det}\,g_{ab}$ of a 3-metric is generally regarded as a physical dof: the local scale of the 3-geometry. The two remaining dofs define the conformal geometry and determine angles between intersecting curves in the manifold. In SD, the dimensionful $g$ is a gauge dof; only the two angle-determining dofs are physical.\footnote{By virtue of the ADM constraints (\ref{admham})--(\ref{admmtm}), there was never any doubt that gravity has only two physical degrees of freedom, but the relativity of simultaneity (refoliation invariance) made it impossible to identify them among the three dofs in a 3-geometry. In SD the physical degrees of freedom are identified and have a simple
geometric characterization as the angle-determining part of the metric.}

We note here an important difference between dilatations and 3D conformal transformations. The former merely change a single global scale, while the latter do two things. The 1D subgroup $\phi = \textrm{const}$ contains transformations that are like the dilatations and change the local scales ($\textrm{det}\,g$) by a common factor and thus change the volume $V=\int_\Sigma {\textrm  d}^3x\sqrt g$ without altering the relative distribution of scale. The infinitely many remaining transformations have no particle counterpart and redistribute the local scales freely while leaving $V$ unchanged. These are \e{volume-preserving} conformal transformations (VPCTs) \cite{barbour_el_al:physical_dof}. 

It is now time to describe Shape Dynamics proper. SD provides a dual representation of GR by replacing\footnote{For readers familiar
with gauge theory, by `replacing' we mean the following: first a gauge-fixing which leads to ADM gravity in CMC gauge. This is
followed by the observation that the gauge-fixed system can be obtained as a gauge-fixing of a different theory which has Weyl (conformal) gauge
symmetries. The precise meaning of `replacing' is to be found in the more advanced concept
of `symmetry trading' \cite{gryb:shape_dyn} or, in a BRST setting, `symmetry doubling' \cite{Gomes:2012uq}.} \emph{almost} all of the
ADM-Hamiltonian constraints  (\ref{admham}) with the following linear constraint:
\bq
{p\over\sqrt g} = \langle p \rangle =  V^{-1}\int_\Sigma {\textrm d}^3 x \, p = \textstyle{ \frac{3}{2}}  \tau = \textrm{const}.\label{concon}
\ee
This constraint generates  VPCTs \cite{gryb:shape_dyn}. In fact $p = g_{ab} \, p^{ab}$ generates 
full conformal transformations, but removing its average $p - \langle p \rangle \, \sqrt g$  deprives
the constraint of its ability to change the global volume.  Besides this simple geometrical interpretation, (\ref{concon})  has also an interpretation in spacetime terms: it foliates $\mathcal M$ with spacelike hypersurfaces of spatially constant mean extrinsic curvature $K = \text{const}$ (called CMC surfaces). The CMC constraint (\ref{concon}) replaces \emph{almost} all of (\ref{admham})  precisely because of its volume-preserving property: 
one single \emph{global} linear combination of the Hamiltonian constraints (\ref{admham}), which
we will call $\mathcal H_\st{global}=0$, is kept among the constraints. Now, the meaning of $\mathcal H_\st{global}=0$ is perfectly analogous to that of the Hamiltonian constraint (\ref{ham})  of the \nbn:
it generates \emph{reparametrizations} of the time label. 

Through its preferred foliation, SD restores simultaneity and with it history to the Universe: the solutions of SD are
arbitrarily parametrized curves in the reduced configuration space $\CSup (\Sigma)\times \mathbbm R^+$. 
Adoption of the VPCT gauge group makes all the local scales into gauge dofs, while the volume $V$, the $\mathbbm R^+$ part of the configuration sapce, can initially be retained as a physical dof just like the moment of inertia in the particle model.   The momentum conjugate to $V$ is  the so-called \emph{York time} $\yorkt = {2 \over 3} \, \langle p \rangle$.  

The pair of variables $V$, $\yorkt$ is  closely analogous to the $I_\st{cm},D$ pair in the particle model. In particular the York time is \emph{monotonic} whenever 
the spacetime is CMC foliable, as we prove in Appendix~\ref{MonotonicityYorkTime}.
Moreover, due to dynamical similarity \cite{BKMpaper}, $V$ and $\tau$ form only a single Hamiltonian dof, i.e., \emph{half} a Lagrangian
dof, just like \m and $D$ in the \nb. Thus, as there we can deparametrize wrt the York time $\yorkt$ \cite{BKMpaper}, transforming the volume $V$ into a physical Hamiltonian and $\yorkt$ into a monotonic time variable. The reduced configuration space is then $\CSup(\Sigma)$. In this respect, the parallel with the particle model is essentially perfect. What makes conformal geometrodynamics so much more interesting is the added richness that the local scales introduce. We now turn to the details.

\subsection{Shape Dynamics in conformal superspace} \label{SDinCS}

In exact analogy with the particle model, we seek to formulate a theory in $\CSup(\Sigma)$ in which an initial shape\id a conformal 3-geometry, and a shape velocity uniquely determine the evolution. For the moment, we restrict ourselves to the matter free case.
The theory is encoded in the two constraints
\begin{equation} \label{SDconstraints}
\nabla_b p^{ab} = 0 \,, \qquad p = 0 \,
\end{equation}
together with the analogue of (\ref{ShapeHamiltonian}), the \emph{Shape Dynamics Hamiltonian,} which generates evolution with respect to $\tau$. It is defined in \cite{Koslowski:ObservableEquivalence}, used in \cite{BKMpaper} and is
\begin{equation}
\mathcal H_\st{SD}[g_{ab}, \dimp^{cd} , \yorkt] = \int_\Sigma \textrm d^3 x \, \sqrt g \, \phi^6[g_{ab}, \dimp^{cd} , \yorkt; x) \,, \qquad  \label{SDHamiltonian}
\end{equation}
where $ \phi^6[h_{ab}, \pi^{cd} , \tau; x) $ is the (unique) positive solution to the \emph{Lichnerowicz--York} (LY) equation:
\begin{equation}
{\textstyle {\phi^{-12}} \over g } \, g_{ac} \, g_{bd} \, ( p^{ab} - \textstyle{ 1 \over 3 } p \, g^{ab})( p^{cd} - \textstyle{ 1 \over 3 } p \, g^{cd}) -\phi^{-4} ( R - 8 \, \phi^{-1} \nabla^2 \phi ) - \textstyle{3 \over {8} } \yorkt^2 = 0 \,. \label{LYEquation}
\end{equation}
The Hamiltonian (\ref{SDHamiltonian})  is invariant under infinitesimal diffeomorphisms, which act on the 
metric and the momenta as $\delta g_{ab} = \nabla_a \xi_b + \nabla_b \xi_a$, $\delta p^{cd} = \nabla_c (\xi^c p^{ab}) - \nabla_c \xi^a p^{cb} - \nabla_c \xi^b p^{ac}$
\begin{equation}
\begin{aligned}
\phi [g_{ab}+ \delta g_{ab}, p^{cd}+\delta p^{cd}, \yorkt ; x) &=  \phi [g_{ab}, p^{cd} , \yorkt ; x) + \xi^c \nabla_c\phi [g_{ab}, p^{cd} , \yorkt ; x) 
\\
&\Downarrow
\\
\mathcal H_\st{SD}[g_{ab}+ \delta g_{ab}, p^{cd}+\delta p^{cd} , \yorkt] &=\mathcal H_\st{SD}[g_{ab}, p^{cd} , \yorkt]\,,
\end{aligned}
\end{equation}
 and therefore commutes with the constraint $\nabla_b p^{ab}$. Moreover, it is invariant under
conformal transformations $ g_{ab} \to \omega^4 g_{ab}$, $ p^{cd} \to \omega^{-4} p^{ab} $:
\begin{equation}
\begin{aligned}
\phi [\omega^4 g_{ab},\omega^{-4} p^{ab}, \yorkt ; x) &=  \omega^{-1} \, \phi [g_{ab}, p^{cd} , \yorkt ; x) 
\\
&\Downarrow
\\
\mathcal H_\st{SD}[\omega^4 g_{ab},\omega^{-4} p^{ab}, \yorkt] &=\mathcal H_\st{SD}[g_{ab}, p^{cd} , \yorkt]\,,
\end{aligned}
\end{equation}
and therefore commutes with the conformal constraint $g_{ab}p^{ab}$. We have a conformally- and diffeo-invariant Hamiltonian
generating evolution with respect to the York time $\yorkt$. If we choose initial data satisfying the constraints (\ref{SDconstraints}),
then $\mathcal H_\st{SD}$ will evolve them in a way that preserves the constraints. $\mathcal H_\st{SD}$  generates 
a curve in conformal superspace $\CSup(\Sigma)$.

A note on dimensional analysis: we follow Dicke's convention \cite{DickeDimensions}, in which the 3-metric has dimensions of an area $[g_{ab}] = \ell^2$, while the coordinates (and, accordingly, space
derivatives) are dimensionless labels for points. The momenta are dimensionless, $[p^{ab}] =1$,
and the York time  is $[\yorkt] = \ell^{-1}$. So the phase-space variables are dimensionful, and the Hamiltonian (\ref{SDHamiltonian}) is time-dependent. As in the particle model, we can rectify these two defects simultaneously by changing variables to the dimensionless \emph{unimodular} metric
$[h_{ab}]=1$, with inverse $h^{ab} $, and traceless momenta $[\p^{ab}] =1$ rescaled by the York time,
\begin{equation}
h_{ab} = g^{-1/3} \, g_{ab} \,, \qquad h^{ab} = g^{1/3} \, g^{ab} \,, \qquad \p^{ab} = \tau^2 \, g^{1/3} (p^{ab} - \textstyle{ 1 \over 3 } p \, g^{ab})\,.
\end{equation}
In terms of those variables, the Lichnerowicz--York equation reads
\begin{equation}
 \Theta^{-12}  \, h_{ac} \, h_{bd} \,\p^{ab} \, \p^{cd} - \Theta^{-4} ( R - 8 \, \Theta^{-1} \nabla^2_h \Theta ) - \textstyle{3 \over {8}  } = 0 \,, \label{DimensionlessLYEquation}
\end{equation}
where now $\Theta = \yorkt^{1/2} \, \phi \, g^{1/12}$ is a density of weight $1/6$ (so $\Theta^6$ is a scalar density which can be used as integration measure),
$\nabla_h^a$ is the covariant derivative associated to $h_{ab}$ and $\Theta$ is related to the SD Hamiltonian by
\begin{equation}
\mathcal H_\st{SD}[h_{ab}, \p^{cd} , \yorkt] = \yorkt^{-3} \, H_0 [h_{ab}, \p^{cd}] = \yorkt^{-3} \, \int_\Sigma \textrm d^3 x \, \Theta^6[h_{ab}, \p^{cd} ; x) \,, \qquad  \label{dimensionlessSDHamiltonian}
\end{equation}
We have thus introduced the \emph{dimensionless SD Hamiltonian} $H_0 = H_0 [h_{ab}, \p^{cd}] $, which does not depend
on the York time. Everything can now be described intrinsically in $\CSup(\Sigma)$. The introduction of the
dimensionless Poisson brackets
\begin{equation}
\ldm\, \cdot , \cdot \, \rdm  = \frac 1 {\yorkt^2} \{\, \cdot , \cdot \, \} \,
\end{equation}
makes the equations of motion autonomous and dissipative if expressed in terms of the logarithm of the York time $\lambda = \log \yorkt/\yorkt_0$,
\begin{equation}
\frac{d h_{ab}}{d \lambda} = \ldm\, H_0 , h_{ab} \rdm  \,, \qquad \frac{d \p^{ab}}{d \lambda} = 2 \, \p^{ab} + \ldm\, H_0 , \p^{ab} \rdm  \,.
\end{equation}
W have obtained a theory that, given a point in $\CSup(\Sigma)$ and a tangent vector to it, generates a curve in $\CSup(\Sigma)$.
But now, from this curve, we can reconstruct a spacetime.

\subsection{Spacetime construction \label{spcon}}

We have shown above how to generate a curve on $\CSup(\Sigma)$, parametrized by the dimensionless
label $\lambda$, starting from purely dimensionless shape
degrees of freedom, namely a unimodular metric $h_{ab}$ and a dimensionless TT-tensor density $\p^{ab}$. These data are sufficient
to construct a whole spacetime. In fact the solution $\Theta$ of the dimensionless version (\ref{DimensionlessLYEquation}) of the  LY equation 
produces a local notion of size from the dimensionless conformally-invariant data $h_{ab}$, $\p^{ab}$.  To give everything its dimensions, we introduce a $\lambda$-dependent spatial constant $\yorkt =\yorkt_0 \, e^\lambda$ with dimensions of length$^{-1}$ and define the dimensionful 3-metric\red{\footnote{Here $G_{ab}$ is not to be confused with the Einstein tensor.}}
\begin{equation}
G_{ab} = \yorkt^{-2} \, \Theta^4  h_{ab} \,, \qquad G^{ab} = \yorkt^{2} \, \Theta^{-4}  h^{ab} \,, \qquad \sqrt G = \yorkt^{-3} \, \Theta^6  \,,
\end{equation}
the (still dimensionless) constant-trace momentum
\begin{equation}
\Pi^{ab} =  \Theta^{-4}  \p^{ab} + {\textstyle \frac 1 2 }\, \Theta^2 \, h^{ab} \,, \qquad \Pi = G_{ab}\, \Pi^{ab} = \textstyle{3 \over 2 } \yorkt \, \sqrt G \,,
\end{equation}
and the constant-trace extrinsic curvature $K^{ab}=\frac 1 {\sqrt{G}} (\frac 1 2 \, \Pi \, G^{ab} -\Pi ^{ab})$,
\begin{equation}
K^{ab} =  {\textstyle \frac 1 4} \, \yorkt^3 \, \Theta^{-4} \, h^{ab} - \Theta^{-10} \, \yorkt^3 \, \p^{ab}  \,, \qquad K = G_{ab}\, K^{ab} = \textstyle{ 1 \over 2} \yorkt \,.
\end{equation}
These derived quantities automatically solve the first Gauss--Codazzi equation,
\begin{equation}
 G_{ac} \, G_{bd} \, K^{ab} \, K^{cd} - {\textstyle \frac 1 2} K^2  -  R[G;x)  = 0\,.
\end{equation}
The transversality of $\p^{ab}$ wrt $h_{ab}$ translates into transversality
of $K^{ab} - K \, G^{ab}$ wrt $G_{ab}$,
\begin{equation} \label{Gauss}
\nabla^G_b (K^{ab} - K \, G^{ab})  = 0\,,
\end{equation}
which is the second Gauss--Codazzi equation. These two equations guarantee 
 that the 3-metric $g_{ab}$ and the extrinsic curvature can be embedded as initial data on a spacelike hypersurface of constant mean extrinsic curvature (CMC) in a 4-dimensional Lorentzian metric whose 4D Einstein tensor is zero (or determined by the matter terms if present).
 
We obtain the 4D metric by solving two other equations, the  \emph{lapse-fixing equation}:
\begin{eqnarray}
 \label{LFE}
 &\nabla^2_G  N  - \left(   R[G;x)  + {\textstyle \frac 9 {16} } \, \tau^2 \right) N 
 - \frac{ \int_\Sigma {\textrm d}^3 x \sqrt G \left(  
 \nabla^2_G  N  - \left(   R[G;x) + {\textstyle \frac  9 {16} }  \, \tau^2 \right) N \right) }{\int_\Sigma  {\textrm  d}^3 x \sqrt G } =0 \,,&
\end{eqnarray}
and the equation for the shift $N_a$, found by York to solve the diffeo constraint \cite{York1973,FlaviosSDtutorial}:
\begin{equation}
\nabla_b^G \left( \nabla^a_G N^b + \nabla^b_G N^a - {\textstyle \frac 2 3} \, g^{ab} \, \nabla_c^G N^c \right) = \nabla_b^G \left( K^{ab} - {\textstyle \frac 1 3} \, K \, g^{ab}  \right) \,.
\end{equation}
These two equations can be used only after the LY equation has been solved for $\Theta$ to obtain $G_{ab}$ from $h_{ab}$.
Like the LY equation, the two equations above have a unique solution $N$ and $N_a$ (the latter modulo
conformal Killing vectors of $G_{ab}$, but this is unimportant here).

Thus, starting only from $h_{ab}$ and $\p^{ab}$, and having deduced $\Theta[h_{ab},\sigma^{ab};x)$, $N[h_{ab},\sigma^{ab},\Theta;x)$ and $N^a[h_{ab},\sigma^{ab};x)$, we can
build the 4-dimensional (dimensionful) spacetime metric
\begin{equation}\label{Reconstructed4Dmetric}
g_{\mu\nu} =  \left( 
\begin{array}{cc}
-N^2 + G_{ab} \, N^a N^b & N_a \\
N_b & G_{ab}
\end{array}
\right) =
 \left( 
\begin{array}{cc}
-N^2 + \yorkt^{-2} \, \Theta^4  h_{ab} \, N^a N^b & N_a \\
N_b &  \yorkt^{-2}\, \Theta^4  h_{ab}
\end{array}
\right),
\end{equation}
which is defined in a open neighbourhood of the initial Cauchy hypersurface. In fact, using the full set of Einstein equations, one can always generate a `slab' of spacetime in CMC foliation once the conformal data $h_{ab}$ and $\p^{ab}$ have been produced. We are not in a position to say how far such data can be evolved since that depends on difficult issues of long-term evolution, but for the purposes of this paper we shall make the `physicist's assumption' that such evolution is possible.

We conclude this part of the discussion by noting that the above `construction of spacetime' is by no means an essential part of SD, which at the fundamental ontological level is solely concerned with the evolution curve in $\CSup(\Sigma)$. Spacetime is emergent, as are rods and clocks, which we now consider. We will return to the status of spacetime  in Sec.~\ref{mhylo}

\subsection{The emergence of rods and clocks \label{erc}}

In Sec.~(2), we saw Machian constraints and dynamics create a (Newtonian) spacetime, in which lengths and times are always defined, but how only in the asymptotic regime is there emergence of well-defined Kepler pairs that physically realize the lengths and times. In the light of the above equations, we now consider the situation in dynamical geometry. This will highlight the way in which \e{local} lengths and times emerge.

We start with conformal 3-geometries and 3D matter fields defined in them. In this ontology, there is no notion of distance, time or equilocality. Among the normally accepted attributes of spacetime geometry, only spatial angles are present. What we find remarkable and just showed is this: given a shape and shape velocity (or momentum), the hidden conformal law in Einstein's equations creates all the additional spacetime attributes: local proper time, local proper distance and equilocality\footnote{In the context of conformal dynamics, this means a given point in one conformal 3-geometry can be said to be \e{at the same position} as a uniquely defined point in another conformal 3-geometry. For once the 4D spacetime has been constructed, equilocal points are determined by the spacetime normals to the CMC spacelike hypersurfaces.} all have their origin in \e{law}. There is no need to presuppose spacetime ontology. We can rely on conformal dynamics to \e{create} a structure in which length and duration (proper time) are `there' to be measured. But we do not yet have rods and clocks to measure the distances and times.

In footnote~(\ref{auto}), we commented on Einstein's `sin' in not creating a proper theory of rods and clocks. It is worth citing the passage \cite{EinsteinNote}:

\begin{quotation}
\noindent \it It is striking that the theory (except for four-dimensional space) introduces two kinds of physical things, i.e., (1) measuring rods and clocks, (2) all other things, e.g., the electromagnetic field, material point, etc. This, in a certain sense, is inconsistent; strictly speaking measuring rods and clocks would have to be represented as solutions of the basic equations... not, as it were, as theoretically self-sufficient entities. 
\end{quotation}

Probably because he was convinced quantum mechanics should play a central role, Einstein never  attempted to rectify the `sin' and develop a proper theory of rods and clocks.  However, at the non-quantum level our $N$-body model
does precisely this: the generic orbit, far from the $D=0$ point, spontaneously forms measuring rods and clocks in the way we described earlier. We now want to consider what we can say about the situation in GR.

First, all metrology relies on the degrees of freedom provided by \e{matter}. Gravitational dofs are not suited to `make' rods and clocks, as we shall see below. Therefore, the theory that Einstein did not supply will certainly need to include matter fields and seek solutions in which the requisite objects emerge as they do in the \nbn. This is the crucial process, about which we can unfortunately say little at present because of the intricate manner in which matter fields interact with geometry. However, the Universe does seem to be very well described by GR and we do know that natural rods and clocks, for example the Earth, which has a diameter and a rotation period, have been created. Thus, it seems that a solution at the classical level is in principle possible.

To some extent it does already exist, namely one knows how stable objects, once formed, will interact with 4D gravity. Assume we have a solution of the equations of motion $h_{ab}(\lambda)$ in CS, with the inclusion of matter. This describes the real physics, from which, in the manner described above, we can construct a complete spacetime. If a `test rod--clock' system  {like a Kepler pair} does form and is sufficiently light that its backreaction on the geometry can be ignored,
one  can show that it moves  along the
geodesics of the 4D metric $g_{\mu\nu}$ shown in  (\ref{Reconstructed4Dmetric}), and the proper time
\begin{equation}
\textrm d s^2 =  
(-N^2 + \tau^{-2} \, \Theta^4 \, h_{ab} \, N^a N^b ) \, (\textrm d x^0)^2 + 2 \, N_a \, \textrm d  x^a \, \textrm d  x^0 + \tau^{-2} \, \Theta^4 \, h_{ab} \, \textrm d  x^a \, \textrm d  x^b \,,
\end{equation}
turns out to be the time it ticks along its worldline. A collection of such test systems will
also keep mutual congruence in the way the Kepler pairs do in the \nbn. In fact, because  gravity is so vastly weaker than the other forces, relatively massive subsystems of the Universe (planets, stars, galaxies, and even cluster of galaxies) will exert only a small backreaction on the underlying conformal geometry and conspire to form a mutually consistent picture of a background spacetime
in which these systems exist.

This picture of the reaction of test objects to a pre-existing spacetime is non-trivial and was, of course, known to relativists, including Einstein, from the 1920s. The new element in the above account, implicit but not explicit in York's work, is the demonstration of the emergence of spacetime itself from conformal data alone. Two things remain to be done to complete the theory of rods and clocks: 1) At the classical level, show how test rods and clocks emerge as solutions of the full conformal equations with matter terms included. This will parallel the classical emergence of Kepler pairs in the \nbn. 2) Create a truly modern theory of metrology based on quantum mechanics. The gap that needs to be filled is the lack of rods and clocks with \e{identical} lengths and periods, respectively. In the \nbn, we obtain rods and clocks that remain mutually concordant but do not have a common length or period. Only quantum mechanics applied to systems with quantized charges can do that. Modern metrology relies on the 
caesium clock and the selection of a transition with fixed frequency that can be reproduced everywhere and at all epochs. Thus, the final step to a satisfactory theory of rods and clocks, together with the creation of information that can be read by them, awaits the unification of quantum mechanics with the conformal dynamics of geometry and matter fields. We make a tentative -- and very speculative -- first step in that direction in Sec.~\ref{sec:QuantumGravity}.

\subsection{The Mixmaster model}

We here explore the similarities and differences of the particle model and the shape-dynamics description of gravity in the simplest non-trivial model that can be worked out in detail: the vacuum Bianchi IX model, also known as the mixmaster Universe \cite{Misner:Mixmaster}. Being spatially homogeneous, it allows us to perform all steps of our construction analytically, and to build the  shape-dynamic description in analogy with the $N$-body problem for $E=0$. 

The model describes the evolution of diagonal homogeneous metrics on $S^3$. Using coordinates $(0\le\theta\le\pi,0\le\phi\le2\pi,0\le\chi\le4\pi)$, we can parametrize the homogeneous metrics on $S^3$ as
\begin{equation}
 \begin{array}{rcl}
   ds^2&=&\left(\alpha \cos^2\chi+\beta\sin^2\chi\right)d\theta^2+\gamma\cos\theta\, d\phi  \,d\chi
   \vspace{6pt}   \\
   &&+\left[(\alpha \sin^2\chi+\beta\cos^2\chi) \sin^2\theta+\gamma\cos^2\theta\right] d\phi^2
      \vspace{6pt}   \\
   &&+\gamma\,d\chi^2+2(\alpha-\beta)\cos\chi\sin\chi\sin\theta\,d\theta \, d\phi \,.
 \end{array}
\end{equation}
where $\alpha,\beta$ and $\gamma$ parametrize the three components of the metric. The total volume of the Cauchy surface $S^3$ is $V= 16 \, \pi^2 \, v = 16 \, \pi^2 \, \sqrt{\alpha\beta\gamma}$, so it is useful to split the metric degrees of freedom into $v$ and the shape degrees of freedom\footnote{They determine the extent to which space has a `cigar' or `pancake' shape.}
\begin{equation}
 x =\frac 1 2 \log \alpha\beta \,, \qquad   y =\sqrt{12} \, \log\frac \alpha \beta \,,
\end{equation}
and denote the momenta conjugate to $x,y$ by $\dimp_x,\dimp_y$\footnote{Notice that these momenta are not $p^{ab}$, the ones conjugate to the metric, which are dimensionless. $\dimp_i$  correspond to the quantities $g^{1/3} (p^{ab} - {1 \over 3} p \, g^{ab})$ which
are conjugate to the unimodular part of the metric $h_{ab} = g^{-1/3} g_{ab}$.} and the momentum conjugate to $v$ by $\tau$. 
This allows us to write the Hamilton constraint as
\begin{equation}\label{equ:B9constraint}
 \mathcal H =\frac 1 2\left(\dimp_x^2+\dimp_y^2-\frac 9 4 \tau^2\, v^2\right)+12\,v^{4/3}\,V_\st{S}\approx 0,
\end{equation}
(notice the minus sign in front of the dilatational kinetic energy $\frac 9 4 \tau^2\, v^2$), where the shape potential $V_\st{S}$ depends only on $x$ and  $y$,
\begin{equation}\label{ShapePolentialB9}
 V_\st{S}=\frac 1 2 \left[ e^{-4 \, x} - 4 \, e^{-x}\cosh(\sqrt{3}\, y)+2 \, e^{2 \, x}\left(\cosh(\sqrt{12} \, y)-1\right)\right] \, .
\end{equation}
The shape potential is a multiple of the three-dimensional Yamabe invariant,\footnote{We shall discuss its interesting properties below.}
\begin{equation}
\mathcal Y = \inf_{\phi} \left\{  \frac{\int d^3 x \sqrt g \left( \phi^2 \, R - 8 \, \phi \nabla^2 \phi \right)}{\int d^3 x \sqrt g \, \phi^6} \right\} \,,
\end{equation}
which, at least in this model, provides a natural candidate for a measure of shape complexity in geometrodynamics.
The potential (\ref{ShapePolentialB9}) is plotted in Fig.~\ref{BianchiIXpicture}.

\begin{figure}[t]
\begin{center}
\includegraphics[width=0.6\textwidth]{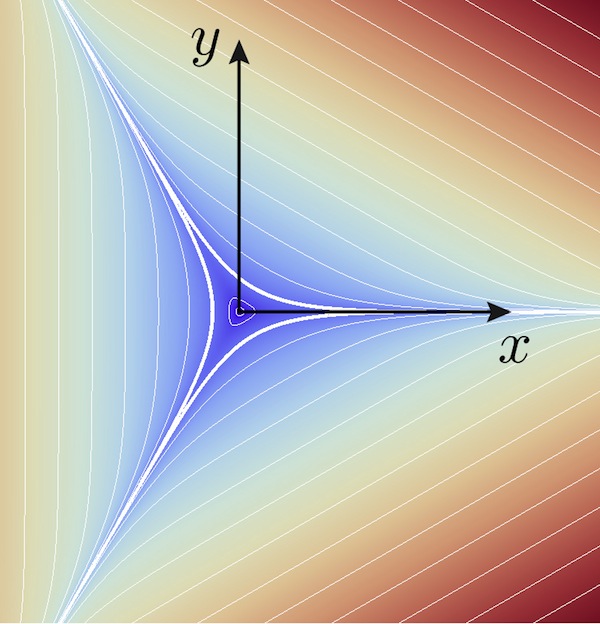}
\end{center}
\caption{\it \small Bianchi IX shape potential in the  $x,y$-plane. The thick white line delimits the region where the 
potential becomes negative. The circle at the centre surrounds the point $x=y=0$.}\label{BianchiIXpicture}
\end{figure}

We prove now that York time $\tau$ is monotonic. Its derivative wrt parameter time\footnote{With `parameter time' we mean the
foliation {label} obtained by choosing a constant unit lapse.} 
\begin{equation}
\dot \tau = \{ \tau , \mathcal H \} =  \frac 9 4 \, \tau^2 v - 16  \, v^{\frac 1 3} \,  V_\st{S} \,,
\end{equation}
can be shown to be positive by using the Hamiltonian constraint (\ref{equ:B9constraint}):
\begin{equation}
\dot \tau =   \frac 3 4 \, \tau^2 v  + \frac 2 3  v^{-1} \left( \dimp_x^2  +  \dimp_y^2 \right) \geq 0 \,.
\end{equation}
Now comes the important difference between Bianchi IX and the $3$-body problem, which carries forward to full GR and the \nbn: the derivative of $v$ is
\begin{equation}\label{VdecreasesWithTau}
\dot v = \{ v , \mathcal H \} =  - \frac 9 4 \, \tau \, v^2   ~~\Longrightarrow~~  \dot {(v^{-1})} = - \frac{\dot v }{v^2} =  \frac 9 4 \, \tau \,,
\end{equation}
so it is $v^{-1}$ that is concave upwards, and not $v$, as the parallel with the $N$-body problem would suggest. This obviously means that $v$  is concave \emph{downwards}. The reason for this {difference, which we discuss in more detail below,} is ultimately traced back to the negative sign of the dilatational kinetic
energy in the Hamiltonian constraint (\ref{equ:B9constraint}) {(which in turn derives from the form of the Einstein--Hilbert action)}. The corresponding term in the $N$-body 
Hamiltonian constraint  has the opposite (positive) sign.

Now recall the dimensional analysis shown in Sec. \ref{SDinCS}: 
the 3D metric has dimensions of an area $[g_{ab}]=\ell^2$.
Then the volume obviously has dimensions $[v] =\ell^3$, and the shape momenta and York time, as can be read off Eq. (\ref{equ:B9constraint}), have
\begin{equation}
[\pi_x] = [\pi_y] = \ell^2 \,, \qquad [\tau]=  \ell^{-1} \,.
\end{equation}
The shape potential is of course dimensionless. 

As in the particle model, we can now pass to an autonomous Hamiltonian with, however, \e{anti-dissipation}. The SD Hamiltonian $H_{\st{SD}}$ is the conformal York Hamiltonian, which is the positive root of (\ref{equ:B9constraint}) if considered as an equation for $v$. It generates evolution in York time $\tau$, which we treat as our time variable from now on. Using the dimensionless quantity $z:=\tau^2 \, v^{2/3}$, we obtain $H_{\st{SD}}= \tau^{-3} \, z_+^{3/2}$, where $z_+$ is the positive root of
\begin{equation}\label{equ:scaling}
 z^3-u\, z^2- k =0 \,, \qquad u={\textstyle \frac{32}{3}}  V_\st{S} \,,   \qquad k = {\textstyle  \frac{8}{9}} \tau^{4}(\dimp_x^2+\dimp_x^2) \,.
\end{equation}
The discriminant of Eq. (\ref{equ:scaling}) is
\begin{equation} \label{discriminant}
\Delta = - k \, \left( 27 \, k + 4 \, u^3  \right) = - k \, \delta \,,
\end{equation}
and $u$ is bounded from below by $u\geq - 16$. Therefore, whenever $u\geq 0$ or $u$ is negative but $27 \, k > - 4 \, u^3$ we have a unique real and positive root,
\begin{equation}\label{equ:B9SDHamiltonian}
\mathcal  H_{\st{SD}} = \frac{1}{3 \tau ^3} \left[ u +   \left(1- {\textstyle \frac 1 2}  \delta \, u^3+ {\textstyle \frac {3 \sqrt{3}} 2} \sqrt{ \delta } \right)^{\frac 1 3}  +  \left(1 - {\textstyle \frac 1 2}  \delta \, u^3 + {\textstyle \frac {3 \sqrt{3}} 2} \sqrt{\delta } \right)^{-\frac 1 3} \right]\,.
\end{equation}
If $u$ is negative and  $27 \, k \leq - 4 \, u^3$ we have three real roots, but two of them are negative, and the positive one has still the form (\ref{equ:B9SDHamiltonian}).

Dimensional analysis shows that equation (\ref{equ:scaling}) still holds after the simultaneous rescalings $z\to\lambda^2\,x$, $u\to\lambda^2\,u$, $k\to\lambda^6\,k$. This implies that the non-canonical transformation
\begin{equation}\label{equ:nonCanonicalTrf}
 \p_x:=\tau^2\,\dimp_x \,, \qquad  \p_y :=\tau^2\,\dimp_y \,,
\end{equation}
allows us to express the Hamiltonian in time-independent form as
\begin{equation}\label{equ:unnamed}
  \begin{array}{rcl}
    \mathcal  H_{\st{SD}}(x,y,\dimp_x,\dimp_y;\tau)&=&\tau^{-3} \mathcal  H_{\st{SD}}(x,y,\tau^2 \, \dimp_x, \tau^2 \, \dimp_y;1) \vspace{6pt} \\
    & =:&\tau^{-3}H_0(x,y,\p_x,\p_y) \,.
  \end{array}
\end{equation}
Using $H_0$, the dimensionless Poisson-bracket $ \ldm f,g \rdm:=\tau^{-2}\{f,g\}$ and the logarithmic time $\lambda:=\ln(\tau/\tau_0)$, we find the non-canonical equations of motion
\begin{equation}\label{AutonomousDissipativeEqnB9}
 \begin{array}{rcl}
   \frac{d}{d \lambda} f(x,y)&=&\ldm f(x,y),H_0 \rdm \,, \vspace{6pt}\\
   \frac{d \p^i}{d \lambda} &=&\ldm \p^i,H_0 \rdm+2 \,\p^i \,.
 \end{array}
\end{equation}

\begin{figure}[t]
        \centering
        \begin{subfigure}[t]{0.4\textwidth}
                \centering
                \includegraphics[width=\textwidth]{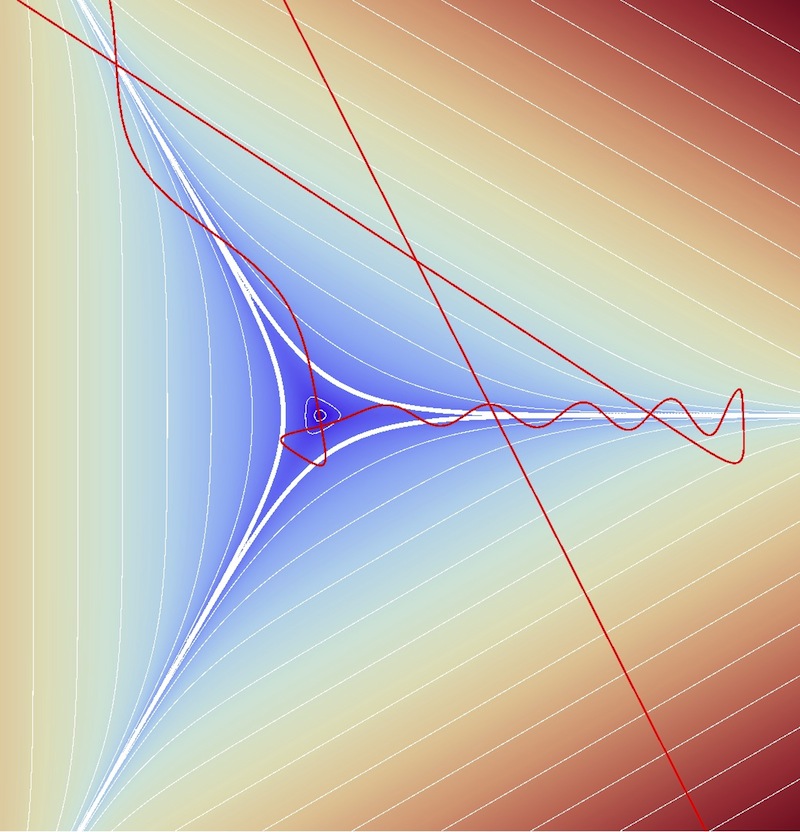}
                \caption{\it }
                \label{BianchiIXorbit}
        \end{subfigure}
~~~~~~~
          \begin{subfigure}[t]{0.5\textwidth}
                \centering
{\includegraphics[width=\textwidth]{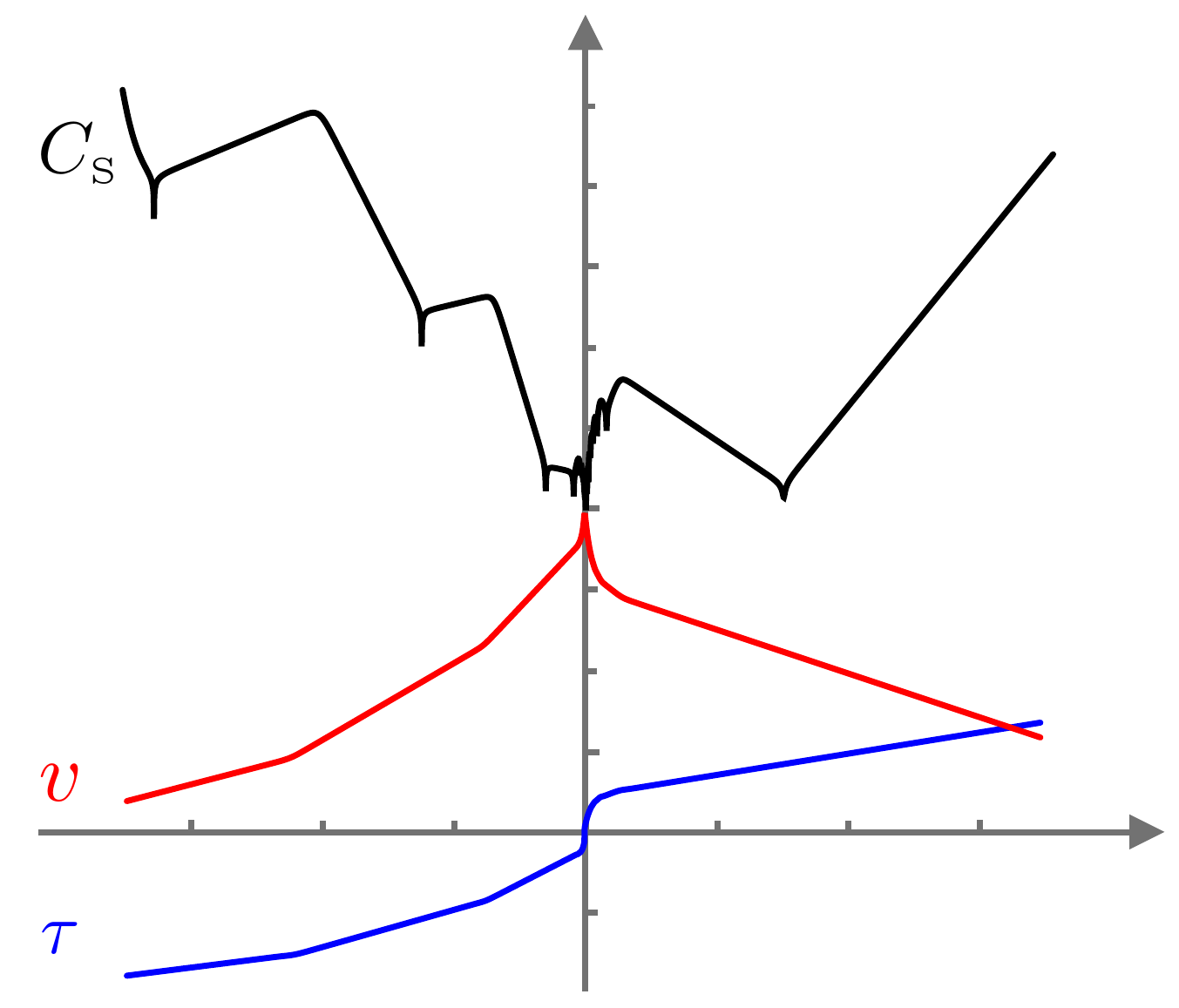}}
                \caption{\it }
                \label{V-tau-CoordinatizedTime}
        \end{subfigure}
        \caption{\\ \it \small (a) Plot of the trajectory in shape space (the $x$,$y$-plane) of a typical Bianchi IX orbit. Near the $\tau=0$ point
        the motion remains close to the minimum of the potential and within the three `gorges'. Far from $\tau=0$, when the momenta
        grow at an exponential rate, the motion is mostly on straight lines (because the momenta are so large that they are not
        affected by the potential walls), with sudden `bounces' when the representative point climbs far enough up the potential walls.\\
        (b) Plot of the dilatational momentum $\tau$ (in blue), which grows monotonically, together with the spatial volume
        $v$ (in red), which is `pyramid-shaped', and the `complexity function' $C_\st{S}$ (in black), which grows (on average) away from the  point $\tau =0 $. }\label{BianchiIXyorktime}
\end{figure}

Study of the solutions of these equations shows that the behaviour of Bianchi IX (and full GR) differs in important ways from Newtonian gravity. 
First, the potential (Fig.~\ref{BianchiIXpicture}) is very different from $V_\st{S}$ (Fig.~\ref{Figura}). It has a single minimum 
at the most homogeneous shape $x=y=0$ (the round metric on the $3$-sphere)
and diverges at infinity in all directions, apart from the three symmetry axes with polar
angles $0$, $2 \pi/3$ and $4\pi/3$ (in the $x,y$-plane).  It has 
triangular symmetry and is negative
inside the {thick white line}. The `walls' around the centre
are extremely steep, as can be seen in the section at $y=0$, 
where $ V_\st{S}(x,0) = \frac 1 2 \left( e^{-4\,x} - 4 \, e^{-x} \right)$. It
grows as $e^{4\,|x|}$ for negative $x$ and shrinks as $-e^{-|x|}$ for positive $x$.
The walls around the central region all grow exponentially,
apart from three `gorges' along the three symmetry axes, which
tend to zero exponentially. As a dynamical system on shape space, this is like
a particle in a triangular funnel with three thin inlets whose
bottoms remain at zero elevation.

The behaviour in parameter time is  \emph{anti-dissipative}: the term $+ 2 \, \p^i$ in Eq.~(\ref{AutonomousDissipativeEqnB9}) has the opposite sign wrt the analogous term $- \p^i$ in (\ref{DissipativeAutonomousEquations3BP}).
Therefore, the momenta tend to grow when the system leaves the turning point $\tau = 0$. The typical orbit has a central region near $\tau = 0$, where the system spends most of the time near the bottom of the potential (close to
the most homogeneous shape), and two branches `before' and `after' the central region (in parameter time),
where the kinetic energy grows exponentially. The representative point of the system behaves like a particle that initially has little kinetic energy and remains near the bottom of the funnel. It is then accelerated exponentially and bounces repeatedly off the funnel walls, reaching ever greater elevations and speeds. It can happen (as in the orbit depicted in Fig.~\ref{BianchiIXorbit}) that the particle, when in the central region of the orbit, gets trapped in an inlet and bounces back and forth between its walls, but it will eventually escape into another arm due to its ever-increasing kinetic energy.




{As we have noted, Bianchi IX (and with it, empty-space GR) differs substantially from the Newtonian 3-body (and $N$-body) problem in crucial signs. It will be helpful to review them together with other differences.}

The dilatational kinetic energy is non-positive in Bianchi IX, non-negative in the $3$-body problem. The shape potential is bounded from below in Bianchi IX, from above in the $3$-body problem. Finally, the shape momenta have the dimensions of the inverse of the dilatational momentum squared in Bianchi IX but of the dilatational momentum in the 3-body problem. Taken together, these differences make the
behaviour anti-dissipative in the direction of growing $\tau$ in Bianchi IX, dissipative in the $N$-body problem. Table~\ref{TableBianchiVS3body} summarizes the differences.

\begin{table}[h!] \begin{center}
\begin{tabular}{lccc}
\hline \hline
& \textbf{Bianchi IX} && \textbf{3-body problem}
 \\
 \hline 
 \\
 Kinetic term & $ \frac 1 2 \left(\dimp_x^2+\dimp_y^2-\frac 9 4 \tau^2\, v^2\right)$ &&  $\frac 1 2  \, R^{-2} \left( p_\theta^2 + \sin^2 \theta \, p_\phi^2+ \frac 1 4 \, D^2 \right)$
  \vspace{6pt}  \\ \hline  \\
 Potential & $V_\st{S} \in [-\frac 3 2 ,\infty) $ &&  $V_\st{S} \in (-\infty ,0) $
   \vspace{6pt}  \\ \hline \\ 
 Dissipation &  $ \frac{d \p^i}{d \lambda} $= $\ldm \p^i,H_0 \rdm+2 \,\p^i$  &&  $  \frac{d \p^i}{d \lambda} = \ldm \p^i,H_0 \rdm- \,\p^i $
    \vspace{6pt} \\
 \hline \hline
\end{tabular}
\caption{\it \small Sign differences between Bianchi IX and the 3-body problem.}\label{TableBianchiVS3body}
\end{center}
\end{table}

The upshot is that in both models complexity grows with $\tau$, but in coordinatized Bianchi IX this is accompanied by a shrinking volume; in the 3-body problem the size (\mn) grows with $\tau$. A further difference is that the complexity fluctuations in the \tb are bounded below by a monotonically growing function, but in Bianchi IX only by zero. 


As a final difference, we note that, besides the anti-dissipation, Bianchi IX fails to parallel the \tb in that there is no emergence and `storing' of dynamical information like that associated with the formation and stabilization of Kepler pairs. A problem here is that, unlike the passage from the \tb to the \nbn, there is no easy generalization from Bianchi IX to full vacuum SD. Even if there were, we do not think stable storage of dynamical information should emerge. For this we believe that matter must be present, but here too we are thwarted by the sheer difficulty of finding non-trivial solutions in GR. However, we think we can make a plausible suggestion for the definition of complexity. To this we now turn.

\subsection{Complexity in geometrodynamics}

A major achievement of 20th-century mathematics was the proof of the Yamabe conjecture: any Riemannian metric $g_{ij}$ on a compact manifold $\Sigma$ can be conformally transformed to a metric of constant scalar curvature $R$. This enables one to define the \e{Yamabe invariant} of any conformal 3-geometry on $\Sigma$. To do this, one considers all 3-metrics in the equivalence class it defines and finds the infimum of $R$ wrt all volume-preserving (ensured by the 1/3rd power of the volume in the denominator of (\ref{yam})) conformal transformations:
\begin{equation}
\mathcal Y = \inf_\phi \frac{ \,  \int{\textrm  d}^3x  \sqrt g \,\phi ( R \, \phi - 8 \, \nabla^2 \phi)}{ \left(\int d^3 x \sqrt g \, \phi^6 \right)^{\frac 1 3} } \,.\label{yam}
\end{equation}

From our perspective, it is encouraging that there is at least a partial analogy between $C_\st{S}$ in particle dynamics and $\mathcal Y$ in vacuum GR:
\begin{enumerate}
 \item $\mathcal Y$ is a function of the conformal geometry only.
 \item $\mathcal Y$ appears in the shape-dynamic description of GR in a way that \e{partially} resembles the way $V_\st{S}$ does in the $N$-body problem: the Hamiltonian constraints (\ref{ham}) and (\ref{admham}) reveal a clear analogy between $V_\st{New}$ and $R$ and hence between $V_\st{S}$ and $\mathcal Y$. We will expand on `partially' below.
  \item $\mathcal Y$ takes all values from $-\infty$ to its absolute maximum at the round metric (the most uniform shape) on $S^3$ and is therefore a measure of the `distance' from the round metric. This analogy with $C_\st{S}$ suggests  that $\mathcal Y$ could serve as a measure of `how structured a conformal geometry' is.
\end{enumerate}

The caveat `partially'  in point 2. is necessary because the Hamiltonian constraint (\ref{admham}) is \e{local}, in contrast to the global Hamiltonian (\ref{partham}) of the particle model. Thus, whereas $V_\st{S}$ is truly the potential that governs the particle dynamics, the local nature of the constraints in conformal geometry makes it impossible{, except in homogeneous models like Bianchi IX,} to separate potential and kinetic energy for the system as a whole. We can at best say that $\mathcal Y$ is an `average' gravitational shape potential energy.

Despite the caveat, we think the analogies between $C_\st{S}$ and $\mathcal Y$ are sufficiently close to warrant further investigation. The most important thing will be to generalize $\mathcal Y$ to include matter degrees of freedom. The way to attempt this is clear, since the local Hamiltonian constraints are then augmented by potential-type matter terms, for example, the square of the curl of the three-dimensional vector potential in the case of coupling to the Maxwell field. The generalization of (\ref{yam}) to include matter terms is therefore straightfoward, though the existence of an infimum must of course be studied.

In fact, since up to the present epoch gravitational waves appear to have been very weak, at least far from the most violent gravitational events, it seems possible that an expression of the kind suggested at the end of Sec.~\ref{exc} will be a relatively good measure of the complexity of the Universe and reflect the formation and increasing relative separation of gravitationally bound systems.

In this connection, our inclination is to keep identifying the arrow of time with the direction of growing complexity \emph{in the matter sector}, because any measuring device, including one to observe geometry, is built from matter. We deduce our notion of time from material records of the past, not from gravitational waves.

One thing at least is secured by observation: the Universe has been getting ever more structured at least since the surface of last scattering. Now in Newtonian theory, which is a good model of GR for virtually everything that has happened gravitationally since the CMB epoch, we can understand very well how this happens. All generic solutions will seem to have evolved from near the most uniform shapes the Universe can have (at which $C_\st{S}$ has its minimum), into the kind of universe we currently observe. This suggests to us that for much of the history of the Universe \e{dissipative} gravitational behaviour of matter outweighs the anti-dissipative behaviour of vacuum gravity.

We conclude this section with a comparison of our approach to the various attempts that have been made to define gravitational entropy since Penrose's initial suggestion \cite{Penrose1979,Penrose1989} that near the big bang the Universe had near zero Weyl curvature. In one way or another, they all invoke gravitationally induced `clumpiness', defined through a 4D scalar,\footnote{An obvious candidate is the ratio of the Weyl and Ricci curvatures considered in \cite{Wainwright1984}.} which may be integrated over a 4D volume or a 3D hypersurface. This is very natural in the light of Penrose's suggestion. All such proposals to use a 4D scalar clearly respect the spirit of 4D general covariance. There is a useful discussion of the proposals (none of which are found to be fully satisfactory) of \cite{Wainwright1984, Hervik2001, Sussman2013, Clifton2013} in \cite{Bolejko2013}.} Although the `clumpiness' idea is obviously shared, our proposal differs in several respects. 

First, for the reasons given in Sec.~\ref{doubts}, we question whether it is appropriate to try to define entropy  for the Universe, as opposed to subsystems of it. However, we do not rule out some connection with an entropy-like quantity. For this, we would regard the complexity (\cs or the Yamabe invariant) as a macroscopic variable analogous to pressure or temperature for which, given a suitably defined metric and coarse graining on shape space, one could define an `entropy' through the logarithm of the corresponding number of shapes that have the given complexity. However, as of now, we regard complexity as the primary quantity of interest. 

Second, where we differ most strongly, our proposal is three dimensional. We do not need to repeat the arguments that led us to this position, but we would like to point out that it yields a well-defined integral quantity at each instant of cosmic history, whereas integrals of 4D scalars inevitably depend on arbitrary choices of a 4D volume or a 3D spacelike hypersurface.

There is one respect in which our ideas, extended to GR with matter, will 
be rather close to Penrose's original suggestion, which was that near the big 
bang the geometry had exceptionally low entropy but the matter was in thermal 
equilibrium. Now the topographical ideas we took over above from the \nb to 
vacuum GR clearly need extension to include matter. Then uniform geometry on 
$S^3$ will appear in shape space alongside structure that represents matter. 
If our ideas are a move in the right direction, this further structure must 
give rise to matter thermal equilibrium while the 3D conformal geometry still 
remains `on the plateau' of very uniform geometry.

Finally, we are not aware of any attempt to use the \nb as heuristic guide to the arrows of time in GR, though it is in \cite{GibbonsEllis} as a cosmology model. We think the close parallels in their key architectonic features, in accordance with which two shapes and just one dimensionless number fix a solution, give strong support to our approach. It suggests that the entire incredibly rich structure of the Universe arises from the presence of that one extra number. We now want to consider the role it might play in quantum gravity.

\section{Possible Quantum Implications}

\subsection{A time-dependent Planck constant}
\label{sec:QuantumGravity} 

Shape space quantities are dimensionless. As shown above, gravitational dynamics can be equivalently formulated in 
$\shs$ as a theory of evolving dimensionless conformally-invariant degrees of freedom. The dynamics on $\shs$ is
naturally described in terms of dimensionless Poisson brackets $\ldm \cdot \, , \, \cdot \rdm$, and the coordinatized,
dimensionful description is recovered, both in the Newtonian and the GR case, in the form
\bq \label{DimensionlessPBdefinition}
\{f,g\}  = D^a \, \ldm f,g \rdm \,,  \qquad a= -1 ~~\text{\it \rm{(Newton)}} \,, ~~~ a = 2 ~~ \text{\it \rm{(GR)}}\,,
\ee
where $D$ is a fixed function of the dissipative time $\lambda$,  $D = D_0 \, \exp \, \lambda $, and $D_0$ is a conventional dimensionful constant with dimensions (action)$^{\frac 1 a}$. It can be seen immediately that $D^{a}$ plays a role analogous to that of $\hbar$, in that it transforms dimensionless 
brackets (like the commutators of quantum mechanics) into the familiar, dimensionful Poisson brackets.

Many observations indicate that the Planck constant is (over a large time interval, at least) an epoch-independent constant of nature. This means that in the \e{coordinatized} description the quantization rule for a set of elementary phase space functions is
\begin{equation}\label{QuantizationRule}
\{f,g\} \to \frac 1{ i \, \hbar} \, [\hat f , \hat g] \,.
\end{equation}
This gives quantum mechanics an intrinsic scale. But scale has only a relative meaning, and the dynamics of the Universe unfolds on shape space. Therefore there should be a law acting on a deeper level than (\ref{QuantizationRule}) that determines the magnitude of quantum effects \emph{in shape space} and such that they appear to be controlled by an unchanging scale in the coordinatized description.

To construct a quantum theory on $\shs$, it is natural to start with the dimensionless Poisson
brackets  $\ldm \cdot \, , \, \cdot \rdm$ and the quantization rule
\begin{equation}
\ldm f , g \rdm \to \frac 1 { i \, \hbar_\st{S}} \,  [\hat f , \hat g] \,,
\end{equation}
where $\hbar_\st{S}$ is a \emph{dimensionless} c-number.\footnote{M. Lostaglio and two of us already noted \cite{BLMpaper} that SD applied consistently requires the magnitude of the quantum effects on $\shs$ to be controlled by a dimensionless `Planck constant' $\hbar_\st{S}$.} Using now Eq. (\ref{DimensionlessPBdefinition})
\begin{equation}
\{ f , g \} \to \frac{D^a} {i \,\hbar_\st{S}} \,  [\hat f , \hat g] \,,
\end{equation}
and comparing with Eq.  (\ref{QuantizationRule}) we get 
\begin{equation}
\hbar = \frac{\hbar_\st{S}}{D^a} \,. 
\end{equation}
Thus, since $D^a$ is epoch dependent, the dimensionless constant in the quantization rule in $\shs$ must be time dependent if the Planck constant in the coordinatized description is to be (at least {to the observed accuracy}) unchanging. If the shape-dynamic conceptual framework is accepted, this is a direct consequence of dimensional analysis and suggests a simple unified picture of evolution of the Universe, which we shall now briefly outline.

We note first that a commonly employed measure of the `size of the Universe' is the ratio of the Hubble radius, assumed to be epoch dependent, divided by the Planck length, assumed to be constant. This ratio is currently $\approx 10^{61}$ and is said to be `increasing because the Universe is expanding'. But only the dimensionless ratio $\approx 10^{61}$ corresponds to an objective fact, so the interpretation in terms of expansion is questionable. Shape Dynamics suggests an alternative, namely, that \e{all} change in the Universe, including its `expansion', arises for just one reason: a dimensionless \e{epoch-dependent} quantity $\hbar_\st{S}$ determines the strength of quantum effects on $\shs$.

We can give heuristic arguments for this based on our toy model: the \nbn. We have seen that, classically, clustering occurs spontaneously and causes $C_\st{S}$ to grow, typically linearly with time. This corresponds rather accurately to `expansion' of the Newtonian Universe as it inescapably sinks ever lower into its shape potential $V_\st{S}$. The point is that bound systems like Kepler pairs keep the same size relative to each other while the distance between them, measured by these `rods', grows linearly. This very closely matches what we observe in our Universe.

Now let us consider quantum mechanics. The quantization procedure is by no means unique: there is a large class of quantum theories that admit the same classical limit. Reducing this ambiguity to the bare minimum, one is left with a one-parameter family of quantum theories that generate the same semiclassical trajectory: they are the ones in which the momenta are represented as $p^i = - i \, \hbar_\st{S} \frac{\partial}{\partial s_i}$ with different values of $\hbar_\st{S}$. These theories generate different quantum wavefunctions that spread out from the classical trajectory to different degrees. This fact is inevitable, and is present even if one ignores all other quantization ambiguities (like ordering issues). It is related to the fact that to generalize
a curve to a wavefunction we need at least one `scale'.\footnote{From a wavefunction one can calculate variances around the mean values, like $\langle (s_i - \langle s_i\rangle)^2\rangle$, which have a magnitude related to $\hbar_\st{S}^2$. With `scale' in this
particular model we mean a dimensionless portion of shape space, as measured by the natural metric on $\shs$.} Our $\hbar_\st{S}^2$ determines the analogue of a `Bohr radius' on $\shs$; it is still a dimensionless and relational concept, but it expresses `over what portion of shape space the wavefunction is spread out'.


In \cite{BLMpaper}, it was pointed out that there seems to be no reason why the dimensionless number $\hbar_\st{S}$ should have one value rather than another. We have also noted that there is an ambiguity inherent in quantum theories on $\shs$. We now suggest tentatively that attempting to fix this ambiguity might be a mistake, that \e{all} values are realized successively, and that such change in the value of $\hbar_\st{S}$ is the sole cause of all change in the Universe.

 Let us first consider, for example, the model of \cite{BLMpaper}, which is the simplest theory one can
build on $\shs$: the Hamiltonian is
\begin{equation}
\mathcal H_\st{S} = p_\theta^2 + \sin^{-2} \theta \, p_\phi^2 - C_\st{S} (\theta,\phi) \,.
\end{equation}
The natural quantization of this Hamiltonian is
\begin{equation}
\hat {\mathcal H}_\st{S} = - \hbar_\st{S}^2 \, \nabla^2 - C_\st{S} (\theta,\phi) \,,\label{BLMS}
\end{equation}
where $\nabla^2$ is the Laplacian on the 2-sphere. We see that $\hbar_\st{S}$ determines
the relative weight of the kinetic vs.~the potential term. If $\hbar_\st{S}$ is allowed to change,
it goes from $\hbar_\st{S} \to \infty$ where the kinetic term dominates, and the operator is basically identical to the Laplacian, with its positive-definite spectrum $\hbar_\st{S}^2\, l \, (l+1)$, $l \in \mathbbm{N}$, to $\hbar_\st{S} \to 0$, where the Hamiltonian becomes the multiplicative operator $- C_\st{S} (\theta,\phi)$, which has a negative-definite continuum spectrum that is bounded from above but not below.

In the above we started from the simplest (geodesic) quantum theory on $\shs$ with a dimensionless Planck constant $\hbar_\st{S}$. Let us now, more realistically, consider the Newtonian (non-geodesic) 3-body problem on $\shs$. Its quantization could start from the time-dependent Hamiltonian (\ref{ShapeHamiltonian}), which we reproduce here with $D$ explicitly shown:
\begin{equation}
\mathcal H = \log \left( \frac 1 2 \frac{p_\theta^2 + \sin^{-2} \theta \, p_\phi^2 + \frac 1 4  D^2 }{D_0 \, C_\st{S}  (\theta,\phi)} \right) \,. \label{ShapeHamiltonianREPRO}
\end{equation}
$\mathcal H $ generates $D$-translations and $ p_s^i $ generate $s_i$-translations. Recall that $ p_s^i $ are still dimensionful at this stage, and the dynamics is generated by dimensionful Poisson brackets. We might naively quantize this system with
the rules
\begin{equation}
\hat{ \mathcal H} =  - i \, \hbar  \frac{\partial }{\partial D}\,  \qquad  \hat p_s^i =   - i \, \hbar  \frac{\partial }{\partial s_i} \,,
\end{equation}
where $\hbar$ is the \e{dimensionful} Planck constant, which, by what we said above, should be expected to be
constant in time. Then the $D$-evolution of the wavefunction on $\shs$ would be generated by the Schr\"odinger equation
(notice that we chose a definite particularly simple ordering choice)
\begin{equation}
-i \, \hbar \frac{\partial \psi}{\partial D} = \log \left(\frac{- \, \hbar^2 \nabla^2 + \frac 1 4 D^2 }{2 \,D_0}\right) \psi -\log\left(\, C_\st{S}  (\theta,\phi) \right)  \psi \,.
\end{equation}
Here, $D_0$ is conventional and has the same dimensions as $\hbar$. Thus, it is mere convention to choose $D_0=\hbar$. Now, calling $\dimt = D/D_0 = D/\hbar$, we obtain
\begin{equation}
-i \, \frac{\partial \psi}{\partial \dimt} = \log \left(- \frac 1 2 \nabla^2 + \frac 1 8  \dimt^2  \right)  \psi  - \log (C_\st{S} )  \psi \,.\label{QU}
\end{equation}
We obtain a kind of `time-dependent Schr\"odinger equation' that contains \e{no arbitrary constant}, and, through the presence of the $\dimt^2/8$ term, has an unconventional form.

Despite this, we should like to explain why we find the idea of a single dimensionless independent variable which generates all change and, in a certain sense, simultaneously plays the role of both time and the strength of quantum effects appealing.
It is well known that quantum probability tends to collect in \e{dimensionful} potential wells until the Heisenberg uncertainty principle halts too deep a descent (which is the reason why
the hydrogen atom is stable). {If we examine (\ref{BLMS}), the simpler of the two `quantum' equations considered above, we see that $1/\hbar_\st{S}^2$ measures the strength of the potential $C_\st{S}$ relative to the kinetic Laplacian.} For greater strengths of the potential, we would expect the quantum probablity density of the wave function of the Universe to sink deeper into the wells of the \e{dimensionless} shape potential $V_\st{S}$. This matches what happens in the classical theory, in which the evolution trajectories pass ever further down the potential wells. Increasing the value of $1/\hbar_\st{S}$ is like increasing the resolution of a dimensionless analogue of a `microscope' used to examine shape space $\shs$ on ever finer scales.


This would then be the reason why the ratio of the distance between hydrogen atoms in the Galaxy and a very distant galaxy divided by the Bohr radius is increasing. Moreover, $1/\hbar_\st{S}$ is then not a function of time but time itself. All change in the Universe, manifested as increasing complexity and growth in the ratio of certain length scales,\footnote{In the \nbn, the ratio of the distance between Kepler pairs in subsystems (models of clusters of galaxies as Saari suggests \cite{Saari2011}) to the distance between subsystems.} arises from its monotonic increase. The so-called expansion of the Universe would not be a classical effect but the dominant macroscopic manifestation of the fundamental quantum nature of the Universe.



\subsection{Metrogenesis and hylogenesis \label{mhylo}}

In Sec.~\ref{hylo} we intoduced the notions of metrogenesis and hylogenesis. In geometrodynamics, the same notions are appropriate and arise in a much more sophisticated way than in particle dynamics. We saw that for metrogenesis in Sec.~\ref{spcon}. A point we should like to make here is that we live in a universe in which hylogenesis has taken place with massive consequences. Near and far, all the structures that we see around us in the Universe owe their origin to it. This leads us to ask how safe it is to apply techniques and results obtained in our epoch to epochs significantly before the surface of last scattering, which reveals to us the CMB. We suspect the methods in which we have gained trust will fail somewhere towards the `past' of our curve on shape space. In the $N$-body problem
this happens when the universe is very homogeneous and no rods and clocks can be formed. We see no reason to suppose it will be any different in GR. The need to develop new concepts and techniques might arise much closer to us than the Planck era. Indeed, we think the very definition of the Planck era may need re-examination. For example, how is one to think about the Planck units in shape space? 

The shape-dynamic approach may also be important for our ideas about singularities, at least some of which can be seen to be mere artifacts of taking spacetime as the fundamental ontology.

\section{Conclusions and Outlook}
\label{sec:Conclusion}

Numerous arrows of time can be identified. Perhaps the four most fundamental are entropy growth, retarded potentials in electrodynamics, wave-function collapse in quantum mechanics, and complexity growth. We do not wish to make any exaggerated claims for this paper. What we have done is show that two-sided growth of complexity and information is generic in NG and used that result to identify a candidate measure of complexity in GR with and without matter. We think we have also made a plausible case that irreversibility in the Universe should be studied in the first place through the evolution of its three-dimensional shapes. The shape-dynamic identification of a distinguished notion of simultaneity in Einstein gravity makes this a defensible position. But there is clearly a long way to go. In this final section, we give some indication of how we look to explain the most fundamental arrows in an approach based on shape-space ontology.

1. Because all the arrows are observed to point in the same direction everywhere and at all epochs, many earlier discussions have conjectured that they have a common cause in the expansion of the Universe. We seek a deeper explanation. On $\shs$ the most predictive law, with minimal initial data, is geodesic: a shape and direction in $\shs$ determine the evolution. But such a theory is dull; it leads to no secular growth of complexity. Next in predictive power are the theories we have considered; for them the initial data are a shape and a shape velocity. At least in NG, either two- or one-sided complexity growth is inevitable. One of its manifestations is emergent `expansion' of the Universe as measured by equally emergent rods and clocks. Thus, expansion, assumed as a primary cause in many earlier discussions, is one and the same thing as growing complexity, and both emerge solely through the law.

2. Our approach may cast light on cosmology. Ellis and Gibbons \cite{GibbonsEllis} note that $N$-body homothetic  solutions are models of FLRW solutions. Both belong to measure-zero sets and pose ``an intriguing fine-tuning problem'' (\cite{GibbonsEllis}, Sec.~3.4) because ``we currently see a FLRW type homothetic expansion. But in order to get such a flow, the initial positions of the particles must be constrained to satisfy [the central configuration condition].'' The explanation ``for such a fine tuning [is] presumably to be sought in an initial relativistic state that results at late times in such a Newtonian configuration.'' This is the standard past-hypothesis approach: a special initial condition creates a FLRW solution that is then perturbed into the kind of universe we observe today. If our conjecture is correct, the past hypothesis is redundant: the generic solutions of the law will all resemble perturbed FLRW.\footnote{We also wonder about the proposal in \cite{GibbonsEllis} to consider, in a follow-up paper, perturbation of homothetic solutions. The problem we see is that in $\shs$ both the homothetic $N$-body solutions and the FLRW solutions are mere points. There cannot be a small perturbation from a point to a complete solution curve. Moreover, all curves that emanate from a central configuration, which is an \e{unstable} equilibrium point, rapidly become effectively indistinguishable from generic solutions and inevitably tend to FLRW behaviour.} We see this as an indication of how a shape-dynamic perspective could inform a new approach to cosmology, shifting attention from the search for special initial conditions to the structure of shape space and its effect on the solution asymptotics. We have seen how this structure is the true cause of striking time-asymmetric effects.

3. The SD perspective also suggests a new approach to the retarded-potential enigma. The most popular attempts, originated by Wheeler and Feynman \cite{Wheeler1945, Wheeler1949}, rely on expansion of the Universe and conditions which ensure absorbtion of all advanced radiation, but these ideas have not gained wide acceptance. All we will note is that this approach relies on special structure within an arena provided by an individual spacetime. The problem might take a very different form in shape space, as we have already seen for complexity growth.

4. As regards entropy growth, we will have to show that it emerges generically in subsystems of the universe. We cannot begin to attack this problem before our shape-dynamic formalism has been fully developed to include realistic matter fields, whose forces are so vastly greater than gravity. Since the era of primordial nucleosynthesis, there has been significant degradation of energy through fission in stars. We have to explain the creation of the low-entropy hydrogen and helium. Equally pressing: how did the Universe get into a state as close to radiative thermal equilibrium as we observe in the CMB at the surface of last scattering? Our best hope at the moment is the importance of shape-space topography in SD, as noted in the penultimate paragraph of Sec.~\ref{mhylo}. In the discussion of our putative quantum equation (\ref{QU}), we likened increasing
the value of $1/\hbar_\st{S}$, and with it moving time forward, to sharpening the resolution of a `microscope' that examines $\shs$ to ever greater depths of the potential. We also noted that the $N$-body $\shs$ has the topography of an undulating plateau dotted with deep wells. The points on the plateau have low complexity and high uniformity. At least in the \nbn, the first `features' that the `microscope' will discern and assign Born amplititude, i.e., probablility, will be those on the plateau, so the quantum state will correspond to a superposition of many nearly uniform matter distributions. Thus, the isotropy and homogeneity of the CMB, which inflation cannot explain, could  arise of necessity from a quantum law of the form (\ref{QU}) and the topography of $\shs$. Of course, a caveat must here be made: it remains to be shown what topography, if it can be meaningfully defined, exists on the appropriate shape space with matter fields included.  

5. We noted that a classical Machian universe has gauge symmetries, but that in the asymptotic regime subsystems arise for which there are global symmetries (rotations, spatial and time translations. In accordance with Noether's first theorem, the latter lead to conserved quantities, whereas all these must be exactly zero for the Universe. Given the central role conserved quantities play in quantum mechanics (QM), this suggests that, just as with classical physics, the quantum Universe can be expected to behave in a qualitatively very different way compared with its subsystems. We have already made suggestions along these lines in Sec.~\ref{sec:QuantumGravity}. To conclude the paper, we will consider implications for the quantum measurement problem and its time asymmetry.

With Bohm and Bell and in full accord with our shape-space ontology, we assume that configurations are fundamental. We now recall the concept of a \e{time capsule}, introduced in \cite{Barbour1994,Barbour1999}. This is a static configuration whose structure suggests it has arisen through some historical process governed by definite laws. A single `snapshot' of the Universe at the present epoch reveals just such a remarkable structure with many similar substructures embedded within it, for example hundreds of planetary systems around stars in our region of the Galaxy. Already the asymptotic \nb gives rise to instantaneous structures that `call for dynamical explanation', namely statistically unlikely close pairs. 

Quantum mechanics as we know it in the laboratory is capable of creating the most remarkable time capsules. Indeed, Mott's explanation of why $\alpha$-particles make straight tracks in cloud chambers is a famous early example \cite{MottPaper}: Gamow had recently explained radioactivity through QM tunneling, but this led to a wave function spreading out isotropically in 3D space. This seemed to be in the most flagrant contradiction with cloud chamber experiments, which always led to $\alpha$ detection at individual points. Only statistically did an isotropic distribution build up. Mott resolved the problem with the remark ``The difficulty that we have in picturing how it is that a spherical wave can produce a straight track arises from our tendency to picture the wave as existing in ordinary three dimensional space, whereas we are really dealing with wave functions in the multispace formed by the coordinates both of the $\alpha$-particle and of every atom in the Wilson chamber.'' Mott then showed that the complete wave function would be concentrated on time capsules as defined above. Bell's discussion \cite{Bell1987} of Mott's paper in his gloss of Everett's many-worlds interpretation of QM is illuminating.

A remarkable aspect of the Mott example is its `snapshot-within-snapshot' aspect. When the cloud chamber is examined and its wave function `collapsed' (according to the Copenhagen interpretion), it reveals a complete track with multiple ionizations at which the $\alpha$-particle undergoes small-angle deflections in accordance with the appropriate QM probabilities for single-particle scattering. Each ionization can be interpreted as a measurement process that is simultaneously a state-preparation process for the next ionization. The complete track is a snapshot that contains within it multiple records of state measurements preceded by records of the state preparation. Moreover, any complete track exhibits a pronounced arrow of time, since the $\alpha$-particle loses energy on each ionization, and its deflection angles become greater.

Although the Mott example and others like it are striking, they are still not fully satisfactory from our point of view since they rely on a perturbative expansion in which an arrow of time is effectively put into the zeroth approximation by hand: the radioactive atom is excited, all atoms of the cloud chamber are in their ground state, and the cloud chamber itself is in a metastable state.\footnote{To amplify a tiny quantum disturbance into a macroscopic effect, all detectors must be in such a state.} We suspect many mysteries observed in the laboratory arise from special conditions created there by both the Universe, which is responsible for local inertial frames and finite-parameter symmetry groups, and experimentalists. We have already seen how an arrow of time and time capsules in the form of statistically unlikely shapes arise generically in the classical \nbn. We think our toy `wave equation of the universe' could do the same
in the quantum domain. If it does, the ultimate reason will surely be same as in the \nbn: the asymmetric structure of shape space and the fact that a law is defined on it in a way that fully respects its intrinsic structure.

\appendix

 \section{Appendices}
 
 \subsection{Growth of complexity in the $N$-body problem \label{MS}}

Let us summarize the main results of \cite{Marchial:1976fi} concerning the final state of the Newtonian $N$-body problem. For non-negative energy there are a number of interesting non-generic cases, i.e., cases that require initial conditions within lower-dimensional strata in phase space. Non-generic behaviour includes superhyperbolic escape and point particle collisions. We will focus our attention on the generic case. A rough description of it is as follows: The $N$-body system breaks up into subsystems, which are defined as groups of particles whose separations are bounded by $O(t^{2/3})$ in Newtonian time. Let us introduce the index $\mathcal J \subset \mathbbm{N}$ identifying the subsystem. A particle with
label $a \in \mathbbm{N}$ will belong to $\mathcal J$ if $ a \in \mathcal J $.
These subsystems are composed of clusters, whose constituents remain close to each other. Subsystems have a number of interesting properties, including the following:
\begin{enumerate}
 \item The centres of mass of the subsystems separate asymptotically linearly with Newtonian time, giving the subsystem $\mathcal J$ an asymptotically conserved linear momentum ${\bf P}_{\mathcal J}$. This implies that particle $a \in \mathcal J$ satisfies asymptotically
       \begin{equation}\label{equ:MarcialSaariAsymptotic}
         {\bf  r}_a= {\bf c}_{\mathcal J}\,t\,+\,\mathcal O (t^{2/3}), \qquad \text{if}~a \in \mathcal J\,,
       \end{equation}
       where ${\bf c}_{\mathcal J} \in \mathbbm{R}^3$ is a constant vector.
 \item Each subsystem $\mathcal J$ has asymptotically conserved quantities: energy $E_{\mathcal J}$ and angular momentum ${\bf J}_{\mathcal J}$, which satisfy bounds
       \begin{equation}
          E_{\mathcal J} =  E^0_{\mathcal J} + \mathcal O(t^{-5/3}) \,, \qquad  
          {\bf J}_{\mathcal J} =  {\bf J}^0_{\mathcal J} + \mathcal O(t^{-2/3}) \,.
       \end{equation}
 \item A subsystem with $E^0_{\mathcal J}=0$ tends to a central configuration; the deviation from a central configuration is bounded by
       \begin{equation} 
         \frac{m_a}{t^2} \left({\bf r}_a - {\bf R}_{\mathcal J} \right) + \frac{\partial V_{\mathcal J}}{\partial  {\bf r}_a}= o(t^{-4/3}) \,,
       \end{equation}
       where ${\bf R}_{\mathcal J}$ denotes the centre of mass of subsystem ${\mathcal J}$ and $V_{\mathcal J}$ its Newtonian potential.

\end{enumerate}

The generic final state of the $E=0$ Newtonian $N$-body problem (i.e. except for lower--dimensional strata in phase space), develops at least two distinct subsystems (which may contain only one particle or may be clusters). This means in particular that there are at least two distinct $ {\bf c}_{\mathcal J}$. One can thus show that the generic $E=0$ evolution of $C_\st{S}$ exhibits secular growth if at least one negative energy subsystem forms, i.e. if at least one $E^0_\mathcal{J}<0$ for a subsystem with at least two particles. 

To show this we write $C_\st{S}$ as the product of the weighted mean square distance $m_2=\sqrt{\sum_{a<b}w_{ab} r^2_{ab}}$ of the particles and the inverse of their weighted mean harmonic distance $1/m_{-1}=\sum_{a<b}\frac{w_{ab}}{r_{ab}}$. According to (\ref{equ:MarcialSaariAsymptotic}), we find after some time $t^0_\mathcal{J}$ that there is a constant $r_\mathcal{J}$ such that all constituent particles of subsystem $\mathcal J$ are within a ball of center ${\bf c}_{\mathcal J}\, t$ and radius $r_\mathcal{J}\,t^{2/3}$. Hence after time $t_{\mathcal I, \mathcal J}=\max\{8((r_\mathcal{I}+r_\mathcal{J})/\|{\bf c}_\mathcal{I}-{\bf c}_\mathcal{J}\|)^3,t^0_\mathcal{I},t^0_\mathcal{J}\}$ the smallest separation of a particle of subsystem $\mathcal I$ from a particle of subsystem $\mathcal J$, which we denote by $r_{\mathcal I, \mathcal J}$, will be $\frac t 2 \|{\bf c}_\mathcal{I}-{\bf c}_\mathcal{J}\|$. Since we exclude point-particle collisions, we can estimate
\begin{equation}
 m_2 <  \text{\it const.} \,  t \, \|{\bf c}_\mathcal{I}-{\bf c}_\mathcal{J}\| \,.
\end{equation}
To estimate $1/m_{-1}$, we use the assumption that at least one $E^0_\mathcal{I}<0$. Since we exclude superhyperbolic escape, we find the estimate
\begin{equation}
 1/m_{-1} < \frac{w_{\mathcal I , \mathcal I}}{\text{\it const.}},
\end{equation}
where $w_{\mathcal I , \mathcal I}=\min\{w_{ab}:a,b \in \mathcal I\}$, after some time $t^0$. Combination of these two estimates results in a bound
\begin{equation}
 C_\st{S} >  t\,A  ~~~~~\text{after time } T,
\end{equation}
where $A$ is a constant. However, $C_\st{S}$ is finite since we exclude superhyperbolic escape and point-particle collisions for all finite $t$. The bound thus implies secular growth of $C_\st{S}$, i.e., for every $t>T$ there exists in particular $\Delta = \Delta (t)$ such that $C_\st{S} (t+\Delta(t))>C_\st{S} (t)$. 
 
 \subsection{Phase-space reduction for the $3$-body problem}
\label{Appendix3bodyProblem}

In this appendix we will show how the coordinates $\mathbf w = (w_1,w_2,w_3)$ used in section~\ref{3BPsection}
and their conjugate momenta $\mathbf z = (z^1 , z^2 , z^3)$ are defined.

We will follow Montgomery~\cite{InfinitelyManySygizes} in the phase space reduction. The first simplification comes from the fact that, because the total angular momentum $\textbf{J}$ is zero, the problem is planar,\footnote{If the total angular momentum
is nonzero and it is not orthogonal to the plane identified by the three particles, then the motion is not planar.} so we can gauge fix
two out of three angular momentum constraints plus one translational constraint by fixing the plane in which the motion unfolds. We are left with three two-component vectors $\mathbf r_1$,
$\mathbf r_2$ and $\mathbf r_3$ that define the position of the three bodies on this plane.
Then we gauge fix the two remaining translation constraints by going to \emph{mass-weighted
Jacobi coordinates} \cite{BLMpaper,LimBinaryTree}:
 \begin{align}
\bm \rho_1 &= \sqrt{ \frac{m_1 \, m_2 }{m_1 + m_2} }\left( \mathbf  r_2 - \mathbf  r_1 \right) \,, \\
 \bm \rho_2 &= \sqrt{ \frac{m_3 \, (m_1+m_2) }{m_1 + m_2 + m_3} } \left(\mathbf r_3 - \frac{m_1 \, \mathbf r_1 + m_2 \, \mathbf r_2}{m_1+m_2} \right) \,,\\
 \bm \rho_3 &=  \frac{m_1\, \mathbf  r_1 + m_2 \, \mathbf  r_2 + m_3 \, \mathbf  r_3}{\sqrt{m_1+m_2+m_3}} \,.
\end{align}
The transformation is linear and invertible,
\begin{equation}
\bm \rho_a = {M_a}^b \, \mathbf r_b \,,\qquad  \det M = \sqrt{ m_1 m_2 m_3 } \,,
\end{equation}
and therefore it is easy to find a canonical extension for it, which transforms the momenta with the inverse matrix,
\begin{equation}
\bm \kappa^a = {(M^{-1})^a}_b \, \mathbf p^b \,.
\end{equation}
Note that the inverse matrix has the form 
\begin{equation}
{(M^{-1})^a}_b = \frac 1 3 \left(
\begin{array}{ccc}
 \frac{-m_2}{\sqrt{m_1 m_2  (m_1+m_2)}} & -{\frac{m_3}{\sqrt{(m_1+m_2+m_3)(m_1+m_2) m_3}}} & \frac 1 {\sqrt{m_1+m_2+m_3}} \\
 \frac{m_1}{ \sqrt{m_1 m_2  (m_1+m_2)}} &-{\frac{m_3}{\sqrt{(m_1+m_2+m_3)(m_1+m_2) m_3}}}  & \frac 1 {\sqrt{m_1+m_2+m_3}} \\
0 & {\frac{m_1+m_2}{\sqrt{(m_1+m_2+m_3)(m_1+m_2) m_3}}}  & \frac 1 {\sqrt{m_1+m_2+m_3}} \\
\end{array}
\right) \,,
\end{equation}
and has a constant column, namely, the column of $\bm \kappa^3$,
\begin{equation}
\bm \kappa^3 =   \frac 1 {\sqrt{m_1+m_2+m_3}} \sum_{a=1}^3 \mathbf p^a \propto \mathbf P_\st{tot} \approx 0 \,,
\end{equation}
which is therefore proportional to the momentum constraint and vanishes. The coordinates
$\bm \rho_3$ can consequently be discarded (they are the coordinates of the centre of mass).
The other two momenta are
\begin{align}
&\bm \kappa^1 =  \frac{ m_1 \, \mathbf p^2 -m_2 \, \mathbf p^1 }{\sqrt{m_1 m_2  (m_1+m_2)}}\,, \nonumber\\
&\bm \kappa^2 ={\frac{(m_1+m_2)  \mathbf p^3- m_3  \, \mathbf p^1 - m_3 \,  \mathbf p^2}{\sqrt{(m_1+m_2+m_3)(m_1+m_2) m_3}}} \,.
\end{align}
The above is a canonical transformation, leaving the Poisson brackets unchanged:
\begin{equation}
\{ \rho^i_a , \kappa^b_j \} = {\delta^i}_j \,{\delta_b}^a \,,
\end{equation}
and it diagonalizes the kinetic term in terms of the new momenta $\bm \kappa^a$,
\begin{equation}
T =  \sum_{a=1}^3 \frac{\mathbf p^a \cdot \mathbf p^a  }{2 \, m_a} =  \sum_{a=1}^3 \sum_{b,c=1}^3  \frac{{M_a}^b {M_a}^c   }{2 \, m_a}  \bm \kappa^b \cdot \bm \kappa^c \,.
\end{equation}
Remembering that $\bm \kappa^3 \approx 0$, it is easy to verify that
\begin{align}
&\sum_{a=1}^3  \sum_{b,c=1}^3 \frac{{M_a}^b {M_a}^c   }{2 \, m_a}   \bm \kappa^b \cdot \bm \kappa^c  \approx \sum_{a=1}^3  \sum_{b,c=1}^2 \frac{{M_a}^b {M_a}^c   }{2 \, m_a}  \bm \kappa^b \cdot \bm \kappa^c  \nonumber\\
&= \frac 1 2 \left(  \| \bm \kappa^1 \|^2 +  \| \bm \kappa^2 \|^2 \right)  \,.
\end{align}
The centre-of-mass moment of inertia is also diagonal in these coordinates, as can be explicitly verified:
\begin{equation}
I_\st{cm} = \sum_{a=1}^3 m_a \, \|  \mathbf r_a - \mathbf r_\st{cm} \|^2 = \| \bm \rho_1 \|^2 + \| \bm \rho_2 \|^2 \,.
\end{equation}

We are left with four coordinates $\bm \rho_1$, $\bm \rho_2$ and momenta $\bm \kappa^1$, $\bm \kappa^2$, and a single angular momentum constraint,
\begin{equation}
\mathbf{J}_\st{tot} = \sum_{a=1}^3 \, \mathbf r_a \times \mathbf p^a \approx  \sum_{a=1}^2 \, \bm \rho_a \times \bm \kappa^a \approx 0\,,
\end{equation}
 to reduce. The coordinates
 \begin{align} \label{DefinitionOfwCoordinates}
w_1 = \frac 1 2 \left(||\bm \rho_1||^2 - || \bm \rho_2 ||^2 \right) \;, \\
w_2 = \bm \rho_1 \cdot \bm \rho_2 \;,  \qquad   w_3 = \bm \rho_1 \times \bm \rho_2 \;, \nonumber
\end{align}
are invariant under the remaining rotational symmetry and therefore give a complete coordinate
system on the reduced configuration space~\cite{InfinitelyManySygizes}. The norm of the vector $\mathbf w $ is proportional  to the square of the moment of inertia \begin{equation}
||\mathbf w ||^2  = \frac 1 4\left(||\bm \rho_1||^2 + || \bm \rho_2 ||^2 \right)^2 = \frac{I^2_\st{cm}} 4 \,,
\end{equation}
so the angular coordinates in the three-space $(w_1,w_2,w_3)$ coordinatize shape space, which has the topology of a sphere \cite{InfinitelyManySygizes}. 

The momenta conjugate to $\mathbf  w$ are
\begin{align}
&z^1=   \frac{ \bm \rho_1 \cdot \bm \kappa^1 - \bm \rho_2 \cdot \bm \kappa^2 } {  \| \bm \rho_1\|^2 + \| \bm \rho_2\|^2} \,, & & 
z^2= \frac{ \bm \rho_2 \cdot \bm \kappa^1 + \bm \rho_1 \cdot \bm \kappa^2 } {\| \bm \rho_1\|^2 + \| \bm \rho_2\|^2} \,,  & &
z^3= \frac{\bm \rho_1 \times \bm \kappa^2 - \bm \rho_2 \times \bm \kappa^1 } {\| \bm \rho_1\|^2 + \| \bm \rho_2\|^2} \,; 
\end{align}
they are rotationally invariant as well. Notice that the kinetic energy decomposes as
\begin{equation}
T = \frac 1 2 \left(  \| \bm \kappa^1 \|^2 +  \| \bm \kappa^2 \|^2 \right) 
=\frac 1 2 \left( I_\st{cm} \, \| \mathbf z \|^2  + \frac{\| {\mathbf J}_\st{tot} \|^2}{I_\st{cm}} \right) \,,
\end{equation}
so we finally reduced the phase space to the cotangent bundle of $\mathsf{Q}_R^3$, the relative configuration
space.

To obtain the Newton potential, we now write the Hamiltonian in the coordinates $(\mathbf w,\mathbf z)$. To do this we need the
distance formula\footnote{Eq. (4.3.14) of \cite{InfinitelyManySygizes}; its proof is not trivial. A sketch of it
goes like this: a dynamical trajectory starting from any configuration in  $\mathsf{Q}_R^3$ and moving the particles
$a$ and $b$ along the straight line connecting them until they collide is a geodesic in  $\mathsf{Q}_R^3$ and
is everywhere perpendicular to the orbits of the Euclidean group. Then, by another theorem, the corresponding curve in shape space  $\sf{\bf S}^3$ is a geodesic as well, and the length of a geodesic connecting a point in  $\sf{\bf S}^3$ with a binary collision
is that given in the equation above.}
\begin{equation}
\| \mathbf r_a - \mathbf r_b \|^2 = \frac{m_a + m_b}{m_a \, m_b} \left( \| \mathbf w  \|- \mathbf w  \cdot \mathbf b_{ab} \right) \,,
\end{equation}
where $ \mathbf b_{ab} $ is the unit vector corresponding to the ray in  $\sf{\bf Q}_R^3$ associated to the binary collision
between particle $a$ and $b$. Notice that all of these  unit vectors lie in the plane $w_3 = 0$, which is the
plane of collinear configurations (called ``syzygy plane'' in \cite{InfinitelyManySygizes}). Then the equation above
can be rewritten
\begin{equation*}
\| \mathbf r_a - \mathbf r_b \|^2 = \frac{m_a + m_b}{m_a \, m_b} \left( \| \mathbf w  \|  -  w_1  \, \cos \, \phi_{ab} -  w_2  \, \sin \, \phi_{ab} \right) \,,
\end{equation*}
and the Newton potential takes the form
\begin{equation}
V_\st{New} = - \sum_{a<b} \frac{ (m_a \, m_b)^{\frac 3 2}(m_a + m_b)^{-\frac 1 2}}{\sqrt{ \| \mathbf w  \|  -  w_1  \, \cos \, \phi_{ab} -  w_2  \, \sin \, \phi_{ab}}} \,.
\end{equation}
Here, $\phi_{ab}$ are the angles identifying the two-body collisions (between body $a$ and body $b$) on the equator. They  can be explicitly calculated in the
generic-mass case by imposing $\mathbf r_a = \mathbf r_b$ and transforming to the $\mathbf w$ coordinates:

\begin{itemize}
\item

collision $1-2$: 
\begin{align}
\bm \rho_1 = 0 \,, \qquad  \bm \rho_2 = \sqrt{ \frac{m_3 \, (m_1+m_2) }{m_1 + m_2 + m_3} } \left(\mathbf r_3 - \mathbf r_1\right) \in \mathbbm R^2  \,,
\end{align}
\begin{align}
w_1 = -   \frac 1 2 || \bm \rho_2 ||^2 \;,  \qquad
w_2 = w_3 =0 \;,  \qquad \Rightarrow \qquad \theta = \frac \pi 2 \,, ~~~ \phi_{12} = \pi \,,
\end{align}

\item

collision $1-3$:
\begin{equation}
 \sqrt{m_2 \, m_3}  ~ \bm \rho_1   +  \sqrt{m_1 (m_1+m_2+m_3)} ~ \bm \rho_2
=0
\end{equation}
\begin{align*}
&w_1 = \frac 1 2 \, ||\bm \rho_1||^2 \left( 1 -  \frac{m_2 \, m_3}{m_1 (m_1+m_2+m_3)}  \right) \;, & &
w_2 = - ||  \bm \rho_1 ||^2 \, \sqrt{\frac{m_2 \, m_3}{m_1 (m_1+m_2+m_3)} }\;,  & \\
  & w_3 = 0 \;, \nonumber
\end{align*}
\begin{equation}
\Rightarrow \qquad \theta = \frac \pi 2 \,, \qquad \phi_{13}  = - \arctan \left(   2 \frac{\sqrt{m_1 \,m_2 \, m_3 (m_1+m_2+m_3) }}{m_1 (m_1+m_2+m_3) - m_2 \, m_3} \right) \,,
\end{equation}

\item

collision $2-3$:
\begin{equation}
 \sqrt{m_1 \, m_3}  ~ \bm \rho_1   -  \sqrt{m_2 (m_1+m_2+m_3)} ~ \bm \rho_2
=0
\end{equation}
\begin{align*}
&w_1 = \frac 1 2 \, ||\bm \rho_1||^2 \left( 1 -  \frac{m_1 \, m_3}{m_2 (m_1+m_2+m_3)}  \right) \;, & &
w_2 =||  \bm \rho_1 ||^2 \, \sqrt{\frac{m_1 \, m_3}{m_2 (m_1+m_2+m_3)} }\;, & \\
&  w_3 = 0 \;, \nonumber
\end{align*}
\begin{equation}
\Rightarrow \qquad  \theta = \frac \pi 2 \,, \qquad \phi_{23}  =  \arctan \left(   2 \frac{\sqrt{m_1 \,m_2 \, m_3 (m_1+m_2+m_3) }}{m_2 (m_1+m_2+m_3) - m_1 \, m_3} \right) \,.
\end{equation}
\end{itemize}
 
 \subsection{Positive energy $N$-body system}
 \label{TimsAppendix}

The shape-dynamical description of the $N$-body system with $E>0$ is similar to the description of the $N$--body problem with $E=0$, but with one important difference: the dynamics on $\shs$ can be described by an effective shape potential $\tilde C_\st{S}$ that is time-dependent, and is identical to $C_\st{S}$ only at the initial instant. The energy being positive, $C_\st{S}$ becomes weaker with time.
The energy $E$ defines a time scale $E^{-\frac 1 2}$ that controls the weakening of $\tilde C_\st{S}$. This leads to a qualitatively new final state: asymptotic freeze-out on shape space, which can be explained as follows: If the dynamics avoids the valleys of $C_\st{S}$ long enough then the secular flattening of the plateaus of $\tilde C_\st{S}$ `outruns' dissipation. In this case, the dynamics asymptotes to a
generic point in $\shs$, not only to a critical point of $C_\st{S}$. This is because the entire  potential $\tilde C_\st{S}$  becomes critical in the limit of infinite Newtonian time. The interpretation of this phenomenon in the coordinatized Newtonian representation is that the system undergoes homothetic expansion at late Newtonian times.

Let us look at the derivation of the Shape Dynamics Hamiltonian to see how the secular decay of the shape potential comes about. The time-dependent Hamiltonian is
\begin{equation}
 \mathcal H=\log\left[\frac 1 2 \frac{C_\st{S}}{E}\left(1-\sqrt{1+\frac{2 \, D^2+ 4 \, K_\st{S}}{\frac{C_\st{S}}{E}}}\right)\right],
\end{equation}
where $K_\st{S}$ is the shape kinetic energy. We see that the shape potential appears only in the combination $\frac{V_\st{S}}{E}$. We now use dynamical similarity, i.e. the simultaneous rescaling $K_\st{S} \to \tilde K_\st{S} = K_\st{S} \, D^{-2}$, $E \to \tilde E=D^2 \, E$, to make the kinetic term autonomous, i.e. independent of dilatational time $D$, and make the description of the system dimensionless. 
\begin{equation}
 \mathcal H=\log\left[\frac 1 2 \frac{C_\st{S}}{\tilde E}\left(1-\sqrt{1+\frac{2 + 4 \, \tilde K_\st{S}}{\frac{C_\st{S}}{\tilde  E}}}\right)\right],
\end{equation}
The shape potential thus enters the description only through $\frac{C_\st{S}}{D^2 \, E}$, which shows how the secular flattening of the shape potential comes about.

\subsection{The monotonicity of York time}
\label{MonotonicityYorkTime}

We will prove here that the CMC evolution of York time is monotonic. For this purpose it is useful to introduce the quantities
\begin{align}
&V=\int_\Sigma {\textrm d}^3x \sqrt{g} \,,& &\tau = {\textstyle{2 \over 3}} \langle p \rangle  = {\textstyle{2 \over 3}}  V^{-1}\int_\Sigma {\textrm d}^3 x \, p  \,,\\
& \bar g_{ab}=V^{-2/3} g_{ab} \,,& & \bar p^{ab}=V^{2/3}\left(p^{ab}- \textstyle{\frac 1 3} \langle p \rangle \, \sqrt g \, g^{ab}\right) \,.
\end{align}
The York time $\tau$ Poisson-commutes with $\bar g_{ab}$ and $\bar \pi^{ab}$ and is the momentum canonically conjugate to $V$.
The CMC-condition becomes
\begin{equation}
  \bar g_{ab} \, \bar p^{ab} = 0 \,,
\end{equation}
which can be used to simplify the ADM-Hamiltonian constraints to
\begin{equation}\label{ADM-Ham-const-Forappendix}
 \frac 1 {V\sqrt{\bar g}} \, \bar g_{ac} \, \bar g_{cd} \, (\bar p^{ab} - \textstyle{1 \over 3} \, \bar p \, \bar g^{ab}) (\bar p^{cd} - \textstyle{1 \over 3} \, \bar p \, \bar g^{cd} ) - \textstyle{\frac 3 8} \, \tau^2 \, V \, \sqrt{\bar g} - V^{1/3} \,, \bar R  \, \sqrt{\bar g} = 0,
\end{equation}
where $\bar R$ denotes the Ricci scalar derived from $\bar g_{ab}$. Let the solution to the CMC lapse equation (\ref{LFE})  be $N_\st{CMC}$. Then the Hamiltonian constraints imply the identity
\begin{equation}\label{equ:CMCreexpress}
\begin{aligned}
V^{1/3}\int_\Sigma {\textrm d}^3x N_\st{CMC} \bar R \sqrt{\bar{g}} = & V^{-1}\int_\Sigma {\textrm d}^3x \frac{N_\st{CMC}}{\sqrt{\bar g}}(\bar p^{ab} - \textstyle{1 \over 3} \, \bar p \, \bar g^{ab}) (\bar p_{ab} - \textstyle{1 \over 3} \, \bar p \, \bar g_{ab} ) \\
&- {\textstyle{\frac 3 8}} \tau^2 V \int_\Sigma {\textrm d}^3x N_\st{CMC} \sqrt{\bar g} \,.
\end{aligned}
\end{equation}
Using the CMC-Hamiltonian 
\begin{equation}\label{CMCHamConst}
\begin{aligned}
\mathcal H_\st{CMC} =\int _\Sigma {\textrm d}^3x N_\st{CMC}\left(\frac 1 {\sqrt{\bar g}} (\bar p^{ab} - {\textstyle{1 \over 3}} \, \bar p \, \bar g^{ab}) (\bar p_{ab} - {\textstyle{1 \over 3}} \, \bar p \, \bar g_{ab} ) -{\textstyle{\frac 3 8}}  \tau^2 V^2 \sqrt{\bar g}-V^{4/3}\bar R \sqrt{\bar g}\right) \,,
\end{aligned}
\end{equation}
 we find the equations of motion for $\tau$ as
\begin{equation}
 \begin{aligned}
   \dot \tau =& \{\tau,\mathcal H_\st{CMC}\}\\
             =&{\textstyle{\frac 3 4}}  \, V \tau^2\int_\Sigma {\textrm d}^3x N_\st{CMC}\sqrt{\bar g}+{\textstyle{\frac 4 3 }} \, V^{1/3}\int_\Sigma {\textrm d}^3x N_\st{CMC} \bar R \sqrt{\bar g}\\
            & +\int _\Sigma {\textrm d}^3x \{ \tau , N_\st{CMC}\} \left(\frac 1 {\sqrt{\bar g}} (\bar p^{ab} - {\textstyle{1 \over 3}} \, \bar p \, \bar g^{ab}) (\bar p_{ab} - {\textstyle{1 \over 3}} \, \bar p \, \bar g_{ab} ) -  {\textstyle{\frac 3 8}}  \tau^2 V^2 \sqrt{\bar g}-V^{4/3}\bar R \sqrt{\bar g}\right) \\
             =&{\textstyle{\frac 1 4}} V \tau^2\int_\Sigma {\textrm d}^3x N_\st{CMC}\sqrt{\bar g}+V^{-1}\int_\Sigma {\textrm d}^3x \frac{N_{CMC}}{\sqrt{\bar g}}(\bar p^{ab} - {\textstyle{1 \over 3}} \, \bar p \, \bar g^{ab}) (\bar p_{ab} - {\textstyle{1 \over 3}} \, \bar p \, \bar g_{ab} )  ~ \ge 0 \,,\label{mono}
 \end{aligned}
\end{equation}
where in the last line the constraint (\ref{ADM-Ham-const-Forappendix}) implies the vanishing of the term multiplied by $\{ \tau , N_\st{CMC}\} $,
and we used the identity (\ref{equ:CMCreexpress}) and the fact that $N_\st{CMC}>0$ as well as $V>0$, $\tau^2\ge 0$, $(\bar p^{ab} - \textstyle{1 \over 3} \, \bar p \, \bar g^{ab}) (\bar p_{ab} - \textstyle{1 \over 3} \, \bar p \, \bar g_{ab} ) \ge 0$ for any physical configuration.

Thus, Eq.~(\ref{mono}) shows that $\dot\tau$ is positive: York time is monotonic.

\subsection{Metriplectic Formalism: A Definition of Bulk Entropy}
\label{sec:MetriplecticFormalism} 

The metriplectic formalism introduced by Kaufman and Morrison~\cite{Kaufman,Morrison} allows one to describe dissipative systems in a language very close to the canonical formalism. To do this, one extends phase space with a new canonical variable $S$, which is a Poisson Casimir, meaning that it Poisson commutes with everything,
\begin{equation}
\ldm S, f \rdm = 0 \,,
\end{equation}
 and is usually interpreted as a (formal) entropy. To be able to describe conversion of energy into heat, one introduces an internal energy $U(S)$ and  denotes $T:=\frac{\partial U(S)}{\partial S}$, which defines a formal temperature through Clausius' relation. The total Hamiltonian is $H_{\st{tot}}=H_0 +U(S)$. The metriplectic time-evolution equations are
\begin{equation}
 \frac{d f}{d \lambda} =  \lmp f , F \rmp =\ldm f,F\rdm-\lsymb f,F\rsymb \,,
\end{equation}
where $F:=H_{\st{tot}}-S$ denotes the free energy and $\lmp .,. \rmp:=\ldm.,.\rdm-\lsymb.,.\rsymb$ is the metriplectic bracket;
$\lmp .,. \rmp$ is the sum of the dimensionless Poisson bracket and a metric bracket $\lsymb.,.\rsymb$, defined as 
\begin{equation}
\lsymb f,g \rsymb:=G^{AB} \frac{\partial f}{\partial z^A}\frac{\partial g}{\partial z^B} \,,
\end{equation}
where $z^A$ runs over the extended set of canonical variables (including $S$) and $G^{AB}$ denotes a metric on extended phase space that encodes the dissipative nature of the system. Moreover, the formalism requires  the metric bracket to be energy-conserving, that is \begin{equation} \label{RoundBracketHtot}
\lsymb H_{\st{tot}} , f \rsymb = 0  \,, 
\end{equation}
which implies, for $f=S$ (using the fact that $U$ is assumed to be a function of $S$ only),
\begin{equation} \label{RoundBracketHtotS}
\lsymb H_{\st{tot}} , S \rsymb  = \lsymb H_0 , S \rsymb + \lsymb U , S \rsymb  = \lsymb H_0 , S \rsymb+T\, \lsymb S , S \rsymb = 0 \,.
\end{equation}
Then the total energy is conserved and,  if the metric $G^{AB}$ is positive semidefinite  the entropy grows monotonically:
\begin{equation}
 \frac{d  H_{\st{tot}}}{d \lambda}=\lmp H_{\st{tot}},F \rmp= \ldm H_{\st{tot}}, H_{\st{tot}} \rdm =0 \,, \qquad   \frac{d S}{d \lambda}=\lmp S,F \rmp = \lsymb S,S\rsymb \ge 0 \,.
\end{equation}

We can of course also describe any conservative canonical system in the metriplectic formalism by choosing $G^{AB}=0$, which removes all dissipation and ensures that $S$ is non--dynamical, i.e.,  $d S/d\lambda=0$. The metriplectic equations of motion reduce for this choice to the canonical equations of motion,  $ d f/d\lambda=\lmp f,F \rmp =\ldm f,H \rdm $.

To apply this formalism to our prototype system,  we need the symmetric bracket $\lsymb  \, \cdot , \cdot \, \rsymb$
to reproduce the non-canonical term $- a\, \p^i$ in the equations of motion for $\pi^i$. This is achieved by setting
\begin{equation}
 \lsymb \p^i,S \rsymb=-a \,\p^i \,,
\end{equation}
which determines some of the elementary round brackets. In addition,  there is no
non-canonical term for the shape coordinates, so
\begin{equation}
 \lsymb s_i,S \rsymb=0 \,.
\end{equation}
The other defining  condition (\ref{RoundBracketHtot}) can be solved for a subset of the round brackets between the original phase space variables, and the bracket $\lsymb S,S \rsymb$ is determined as $-\frac 1 T \lsymb H_0,S \rsymb$ by Eq.~(\ref{RoundBracketHtotS}). Note that this procedure leaves the metric $G^{AB}$ underdetermined. This is a mild non-uniqueness, since the equations of motion are unaffected by a different choice of the dissipation $G^{AB}$. We have focused on a non-canonical transformation that makes the description of the system dimensionless, scale-invariant and autonomous. This procedure fixes the Hamiltonian. We would have encountered a qualitatively different source of non-uniqueness had we just looked for a pair $H$, $G^{AB}$ that produces the equations of motion.

\noindent\textbf{Metriplectic formalism for the 3- and $N$-body problem}

We now apply what we wrote in the last subsection to the 3-body problem. We introduce a formal entropy $S$ as a Poisson-commuting extension of phase space and postulate the production of internal energy $U(S)$ to compensate the nonconservation of the Hamiltonian:
\begin{equation}
\frac{d  U}{d \lambda} = -\frac{d H_0}{d  \lambda} = \frac{\partial H_0}{\partial \p_\theta}\p_\theta + \frac{\partial H_0}{\partial \p_\phi}\p_\phi= 2 \frac{\p_\theta^2 + \sin^{-2} \theta \,\p_\phi^2}{\p_\theta^2 + \sin^{-2} \theta \,\p_\phi^2 + \frac 1 4 } \,.
\end{equation}
Define $\Xi = \p_\theta^2 + \sin^{-2}\theta \,\p_\phi^2 + \frac 1 4 $.
The formal temperature is $T = \frac{\partial  U }{ \partial S},$ which yields
\begin{equation}
\frac{d  S}{d \lambda} =\frac 2 T \frac{\Xi - \frac 14 }{\Xi} \,.
\end{equation}
We now have to find round brackets that satisfy the requirements of the formalism,
\begin{align}
 \lsymb \p^i,S \rsymb= - \frac { \p^i } T\,, \qquad \lsymb s_i,S \rsymb=  0  \,, \qquad  \lsymb S,S \rsymb = -\frac 1 T \lsymb H_0,S \rsymb\,, \\
  \lsymb H_0 , f \rsymb = -T\, \lsymb S , f  \rsymb  \,, ~~~~  f = f(\theta,\phi,\p_\theta,\p_\phi) \,, \nonumber 
\end{align}
and produce the non--canonical equations of motion. A solution is
\begin{equation}
\hspace{-3pt}G^{AB} = \hspace{-3pt} \left(
\begin{array}{ccccc}
  0  & 0 & 0 & 0 &  0 \\
  0 &  0 & 0 & 0 &  0 \\
  0 & 0 & - \frac {T\, \Xi} 2   & 0 &   -  \p_\theta   \\
  0 & 0 & 0 & - \frac{ T\, \Xi} {2} \, \sin^2 \theta   &  - \p_\phi  \\
  0 & 0 & - \p_\theta &  - \p_\phi   &  \frac{1 - 4 \, \Xi }{2 \,T \,\Xi  } 
\end{array}\right)
\,,
\end{equation}
where the order of the entries is $\theta,\phi,\p_\theta,\p_\phi,S$.

This description of the 3-body problem on shape space can straightforwardly be generalized to the $N$-body problem on pre-shape space (i.e., the Newtonian configuration space quotiented wrt translations and scale transformations). The description of this system as a dissipative dynamics on pre-shape space is completely analogous to the 3-body problem.

\noindent\textbf{Metriplectic formalism for dynamical geometry}

\textbf{Bianchi IX}. Following the mechanism shown above, we absorb the extra $2 \, \p^i$ of Eq.~(\ref{AutonomousDissipativeEqnB9})
in the round bracket
\begin{equation} \label{RoundBracketBianchi-IX}
 \lsymb \p^i,S \rsymb =2 \, \p^i,\,\,\,i=x,y \,.
\end{equation}
The Hamiltonian $H_0$ is not preserved by the equations of motion, instead
\begin{equation}
 \frac{\partial H_0}{\partial \lambda}=2 \sum_{i=x,y} \p^i \frac{\partial H_0}{\partial \p^i}\ne 0,
\end{equation}
which is compensated by the change of internal energy $U(S)$ if we choose the remaining parts of the round bracket such that $\lmp H_{\st{tot}},F \rmp =0$. The change of $H_0$ then differs from the change of entropy only by a factor of formal temperature $T=\frac{\partial U(S)}{\partial S}$.

\textbf{Full GR}. The analogue of Eq. (\ref{RoundBracketBianchi-IX}) in full GR is
\begin{equation}
  \lsymb  \p^a{}_b(x),S \rsymb=2 \, \p^a{}_b(x) \, .
\end{equation}
In complete analogy to Bianchi IX, the Hamiltonian $H_0$ is not conserved, but 
\begin{equation}
 \frac{\partial H_0}{\partial \lambda}= 2\int_\Sigma{\textrm  d}^3x \frac{\delta H_0}{\delta \p^a{}_b(x)}\p^a{}_b(x) \ne 0,
\end{equation}
which is again compensated by the change in internal energy $U(S)$.

\noindent\textbf{Some considerations}

Let us conclude this section with the remark that we do not have a physical interpretation of the formal bulk entropy $S$. We do know that it is forced upon us by the requirement that we want to describe gravity as an autonomous dimensionless evolution of spatial {conformal} geometry, as is dictated by the shape-dynamics ontology. Moreover, we do not know whether this entropy can be related to the familiar black hole entropy of Beckenstein and Hawking. At least formally $S$, which is associated with change in the spatial conformal metric, seems more closely related to Penrose's proposal that the bulk entropy of GR should be related to the conformal geometry, but we want to caution that $S$ could be a completely unrelated entity that simply expresses the irreversibility of dynamics on shape space. It should also be noted that Penrose's proposal relates to the four-dimensional conformal geometry, while ours is tied to three-dimensional conformal geometry in a CMC foliation.

 \section*{Acknowledgements}

JB thanks Alain Albouy, Alain Chenciner, Christian Marchal, and Richard Montgomery for discussions over many years; we all thank AA, AC and CM for recent discussions; TK thanks Viqar Husain; FM and TK thank Niayesh Ashfordi; we all thank Lee Smolin, Phillipp H{\"o}hn and Brendan Foster; finally special thanks to Jerome Barkley for his numerical calculations and permission to show some of his results.
TK was supported in part through NSERC.
Perimeter Institute is supported by the Government of Canada through Industry Canada
and by the Province of Ontario through the Ministry of Economic Development and Innovation. 
This research was also partly supported by grants from FQXi and the John Templeton Foundation.

\providecommand{\href}[2]{#2}\begingroup\raggedright

\end{document}